\documentclass[aps,prl,twocolumn,showpacs,10pt,superscriptaddress,floatfix,longbibliography]{revtex4-2}

\usepackage[usenames, dvipsnames]{color}
\usepackage[normalem]{ulem}
\usepackage{amsmath}
\usepackage{float}
\usepackage{color}
\usepackage{bbold}
\usepackage{amssymb}
\usepackage{amsfonts}
\usepackage{bm}
\usepackage[edges]{forest}

\usepackage{rotating}

\usepackage[normalem]{ulem}
\usepackage{soul}
\usepackage{amssymb}
\usepackage{graphicx}

\usepackage[colorlinks, breaklinks, 
linkcolor=OrangeRed,
citecolor=RoyalBlue,
urlcolor=RoyalBlue]{hyperref}          
\usepackage{multirow}
\usepackage[all]{hypcap} 
\usepackage{xspace}
\usepackage{amssymb}
\usepackage{appendix}
\usepackage{pifont}

\usepackage{etoolbox} 
\usepackage{lipsum} 
\usepackage[capitalize]{cleveref}

\makeatletter
\appto{\appendix}{%
  \@ifstar{\def\theequation@prefix{A.}}%
          {}%
}
\makeatother

\newcommand{\ve}{\varepsilon}

\begin{document}

\title{Role of Disorder in Third-order Anomalous Hall Effect in Time-reversal Symmetric Systems}

\author{Chanchal K. Barman}
\email{arckb2@gmail.com}
\affiliation{Department of Physics, Sungkyunkwan University, Suwon 16419, Republic of Korea}
\affiliation{Dipartimento di Fisica, Universit\`a di Cagliari, Cittadella Universitaria, Monserrato (CA) 09042, Italy}

\author{Arghya Chattopadhyay}
\affiliation{Service de Physique de l’Univers, Champs et Gravitation, Université de Mons 20 Place du Parc, 7000 Mons, Belgium}

\author{Surajit Sarkar}
\email{surajit.phys1991@gmail.com}
\affiliation{Department of Physics, Concordia University, Montreal, QC H4B 1R6, Canada}
\affiliation{Univ. Grenoble Alpes, CEA, Grenoble INP, IRIG, PHELIQS, 38000 Grenoble, France}

\author{Jian-Xin Zhu}
\affiliation{Theoretical Division, Los Alamos National Laboratory, Los Alamos, New Mexico 87545, USA}
\affiliation{Center for Integrated Nanotechnologies, Los Alamos National Laboratory, Los Alamos, New Mexico 87545, USA}

\author{Snehasish Nandy}
\email{snehasish@phy.nits.ac.in}
\affiliation{Department of Physics, National Institute of Technology Silchar, Assam 788010, India}
\affiliation{Theoretical Division, Los Alamos National Laboratory, Los Alamos, New Mexico 87545, USA}

\begin{abstract}
The third-order anomalous Hall effect (TOAHE) driven by Berry connection polarizability in Dirac materials offers a promising avenue for exploring quantum geometric phenomena. We investigate the role of impurity scattering on TOAHE using the semiclassical Boltzmann framework, via a comparison of the intrinsic contributions (stemming from the Berry connection polarizability) with the extrinsic contributions caused by the disorder.
To validate our theoretical findings, we employ a generalized two-dimensional low-energy Dirac model to analytically assess the intrinsic and extrinsic contributions to the TOAHE. Our analysis reveals distinct disorder-mediated effects, including skew-scattering and side-jump contributions. We also elucidate their intriguing dependencies on Fermi surface anisotropy and discuss opportunities for experimental exploration.
\end{abstract}
\maketitle

\textcolor{blue}{\it Introduction:} The family of Hall effects, referring to a transverse voltage in response to a current applied in a sample of metal or semiconductor, have led to striking progress in searching the topological phases of matter and many
practical applications~\cite{Karplus_1954, Klitzing_1980, Haldane_1988, Kane05, Nagaosa_2010, Liu_2016, Das_2021}. Among different kinds of Hall effects, the intrinsic anomalous Hall effect (AHE)~\cite{Nagaosa_2010, Ado_2015} in linear response regime taking place without the external magnetic field has drawn tremendous theoretical and experimental attention. It is because the linear AHE serves as a smoking gun to probe the {\it Berry curvature}, a fundamental ingredient of modern topological band theory derived from the electronic wave function~\cite{Niu_2010}. However, the linear AHE appears only in time-reversal symmetry (TRS) broken systems due to the Onsager reciprocity relation~\cite{Landau_1980}. 

On the other hand, in TRS invariant systems, where linear AHE vanishes, it has been proposed that the nonlinear anomalous Hall effect (response to second-order in an applied electric field) can detect the quantum geometry of the Bloch band. Specifically, it probes the {\it first-order moment} of the Berry curvature, namely, {\it Berry curvature dipole} (BCD)~\cite{Sodemann_2015, Nandy_21}. After vigorous efforts, the BCD-induced nonlinear Hall effect has been observed experimentally as a leading-order response in bilayer~\cite{Ma_2018} and multilayer WTe$_2$~\cite{Kang_2019} and later in different materials~\cite{Ortix_2021,Hai_2021} such as oxide interface~\cite{Lesne_2023}, twisted WSe$_2$~\cite{Huang_2022} and bilayer graphene~\cite{Duan_2022}, Weyl–Kondo semimetal Ce$_3$Bi$_4$Pd$_3$~\cite{Dzsaber_2021}. Interestingly, there exists a large class of nonmagnetic materials, where both the first- and second-order Hall responses vanish, for example, a nonmagnetic material with inversion symmetry (IS) or a twofold rotational symmetry in the transport plane~\cite{Hai_2021}. This fact leads to an immediate question: how to probe the quantum geometry in this class of systems?

Recently, it has been shown within the framework of semiclassical Boltzmann formalism that the third-order anomalous Hall effect (TOAHE) driven by geometrical quantities can appear as a leading-order response in these systems regardless of TRS. However, there are two distinct origins of TOAHE associated with TRS. Specifically, in TRS invariant system, the TOAHE is induced by {\it Berry connection polarizability} (BCP), which is linked to the field-induced Berry connection~\cite{Liu_2022, Wei_2022, Nandy_2022, Xiang_2023}. Interestingly, BCP has been identified as a band-renormalized manifestation of the quantum metric~\cite{Provost_1980, Kaplan_2024}. Remarkably, TOAHE induced by BCP has been discovered in very recent experiments with bulk T$_{d}$-MoTe$_2$~\cite{Lai2021}, few-layer WTe$_2$ flakes~\cite{Min_2022}, and TaIrTe$_4$~\cite{Xiao_2022}.  Conversely, TOAHE in TRS broken system can access the {\it second-order moment of the Berry curvature}, namely, {\it Berry curvature quadrupole}~\cite{Zhang_2023, Sorn_2024}, which has been lately observed experimentally in kagome antiferromagnet FeSn~\cite{Sankar_2023}.

Despite the experimental discovery~\cite{Lai2021, Min_2022, Xiao_2022, Sankar_2023}, the complete theoretical picture of the TOAHE has not been understood yet. A key missing ingredient in the current research on TOAHE is the disorder-mediated (extrinsic) contribution. In the linear response regime, the quantitative agreement between theories and experiments on AHE depicts that the disorder-induced contributions, in particular, side-jump and skew-scattering contributions are comparably important along with the intrinsic part~\cite{Nagaosa_2010, Ado_2015}. In the case of the second-order AHE, disorder is even more important because it is a Fermi-surface quantity (usually, disorder-scattering is dominant at the Fermi surface) which is supported by the recent finding that the disorder scattering is
inevitable and enters the second-order Hall effect even in the leading order~\cite{Nandy_2019, Du_2019, Du_2021, Konig_2019}. Focusing on the nonmagnetic system in this work, it is now of immediate urgency to investigate the disorder-mediated contributions to better understand the recent discovery of BCP induced TOAHE. 

In this work, we investigate the effect of impurity scattering on the TOAHE in time-reversal symmetric systems within the semiclassical Boltzmann framework. We show analytically that both the leading-order intrinsic and extrinsic contributions (i.e., skew-scattering and side-jump scattering) to the TOAHE are linearly proportional to scattering time $\tau$. Utilizing a two-dimensional (2D) gapped Dirac model, we demonstrate that the intrinsic contribution is exclusively governed by BCP (a geometrical quantity closely related to quantum metric) and tilt parameter $t$, leading to a quadratic relationship in the lowest
order of $t$. Our analytical expression for the  skew-scattering contribution shows that it follows  sinusoidal  
angular dependency and vanishes at the band edges. Interestingly, in contrast to the skew-scattering part, we find that the side-jump contribution proportional to $\tau$ is free from the BCP and completely vanishes in TRS invariant systems.

\textcolor{blue}{\it Quasiclassical Framework:} We are mainly interested in two different kinds of contributions of the BCP-driven TOAHE in response to an external electric field: (i) intrinsic contribution; (ii) extrinsic or disorder-mediated contribution containing side-jump and skew-scattering parts.
To derive the general expression of intrinsic and disorder-mediated contributions to TOAHE, we begin with the phenomenological Boltzmann transport equation, which can be written as~\cite{Ashcroft_1976, Ziman_2001,Stephanov2012}:
\begin{equation}
\label{eq:bte}
(\partial_{t}+\mathbf{\dot{r}}.{\mathbf{\nabla_r}}+\mathbf{\dot{k}}.{\mathbf{\nabla_k}})f_{\mathbf{r},\mathbf{k},t} = I_{\text{coll}}\{f_{\mathbf{r},\mathbf{k},t}\},
\end{equation}
where $f_{\mathbf{r},\mathbf{k},t}$ denotes the local non-equilibrium electron distribution function, and $I_{\text{coll}}\{f_{\mathbf{r},\mathbf{k},t}\}$ is the collision integral, \textcolor{black}{
which accounts for various scattering processes that relax the electron distribution, including impurity scattering and intrinsic mechanisms such as electron-electron and electron-phonon interactions. However, in this work, we introduces a phenomenological scattering time $\tau$ within the relaxation time approximation.} 
The scattering time on the Fermi surface can in general have
a momentum dependence but we will ignore this dependence for simplicity. Since we are interested in the steady-state solution, we drop the time-dependence hereafter. In addition, assuming a spatially uniform electric field, we also drop the $\rm{r}$ dependence of the distribution function in Eq.~(\ref{eq:bte}). Therefore, we consider $f_{\mathbf{r},\mathbf{k},t}=f_{l}$ for the rest of the work, where $l = (n, \mathbf{k})$ is a combined index with the band index $n$ and momentum $\mathbf{k}$.

Assuming negligible electron-electron and electron-phonon interactions, the collision integral due to only the electron-impurity (static) scattering can be expressed as $I_{\text{coll}}\{f_{l}\} = - \sum_{l'}\left(\varpi_{l'l}f_l - \varpi_{ll'}f_{l'}\right)$ \cite{sinitsyn2007semiclassical, Niu_2010}. Here, $\varpi_{ll'}$ is the scattering rate from the $l$ state to the $l'$ state, relies on details of the scattering potential, and is derived by using the Fermi golden rule~\cite{sinitsyn2007anomalous} (see Supplemental Material (SM)\,\cite{suppl} for detailed derivation). It is important to note that in a noncentrosymmetric crystal, the scattering rate is not symmetric with respect to the interchange of the initial and final states. Therefore, decomposing the scattering rate into symmetric and antisymmetric parts $(\varpi_{ll'}=\varpi_{ll'}^{\text{sym}}+\varpi_{ll'}^{\text{asym}}$), we note that the antisymmetric component gives rise to the skew-scattering contribution to the TOAHE, where exchanging the incoming and outgoing states leads to a change in sign~\cite{SMIT1955877}. 
On the other hand, the symmetric part $\varpi_{ll'}^{\text{sym}}$ contains both intrinsic and side-jump scattering contributions. In particular, the side-jump scattering arises from alterations in the direction of motion of the wave packet and a shift in coordinates that occur during an impurity scattering process, while intrinsic scattering is a result of the finite non-trivial Berry curvature and its higher-order moments~\cite{Berger_prb70}. Therefore, we decompose the collision term into intrinsic, side-jump, and skew-scattering components: $I_{\text{coll}}\{f_{l}\}=I_{\text{coll}}\{f_{l}\}^{\text{in}}+I_{\text{coll}}\{f_{l}\}^{\text{sj}}+I_{\text{coll}}\{f_{l}\}^{\text{sk}}$ (See SM Note 1) \cite{suppl}.

Since the current work is concentrating on the TOAHE, we employ second-order semiclassical theory that includes a first- (second-) order electric field correction to the Berry curvature (band energy) and modifies the relation between the physical position and crystal momentum of wave-packet with regard to the canonical ones. In this regard, the semiclassical equations of electron motion in the absence of an external magnetic field can be written in the following form~\cite{Niu_1999}:
\begin{eqnarray}
\dot {\mathbf r_l} &=& \frac{1}{\hbar}\nabla_{\bm k}\tilde{\ve}_{l}
-\dot{\mathbf{k}} \times {\bm \tilde{\Omega}_{l}}+\mathbf{v}_l^{\text{sj}},\quad \hbar \dot {\mathbf k}= e {\mathbf E}~.
\label{eq_rdot}
\end{eqnarray} 
Here $e<0$, $\tilde{\ve}_{l}=\sum_{i=0}^{2}{\ve}_{l}^{(i)}$, ${\bm \tilde{\Omega}_{l}}=\bm{\nabla_{\mathbf{k}}} \times \sum_{i=0}^{1}\bm{A}_{l}^{(i)}$, and $\mathbf{v}_l^{\text{sj}}$ denotes the side-jump velocity, which describes the transverse velocity resulting from the transverse coordinate shift of the wave packet in response to scattering by an impurity potential. 
With the unperturbed Bloch  band energy ${\ve}_{l}^{(0)}$ and Bloch eigenfunction $|u^{(0)}_{l}\rangle$ (the cell-periodic part),  the  unperturbed  intraband  
Berry connection is given by $\bm{A}_{l}^{(0)}(\mathbf{k})=\langle u^{(0)}_{l}|i\bm{\nabla_{\mathbf{k}}}|u^{(0)}_{l}\rangle$.  Interestingly, the $a^{th}$ component $a\in\{x,y,z\}$ of the first-order Berry connection $A_{l,a}^{(1)}$ and second-order band energy $\ve^{(2)}_l$ 
are directly related to the purely geometric quantity BCP tensor ($G$) following the relation: $A_{l,a}^{(1)}=G_{l,ab}E_b$ and $\ve^{(2)}_l=e^2 E_a G_{l,ab}E_b/2$ respectively. It is important to note that the first-order correction to the band energy $\ve^{(1)}_l$ is independent of momentum and therefore, acts like potential energy, leading
to an overall shift of the energy~\cite{Gao_2014, Nandy_2022} with no contribution to the velocity.

The first-order correction to the Berry connection, $A_{l,a}^{(1)}$, measuring a shift in its center of
mass position of the wave packet gives the positional shift for the band $l$. It is crucial to emphasize the relationship between BCP ($G_{l,ab}$) and the quantum metric $\mathcal{Q}_{l,ab}$, given their significant contributions to the nonlinear Hall effect \cite{gao2023quantum,wang2023quantum}. The quantum metric tensor ($\mathcal{Q}_{l,ab}= \text{Re}\sum_{l'\neq l}  \bm{A}_{l,l'}^{(0)} \bm{A}_{l',l}^{(0)} $) is intricately connected to the interband Berry connection ($\bm{A}_{l,l'}^{(0)}$) associated with unperturbed states. Remarkably, the BCP is revealed to be a band-renormalized quantity of the quantum metric, expressed succinctly as $G_{l,ab}=2\text{Re}\sum_{l'\neq l}  \frac{\bm{A}_{l,l'}^{(0)} \bm{A}_{l',l}^{(0)}}{\ve^{(0)}_l - \ve^{(0)}_{l'}}$, underlining its crucial connection to energy differences between bands.

 To investigate the disorder-mediated contributions to the TOAHE, we consider static nonmagnetic impurity that involves randomly positioned delta-function scatterers: $V_{\text{imp}}(\mathbf{r}) = \sum_{i} V_i \delta(\mathbf{r-R_i})$ with $\mathbf{R_i}$ random location of the scatterers and $V_i$ is the disorder strength satisfies $\langle V_i\rangle_{\text{dis}} = 0$, $\langle V_i^2\rangle_{\text{dis}} = V_0^2 \neq 0$, $\langle V_i^3\rangle_{\text{dis}} = V_1^3 \neq 0$ \cite{sinitsyn2007anomalous}.
 Now solving the phenomenological Boltzmann transport equation in response to an applied uniform ac electric field $\mathbf{E}_a=\rm Re[\mathcal{\xi}_a e^{i\omega t}]$ ($\mathcal{\xi}$ is the amplitude and $\omega$ is the frequency), the third-order current can be obtained as $j^{(3)}_{a}=\chi_{abcd} E_b E_c E_d$, where the third-order conductivity tensor
 is given by $\chi_{abcd}=\chi_{abcd}^{\text{in}}+\chi_{abcd}^{\text{sj}}+\chi_{abcd}^{\text{sk}}$ (see SM Note) \cite{suppl}. Here, $\chi_{abcd}^{\text{in}}$, $\chi_{abcd}^{\text{sj}}$, and $\chi_{abcd}^{\text{sk}}$ are the contributions arising from the intrinsic, side-jump, and skew scattering of electronic wave-packet with the impurity.

\textcolor{blue}{\it Symmetry Requirements and Candidate Materials:} The general expression for the third-order current in response to an applied electric field (summation over repeated indices is implied), $j^{(3)}_{a}=\chi_{abcd} E_b E_c E_d$, is determined by a fourth-rank tensor $\chi$ containing both longitudinal and transverse
third-order current response. Considering the current and electric fields transform as polar vectors under coordinate changes, the power supplied by the electric field on the electronic fluid is the scalar $P=j_a E_a$ implying that the symmetric part of the conductivity tensor will contribute to the power or dissipation, whereas anti-symmetric part or Hall component remains dissipationless. Since in this work, we are interested in third-order Hall response, we wish to isolate the dissipationless component of the conductivity tensor $\chi_{abcd}$ by antisymmetrizing the first
index with either the second or third or fourth. These three choices of antisymmetrization are equivalent construction. Crystalline symmetries play a crucial role in the observation of third-order Hall conductivity. Specifically, for the rank-4 TOAHE conductivity tensor ($\chi_{abcd}$), the constraints imposed by a symmetry operation $R$ can be expressed as $\chi_{abcd} = R_{aa'}R_{bb'}R_{cc'}R_{dd'}\chi_{a'b'c'd'}$, where $R_{aa'}$ represents the matrix element of the symmetry operation $R$. Notably, this relation indicates that mirror symmetry $\mathcal{M}_a$ along the $a$-axis enforces the condition $\chi_{aaab} = 0 = \chi_{baaa}$. 
Moreover, the components of $\chi_{abcd}^{\gamma}$ ($\gamma\in\{\text{in, sj, sk}\}$) that are allowed or disallowed by symmetry considerations under fundamental symmetries such as mirror $\mathcal{M}$, rotation $C_n$, and inversion $\mathcal{P}$, are summarized in Table\,\ref{table1}.

\begin{table}[t]
\begin{ruledtabular}
\caption {Tensor components governing $\chi_{\perp}^{\gamma}$ under various symmetries. The \text{\ding{51}} (\text{\ding{55}}) symbol indicates that the corresponding tensor component is allowed (forbidden) under the specified symmetry. Here, $\chi_{12}^{\gamma}=\frac{1}{3}(\chi_{xxyy}^{\gamma} + \chi_{xyxy}^{\gamma} + \chi_{xyyx}^{\gamma}),~\chi_{21}^{\gamma}=\chi_{12}^{\gamma}(x\leftrightarrow y)$.}
\label{table1}
\begin{tabular}{c c c c c c }
 & $\mathcal{P} $ & $\mathcal{M}_x, \mathcal{M}_y, \mathcal{M}_z$ & $C_{2x}, C_{2y}$ & $C_{2z},C_{4z}$ & $C_{3z},C_{6z} $ \\ \hline
 $\chi_{abbb}$ ($a,b \in x,y$) & \ding{51} & \ding{55},\ding{55},\ding{51} & \ding{55},\ding{55} & \ding{51},\ding{51} & \ding{51},\ding{51} \\
 $\chi_{aaaa}$ ($a \in x,y$) & \ding{51} & \ding{51},\ding{51},\ding{51} & \ding{51},\ding{51} & \ding{51},\ding{51} & \ding{51},\ding{51} \\
 $\chi_{ab}$ ($a,b \in 1,2$) & \ding{51} & \ding{51},\ding{51},\ding{51} & \ding{51},\ding{51} & \ding{51},\ding{51} & \ding{51},\ding{51} \\
\end{tabular}
\end{ruledtabular}
\end{table}

\textcolor{blue}{\it Third-order Hall Effect in Dirac Fermion:} We consider a generic model of tilted 2D Dirac cones, which captures the low-energy properties of various Dirac materials such as the surface of topological crystalline insulators and strained transition-metal dichalcogenides. The low-energy effective Hamiltonian can be written as~\cite{Sodemann_2015, Nandy_2019}
\begin{eqnarray}\label{diracHam}
H(\mathbf{k}) = v_{x}k_{x}\sigma_{y} + s v_{y} k_{y}\sigma_{x} + t_s k_{x} \sigma_{0} + \Delta \sigma_{z},
\end{eqnarray}
where $s=\pm 1$ \textcolor{black}{denotes the valley index}, $v_x$ and $v_y = \lambda v_x$ are the $x$ and $y$-component of the Fermi velocity, $\Delta$ is the gap, and $t_s=st$ with $\emph{t}$ representing the tilt parameter along the $k_x$ direction. \textcolor{black}{We note that in Eq.\,\eqref{diracHam}, the TRS operation connects one valley to the other, rendering the full Hamiltonian TRS invariant when both valleys are considered. The corresponding TRS operator is given by $\mathcal{T} = \sigma_0 \mathcal{K}$, where $\mathcal{K}$ denotes complex conjugation.}
The energy dispersion is obtained as $\ve^{\pm}_{\mathbf{k}}=s t k_x \pm \sqrt{v_x^2 k_x^2 + v_y^2 k_y^2+ \Delta^2}$ where $\pm$ represents conduction and valence band respectively. These two Dirac cones are related by the TRS. The Hamiltonian breaks the inversion symmetry while preserving the only mirror symmetry $\mathcal{M}_y (y \rightarrow -y)$. Here, $C_3$ symmetry is broken due to the presence of tilt as well as anisotropic velocities. It is important to note that the linear AHE will vanish in this system due to the presence of TRS while the second-order response could be finite~\cite{Sodemann_2015}. Interestingly, it has been shown that in the absence of tilting (i.e., $t=0$), the BCD vanishes and consequently, so does second-order AHE~\cite{Sodemann_2015, Nandy_2019}. Therefore, the third-order anomalous Hall response will emerge as a leading-order response in these systems.


To investigate the BCP induced TOAHE response we consider a planar setup where $\mathbf{E}=E(\cos\theta,\sin\theta)$ field forms a polar angle $\theta$ with the crystal's mirror axis along $x$-direction. 
In this setup, the in-plane third-order anomalous Hall conductivity ($\chi_{\perp}^{\gamma}$) can be calculated as~\cite{Lai2021},
\begin{eqnarray}\label{eq:TOAHE}
   \chi_{\perp}^{\gamma}\left(\theta\right)  &=& \left(-\chi_{xxxx}^{\gamma} + 3\chi_{21}^{\gamma}\right)\cos^3(\theta)\sin(\theta) \notag \\
  && + \left(\chi_{yyyy}^{\gamma} - 3\chi_{12}^{\gamma} \right)\cos(\theta)\sin^3(\theta),
\end{eqnarray}
where $\chi_{12}^{\gamma}=\frac{1}{3}(\chi_{xxyy}^{\gamma} + \chi_{xyxy}^{\gamma} + \chi_{xyyx}^{\gamma}),~\chi_{21}^{\gamma}=\chi_{12}^{\gamma}(x\leftrightarrow y)$. \textcolor{black}{The nonlinear anomalous Hall response in Eq.~\eqref{eq:TOAHE} arises from a Fermi surface distortion induced by an external electric field (see SM \cite{suppl}). Sustaining the resulting steady-state current requires relaxation via scattering, leading to a characteristic $\tau$ or $\tau^3$ dependence. This contrasts with the linear anomalous Hall effect, which originates from the Berry curvature of occupied bands--i.e., Fermi sea contributions--and remains independent of the scattering time $\tau$~\cite{Nagaosa_2010}.}

Considering the 2D Dirac model in Eq.~\eqref{diracHam}, we have conducted an analytical assessment of $\chi_{\perp}^{\gamma}\left(\theta\right)$, as detailed in SM \cite{suppl}. The analytical expression of the intrinsic contribution of TOAHE for each massive Dirac cone is given in Eq.~(\ref{chi:intrin}), where we have renormalized $t$, $\Delta$, and $v_x$ as $\tilde{t}=t/\mu$, $\tilde{\Delta} = \Delta/\mu$, and $\tilde{v}_x = v_x/\mu$ respectively for simplicity and $\mu$ is the chemical potential. 

\begin{widetext}
\begin{equation}\label{chi:intrin}
\begingroup
\fontsize{9.5pt}{11pt}\selectfont  
\begin{aligned}
    \chi_{\perp}^{\text{in}}\left(\theta\right)  &= \frac{\tau}{\lambda  \mu}\left[\left(\lambda^2-1\right) \tilde{v}_x^2 \mathcal{F}_1(\tilde{\Delta}) + \tilde{t}^2 \mathcal{G}_1(\lambda,\tilde{\Delta})\right] \cos^3(\theta)\sin(\theta)  + \frac{\tau}{  \mu}\left[\lambda\left(\lambda^2-1\right) \tilde{v}_x^2 \mathcal{F}_2(\tilde{\Delta}) +  \lambda~\tilde{t}^2 \mathcal{G}_2(\lambda,\tilde{\Delta}) \right]\cos(\theta)\sin^3(\theta)
\end{aligned}
\endgroup
\end{equation}
\end{widetext}

The functional dependencies of $\mathcal{F}_{1,2}(\tilde{\Delta})$, $\mathcal{G}_{1,2}(\lambda,\tilde{\Delta})$ in Eq.~\eqref{chi:intrin} are given in SM Note 8 \cite{suppl}. Since each Dirac cone
produces an identical contribution to $\chi_{\perp}^{\gamma}\left(\theta\right)$, a factor of 2 will be multiplied for the total contribution. Interestingly, the intrinsic part vanishes parallel (with
$\theta=0$) and perpendicular (with $\theta=\pi/2$) to the mirror line. Although $\chi_{\perp}^{\text{in}}$ displays both linear and cubic dependencies on $\tau$ (see SM Note 6--8) \cite{suppl}, the linear in $\tau$ contribution is purely arising from the BCP while the $\tau^3$ contribution originates from the group velocity and is therefore, treated as `semiclassical' contribution. However, we focus on the linear in $\tau$ contribution as it dominates the behavior which is influenced by two key parameters: anisotropic velocity ratio $\lambda = v_y/v_x$ and the tilt parameter $t$. It is important to note that BCP-induced TOAHE is very different from the Berry curvature quadrupole mediated TOAHE which scales as $\tau^2$ in leading order\,\cite{Zhang_2023}. The latter one vanishes in TR symmetric systems and is free from the field-induced correction of band energy and Berry curvature.


Equation~(\ref{chi:intrin}) is one of our main results and has some striking features. {\it First,} the intrinsic contribution {\it vanishes} in the absence of tilt for isotropic case ($\lambda=1$). {\it Second},  $\chi_{\perp}^{\text{in}}\left(\theta\right)$ in the isotropic system is exclusively governed by the tilt parameter, leading to a {\it quadratic} relationship in the lowest order of $t$. {\it Third,} it is clear from the equation that  $\chi_{\perp}^{\text{in}}\left(\theta\right)$ is directly proportional to $(\lambda^2 - 1)$. This would have immense importance in the case of switching applications. This is because in the limit $t \ll v_x$, the sign of $\chi_{\perp}^{\text{in}}\left(\theta\right)$ can be altered by tuning the $\lambda$, in particular, for $\lambda < 1$. Moreover, this fact helps to distinguish TOAHE from first- and second-order AHE~\cite{Nagaosa_2010, Sodemann_2015}.

\begin{eqnarray} \label{chi:skew}	\chi_{\perp}^{\text{sk},1}\left(\theta\right) &=&  \frac{\tau\tilde{t}^2 \tilde{\Delta}^2 \left(\tilde{\Delta}^2 - 1\right)^2 \sin(2\theta)N_i} {1024\pi ~\tilde{v}_x^4 } \times \nonumber \\
&&\left[ \frac{\left(1 - 5\tilde{\Delta}^2 \right)V_1^3} {2 \mu^2} + \frac{ \left(1 - 4\tilde{\Delta}^2 \right)N_i V_0^4} {\mu^3} \right]
\end{eqnarray}

\begin{figure*}[htb!]
\centering
\includegraphics[width=0.95\linewidth]{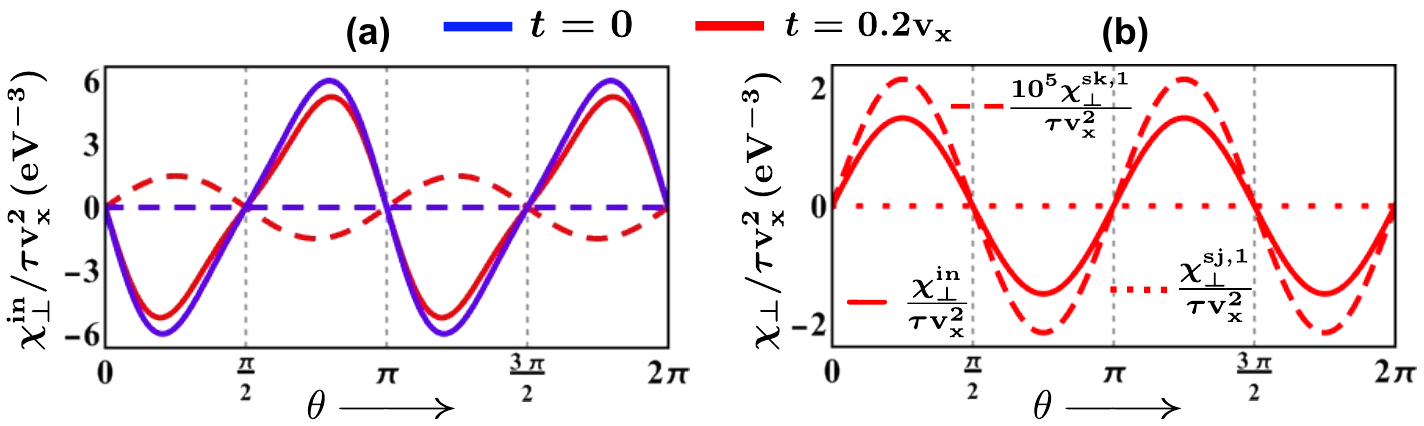}
\caption{Third-order anomalous Hall conductivity ($\chi_{\perp}^{\gamma}$) with a linear dependence on $\tau$. Figure (a) depicts the intrinsic component $\chi_{\perp}^{\text{in}}$ as a function of $\theta$ for $\lambda=1$ (dashed lines), and $\lambda=0.7$ (solid lines). Figure (b) illustrates the skew and side-jump scattering components $\chi_{\perp}^{\text{sk},1}$ and $\chi_{\perp}^{\text{sj},1}$ in comparison with $\chi_{\perp}^{\text{in}}$ for $\lambda=1$ and $t=0.2 v_x$ (see legends). The red (blue) color code in Figs. (a-b) corresponds to $t=0.2 v_x$ ($t=0$). In both (a) and (b) we have taken $\Delta=0.02$ eV, $\mu=0.1$ eV. \textcolor{black}{In panel (b), the skew scattering term is plotted as $\frac{10^{5}\,\chi_{\perp}^{\text{sk},1}}{\tau v_x^2}$ to make its contribution visible, as it is significantly smaller than the intrinsic part.}} 
\label{fig:chi}
\end{figure*}

Turning focus on the disorder-mediated contribution to TOAHE, we first consider the skew-scattering case. Similar to the intrinsic part, the skew-scattering contribution also encompasses both linear ($\chi_{\perp}^{\text{sk},1}$) and cubic ($\chi_{\perp}^{sk,2}$) terms with respect to the scattering time, and both components are contingent on the Gaussian ($V_0$) and non-Gaussian ($V_1$) components of the disorder strength (SM Note 12--13) \cite{suppl}. 
The contribution proportional to $\tau$ due to skew-scattering for isotropic system ($\lambda=1$) is presented in Eq.~(\ref{chi:skew}), representing another principal outcome of this work. It is clear from the equation that $\chi_{\perp}^{\text{sk},1}$ follows $\sin 2\theta$ angular dependence. Our calculation reveals that $\chi_{\perp}^{\text{sk},1}$ is directly proportional to $(\tilde{\Delta}^2 - 1)$ ensuring that the skew scattering contribution vanishes when the chemical potential is in the gap of the massive Dirac fermions. Furthermore, $\chi_{\perp}^{\text{sk},1}$ exhibits $t^2$ relationship similar to the intrinsic case.

Finally, we investigate the side-jump scattering case (see SM Note 9--11 for details) \cite{suppl}. It is important to note that the side-jump contribution stems from two distinct sources: side-jump velocity ($\chi_{\perp}^{\text{sj},1}$) and the side-jump scattering effect ($\chi_{\perp}^{\text{sj},2}$). Notably, $\chi_{\perp}^{\text{sj},1}$ contains both linear and cubic in $\tau$ dependency. 
Interestingly, both the components in $\chi_{\perp}^{\text{sj},1}$ vanishes in TRS invariant system. On the contrary, $\chi_{\perp}^{\text{sj},2}$ shows a quadratic relationship with respect to $\tau$ (see SM Note 11) \cite{suppl}. In contrast to the intrinsic and skew-scattering components, remarkably, we find that the side-jump contribution $\chi_{\perp}^{\text{sj},2}$ is independent of BCP and comes from the unperturbed Berry curvature. It is noted that the BCP contribution can only affect the $\chi_{\perp}^{\text{sj},2}$ component in the fourth-order anomalous Hall effect. However, since we are interested in the linear in $\tau$ contribution, we have disregarded $\chi_{\perp}^{\text{sj},2}$ in this study. 

\textcolor{black}{It is important to note that both the intrinsic and extrinsic contributions to the conductivity comprise two distinct components: a term linear in $\tau$, arising from the BCP, and a term proportional to $\tau^3$, originating from the wave-packet group velocity.} The leading contribution (linear in $\tau$) of the third-order anomalous Hall conductivity ($\chi_{\perp}^{\gamma}$) as a function of $\theta$ is depicted in Fig.\,\ref{fig:chi}.  For a moderate strength of disorder ($N_i V_0^2 = 10^2$ eV$^2$\AA$^2$ and $N_i V_1^3 = 10^4$ eV$^3$\AA$^4$, with $N_i$ denoting the disorder concentration), it is evident that $\chi_{\perp}^{\text{sk},1}/\tau$ is significantly smaller than $\chi_{\perp}^{\text{in}}/\tau$, as illustrated in Fig.~\ref{fig:chi}. For the intrinsic part, our analytical results perfectly match with the numerical results as well as earlier study~\cite{Liu_2022}. We would also like to point out that the Berry curvature-mediated first-order, BCD-induced second-order, and BCP-driven third-order responses can appear simultaneously in experiments for a system with broken TRS and IS. However, these responses can easily be separated from each other via frequency lock-in ac measurements, specifically, by measuring second-harmonic and third-harmonic Hall resistance.

\textcolor{blue}{\it Conclusions:} We have demonstrated the impact of disorder-mediated scattering on the third-order anomalous Hall effect driven by Berry connection polarizability, an aspect not considered in the earlier studies~\cite{Liu_2022, Nandy_2022}. Using the semiclassical Boltzmann formalism, we have investigated skew-scattering and side-jump contribution to the TOAHE along with the intrinsic one. Our analytical calculations elucidate a quadratic dependency of $\chi_{\perp}^{\gamma}/\tau$ on the tilt parameter for a tilted 2D Dirac model.
Interestingly, our findings reveal that the side-jump contribution does not manifest in the linear in $\tau$ order in the time-reversal invariant system. Conversely, the skew-scattering contribution remains finite and exhibits a $\sin 2\theta$ angular dependency.

\textcolor{blue}{\it Acknowledgements:} 
C.K.B. acknowledges the Department of Physics at Sungkyunkwan University for providing the necessary computing facilities.
S.S. was also supported by the Horizon Postdoctoral fellowship from Concordia University.
We acknowledge Dr. Sumanta Tewari and Dr. Hridis K. Pal for insightful discussions.
The work at Los Alamos National Laboratory was carried out under the auspices of the US Department of Energy (DOE) National
Nuclear Security Administration under Contract No.
89233218CNA000001. It was supported by the LANL
LDRD Program, and in part by the Center for Integrated Nanotechnologies, a DOE BES user facility, in partnership with the LANL Institutional Computing Program for computational resources.

\bibliography{TOHE_Disorder}
\end{document}


\title{Supplemental Material for ``Role of Disorder in Third-order Anomalous Hall Effect in Time-reversal Symmetric Systems"}

\author{Chanchal K. Barman}
\email{arckb2@gmail.com}
\affiliation{Department of Physics, Sungkyunkwan University, Suwon 16419, Republic of Korea}
\affiliation{Department of Physics, University of Cagliari, Monserrato, CA 09142-I, Italy}

\author{Arghya Chattopadhyay}
\affiliation{Service de Physique de l’Univers, Champs et Gravitation, Université de Mons 20 Place du Parc, 7000 Mons, Belgium.}

\author{Surajit Sarkar}
\email{surajit.phys1991@gmail.com}
\affiliation{Department of Physics, Concordia University, Montreal, QC H4B 1R6, Canada}
\affiliation{Univ. Grenoble Alpes, CEA, Grenoble INP, IRIG, PHELIQS, 38000 Grenoble, France}

\author{Jian-Xin Zhu}
\affiliation{Theoretical Division, Los Alamos National Laboratory, Los Alamos, New Mexico 87545, USA}
\affiliation{Center for Integrated Nanotechnologies, Los Alamos National Laboratory, Los Alamos, New Mexico 87545, USA}

\author{Snehasish Nandy}
\email{snehasish@phy.nits.ac.in}
\affiliation{Department of Physics, National Institute of Technology Silchar, Assam 788010, India}
\affiliation{Theoretical Division, Los Alamos National Laboratory, Los Alamos, New Mexico 87545, USA}

\date{\today}

\maketitle

\tableofcontents

\clearpage
\subsection{Supplementary Note 1: Decomposition of the collision integral}

The spatially homogeneous Boltzmann transport equation is expressed as:

\begin{eqnarray*}
    \frac{\partial f_l}{\partial t} + \dot{\mathbf{k}}\cdot \frac{\partial f_l}{\partial \mathbf{k}}  = \mathcal{I}_{coll}\left\lbrace f_l\right\rbrace,
\end{eqnarray*}

where the total collision integral $\mathcal{I}_{coll}\left\lbrace f_l\right\rbrace$ describes elastic disorder scattering caused by static defects or impurities. In terms of scattering rate $\varpi_{ll'}$, evaluated using Fermi's golden rule\,\cite{dirac1927quantum} as $\varpi_{ll'} \equiv \frac{2\pi}{\hbar} |T_{ll'}|^2 \delta\left(\varepsilon_l - \varepsilon_{l'} \right)$, the total collision integral can be expressed as

$$\mathcal{I}_{coll}\left\lbrace f_l\right\rbrace = -\sum_{l'}   \left( \varpi_{l'l} f_l - \varpi_{ll'} f_{l'}  \right).$$

This expression captures the impact of elastic disorder scattering on the distribution function $f_l$, where $\varpi_{ll'}$ represents the scattering rate between states 
$l$ and $l'$, and $T_{ll'}$ denotes the transition matrix element \cite{mahan1990many}. Here, $l = (n, \mathbf{k})$ is a combined index with the band index $n$ and momentum $\mathbf{k}$.

The elastic disorder scattering can be further dissected into intrinsic, side-jump, and skew-scattering components as,

$$ \mathcal{I}_{coll}\left\lbrace f_l\right\rbrace  = \mathcal{I}_{coll}^{in}\left\lbrace f_l\right\rbrace + \mathcal{I}_{coll}^{sj}\left\lbrace f_l\right\rbrace + \mathcal{I}_{coll}^{sk}\left\lbrace f_l\right\rbrace.$$

The intrinsic part arises from symmetric scatterings, where incoming and outgoing states are reversible in a scattering event. In contrast, the side-jump component is a consequence of coordinate shifts during scattering processes. Finally, the skew-scattering part is attributed to anti-symmetric scatterings, where exchanging the incoming and outgoing states introduces a minus sign. This decomposition provides a nuanced understanding of the distinct contributions of intrinsic, side-jump, and skew-scattering phenomena in the context of elastic disorder scattering. In terms of symmetric ($\varpi_{ll'}^{sym}$) and anti-symmetric ($\varpi_{ll'}^{asym}$) parts of scattering rate $\varpi_{ll'}$, one can write

\begin{eqnarray*}
 \mathcal{I}_{coll}\left\lbrace f_l\right\rbrace &=& -\sum_{l'}   \left( \varpi_{l'l} f_l - \varpi_{ll'} f_{l'}  \right) = -\sum_{l'}   \left[ \left(\varpi_{ll'}^{sym} + \varpi_{ll'}^{asym}\right) f_l - \left(\varpi_{l'l}^{sym} + \varpi_{l'l}^{asym}\right) f_{l'}  \right] \\
  &=& -\sum_{l'} \varpi_{ll'}^{sym} \left(f_l - f_{l'}\right) -\sum_{l'} \varpi_{ll'}^{asym} \left(f_l + f_{l'}\right) = \mathcal{I}^{sym}_{coll}\left\lbrace f_l\right\rbrace + \mathcal{I}^{sk}_{coll}\left\lbrace f_l\right\rbrace. 
\end{eqnarray*}

Taking into account the work done by the electric field as an electron gets displaced within the unit cell during the collision, the scattering rate is modified as:

\begin{eqnarray*}
\varpi_{ll'}^{sym} &\Rightarrow& \tilde{\varpi}_{ll'}^{sym} = \frac{2\pi}{\hbar} |T_{ll'}|^2 \delta\left(\ve_l - \ve_{l'} + e \mathbf{E}\cdot \delta \mathbf{r}_{ll'}\right).
\end{eqnarray*}

Here, $\delta \mathbf{r}_{ll'}$ represents the coordinate shift, commonly referred to as the side-jump \cite{sinitsyn2007semiclassical}. Expanding the $\delta$-function upto first order in $\delta \mathbf{r}_{ll'}$, we can express it as:
\begin{eqnarray*}
\delta\left(\varepsilon_l - \varepsilon_{l'} + e \mathbf{E}\cdot \delta \mathbf{r}_{ll'} \right) &\simeq& \delta\left(\varepsilon_l - \varepsilon_{l'}\right) + e \mathbf{E}\cdot \delta \mathbf{r}_{ll'} \frac{\partial}{\partial \varepsilon_l} \delta\left(\varepsilon_l - \varepsilon_{l'}\right) \\
&=&  \delta\left(\varepsilon_l - \varepsilon_{l'}\right) - e \mathbf{E}\cdot \delta \mathbf{r}_{ll'} \frac{\partial}{\partial \varepsilon_{l'}} \delta\left(\varepsilon_l - \varepsilon_{l'}\right). 
\end{eqnarray*}

Now, the modified scattering rate, $\tilde{\varpi}_{ll'}^{sym}$, can be re-written as:

\begin{eqnarray*}
\tilde{\varpi}_{ll'}^{sym} &=& \frac{2\pi}{\hbar} |T_{ll'}|^2 \delta\left(\varepsilon_l - \varepsilon_{l'}\right) + e \mathbf{E}\cdot \frac{2\pi}{\hbar} |T_{ll'}|^2 \delta \mathbf{r}_{ll'} \frac{\partial}{\partial \varepsilon_{l}} \delta\left(\varepsilon_l - \varepsilon_{l'}\right) \\
&=& \varpi_{ll'}^{sym} + e \mathbf{E}\cdot \mathbf{O}_{ll'},
\end{eqnarray*}

where $$\mathbf{O}_{ll'} = \frac{2\pi}{\hbar} |T_{ll'}|^2 \delta \mathbf{r}_{ll'} \frac{\partial}{\partial \varepsilon_{l}} \delta\left(\varepsilon_l - \varepsilon_{l'}\right).$$

Thus the symmetric collision term can be written as --
\begin{eqnarray*}
\mathcal{I}^{sym}_{coll}\left\lbrace f_l\right\rbrace &=&  -\sum_{l'} \tilde{\varpi}_{ll'}^{sym} \left(f_l - f_{l'}\right) 
=  -\sum_{l'} \varpi_{ll'}^{sym} \left(f_l - f_{l'}\right) - e \mathbf{E}\cdot\sum_{l'} \mathbf{O}_{ll'} \left(f_l - f_{l'}\right) \notag \\
&=& \mathcal{I}^{in}_{coll}\left\lbrace f_l\right\rbrace + \mathcal{I}^{sj}_{coll}\left\lbrace f_l\right\rbrace.
\end{eqnarray*}
 
 Thus, the elastic collision term ($\mathcal{I}_{coll}\left\lbrace f_l\right\rbrace $) has been approximately decomposed into the intrinsic, side-jump and skew-scattering parts, reflecting the distinct contributions and effects of each component in the presence of the electric field and electron displacement within the unit cell during collisions.

Now, to solve the Boltzmann equations up to the
third order of $\mathbf{E}$, we decompose the distribution function as,

\begin{eqnarray*}
f_l &=& f_l^{in} + \delta f_l^{sj} + \delta f_l^{sk}
\end{eqnarray*}

and thus, the standard Boltzmann equation takes the form:

\begin{eqnarray*}
\left( \partial_t + \dot{\mathbf{k}}\cdot \partial_{\mathbf{k}} \right)  \left( f_l^{in} + \delta f_l^{sj} + \delta f_l^{sk} \right) &=& \mathcal{I}^{in}_{coll} \left\lbrace f_l\right\rbrace + \mathcal{I}^{sj}_{coll} \left\lbrace f_l\right\rbrace + \mathcal{I}^{sk}_{coll} \left\lbrace f_l\right\rbrace. 
\end{eqnarray*}

Neglecting terms with mixed side-jump and skew-scattering contributions, the standard Boltzmann equation can be approximately decomposed into three equations:

\begin{eqnarray}
\left( \partial_t + \dot{\mathbf{k}}\cdot \partial_{\mathbf{k}} \right) f_l^{in} &=& \mathcal{I}^{in}_{coll} \left\lbrace f_l^{in}\right\rbrace \label{BEqintrinsic}\\
\left( \partial_t + \dot{\mathbf{k}}\cdot \partial_{\mathbf{k}} \right) \delta f_l^{sj} &=& \mathcal{I}^{in}_{coll} \left\lbrace \delta f_l^{sj}\right\rbrace + \mathcal{I}^{sj}_{coll} \left\lbrace f_l^{in}\right\rbrace \label{BEqSJ}\\
\left( \partial_t + \dot{\mathbf{k}}\cdot \partial_{\mathbf{k}} \right) \delta f_l^{sk} &=& \mathcal{I}^{in}_{coll} \left\lbrace \delta f_l^{sk}\right\rbrace + \mathcal{I}^{sk}_{coll} \left\lbrace f_l^{in}\right\rbrace. \label{BEqSK}
\end{eqnarray}

These equations account for the intrinsic, side-jump, and skew-scattering components in the distribution function. They will be utilized to derive expressions for the current up to the third-order response to the alternating current (ac) electric field in the subsequent sections, providing a comprehensive understanding of the electron dynamics in the presence of scattering phenomena.

\subsection{Supplementary Note 2: Gapped 2D Dirac Hamiltonian}
In this section we briefly review the scattering matrices and co-ordinate shifts of the scattered wave considering a tilted and gapped 2D Dirac fermionic system as given in Eq.\,\eqref{diracHam}.

\begin{eqnarray}\label{diracHam}
H(\mathbf{k}) = v_{x}k_{x}\sigma_{y} +  s v_{y} k_{y}\sigma_{x} + t_s k_{x} \sigma_{0} + \Delta \sigma_{z},
\end{eqnarray}

where $s=\pm 1$ denotes the valley index, $v_x$ and $v_y = \lambda v_x$ are the $x$ and $y$-component of the Fermi velocity, $\Delta$ is the gap, and $t_s=st$ with $\emph{t}$ representing the tilt parameter along the $k_x$ direction. The Chiral eigenstates that diagonalize the above 2D Dirac Hamiltonian are, $|\psi_{\mathbf{k}}^{\pm}\rangle = \frac{e^{i\mathbf{k}\cdot \mathbf{r}}}{\sqrt{V}}|u_{\mathbf{k}}^{\pm}\rangle$, where $\pm$ indicates the conduction and valence bands, 
$V$ represents the system volume, $|u_{\mathbf{k}}^{\pm}\rangle$ are the Bloch periodic states, defined as:

\begin{eqnarray}\label{eq:basis}
    |u_{k}^{+} \rangle= \begin{pmatrix}
        \cos\frac{\beta}{2} \\
        e^{i\alpha}\sin\frac{\beta}{2}
    \end{pmatrix}, \quad
        |u_{k}^{-}\rangle = \begin{pmatrix}
        \sin\frac{\beta}{2} \\
        -e^{i\alpha}\cos\frac{\beta}{2}
    \end{pmatrix}
\end{eqnarray}
Here, the parameters $\alpha$ and $\beta$ are defined as,
\begin{eqnarray*}
\tan\alpha = \frac{v_y k_y}{v_x k_x}, \quad \text{and~} \cos\beta = \frac{\Delta}{\sqrt{v_x^2 k_x^2 + v_y^2 k_y^2 + \Delta^2}}
\end{eqnarray*}

The Berry curvature for both conduction and valence band is given by,
\begin{eqnarray*}
    \Omega_{\hat{k}_z}^\pm &=&  i\frac{\langle u_k^\pm\vert \partial H/\partial k_x\vert u_k^\mp\rangle \langle u_k^\mp\vert \partial H/\partial k_y\vert u_k^\pm\rangle - \left(k_x \leftrightarrow k_y\right)}{\left(\varepsilon^\pm - \varepsilon^\mp\right)^2} ~\hat{k}_z\\
    &=& \mp \frac{v_x v_y \Delta}{2\left(\Delta^2 + v_{x}^2 k_x^2 + v_{y}^2 k_y^2\right)^{3/2}} ~\hat{k}_z 
\end{eqnarray*}


\subsection{Supplementary Note 3: The Fermi Golden rule and scattering rates }
The Fermi-Golden rule \cite{dirac1927quantum} connects the quantum mechanical scattering matrix to the classical scattering rate. In this context, we briefly discuss how scattering rate $(\varpi_{ll'})$ between states with different quantum numbers 
 \(l'\) and l is associated with the so called $T-$matrix elements. For a weak impurity potential in the lowest Born approximation scattering rate reads,

 \[\varpi_{ll'} = 2\pi |T_{ll'}|^2 \delta\left(\ve_l - \ve_{l'}\right),\]

where $\ve_l$ and $\ve_{l'}$ are the energies of the initial and final states, respectively. The scattering matrix element, which determines the transition probability, is defined as:

\[T_{ll'} = \left\langle l| V_{imp}| \Psi_{l'}\right\rangle. \]

Here $V_{imp}$ is the impurity potential operator representing the scalar potential introduced by the impurity. $\Psi_{l'}$ is the eigenstate of the full Hamiltonian $H_t = H_0 + V_{imp} $, where $H_0$ is the unperturbed Hamiltonian of the system (without the impurity). Using the Lippman-Schwinger equation \cite{lippmann1950variational,merzbacher1998quantum}, 
\begin{eqnarray*}
   \Psi_{l'} &=&  | l' \rangle + \frac{V_{imp}}{\ve_{l'} - H_0 + i\zeta }|\Psi_{l'}\rangle 
\end{eqnarray*}

For weak disorder, the scattering state $|\Psi_{l'}\rangle$ can be approximated as a truncated series expansion in terms of the impurity potential $ V_{ll'} = \left\langle l| V_{imp}| l'\right\rangle$, given by:

\begin{eqnarray}\label{eq:lippmann}
    \Psi_{l'} &=& |l'\rangle + \sum_{l''} \frac{V_{l''l'}}{\ve_{l'}-\ve_{l''}+i\zeta} |l''\rangle + \sum_{l''} \sum_{l'''} \frac{V_{l''l'} V_{l'''l''}}{\left(\ve_{l'}-\ve_{l''}+i\zeta\right)\left(\ve_{l'}-\ve_{l'''}+i\zeta\right)} |l'''\rangle + \cdots
\end{eqnarray}

Now, the $T-$matrix can be expanded using the Lippmann-Schwinger equation as, $T_{ll'} = \langle l |V_{imp}|\Psi_{l'}\rangle$. Substituting $T_{ll'}$ in Eq.\,\eqref{eq:lippmann}, we can get the scattering rates as follows--

\begin{eqnarray*}
    \varpi_{ll'} = 2\pi |T_{ll'}|^2 \delta\left(\ve_l - \ve_{l'}\right) = \varpi_{ll'}^{(2)} + \varpi_{ll'}^{(3)} + \varpi_{ll'}^{(4)} + \cdots,
\end{eqnarray*}

where

\begin{eqnarray}
    \varpi_{ll'}^{(2)} &=& 2\pi \langle V_{ll'} V_{ll'}^\ast \rangle_{dis}\delta\left(\ve_l - \ve_{l'}\right) \\
    \varpi_{ll'}^{(3)} &=& 2\pi \sum_{l''} \left(\frac{\langle V_{l'l} V_{ll''} V_{l''l'} \rangle_{dis}}{\ve_{l'}-\ve_{l''}+i\zeta} + \frac{\langle V_{l'l}^\ast V_{ll''}^\ast V_{l''l'}^\ast \rangle_{dis}}{\ve_{l'}-\ve_{l''}-i\zeta} \right) \delta\left(\ve_l - \ve_{l'}\right) \\
    \varpi_{ll'}^{(4)} &=& 2\pi \sum_{l''}\sum_{l'''} \left( \frac{\langle V_{ll'}^\ast V_{ll'''} V_{l'''l''} V_{l''l'} \rangle_{dis}}{\left(\ve_{l'}-\ve_{l''}+i\zeta\right)\left(\ve_{l'}-\ve_{l'''}+i\zeta\right)} + \frac{\langle V_{ll''} V_{l''l'} V_{ll'''}^\ast V_{l'''l'}^\ast \rangle_{dis}}{\left(\ve_{l'}-\ve_{l''}+i\zeta\right)\left(\ve_{l'}-\ve_{l'''}-i\zeta\right)} \right. \notag\\
    && + \left. \frac{\langle V_{ll'}V_{l'''l''}^\ast V_{l''l'}^\ast V_{ll'''}^\ast \rangle_{dis}}{\left(\ve_{l'}-\ve_{l''}-i\zeta\right)\left(\ve_{l'}-\ve_{l'''}-i\zeta\right)} \right) \delta\left(\ve_l - \ve_{l'}\right)
\end{eqnarray}

One can now express the symmetric and antisymmetric parts of scattering rate as $\varpi_{ll'}^{sy(as)} = \frac{1}{2}\left(\varpi_{ll'} \pm \varpi_{l'l}\right)$. In this context, however, the scattering rate $\varpi_{ll'}^{(2)}$ is specifically a symmetric quantity.

We consider a model with randomly located $\delta-$function scatterers, where the impurity potential is given by $V_{imp}(r) = \sum_{i}V_i\delta\left(r - R_i\right)$, with $R_i$ representing random positions. The strength distributions of the impurity potentials satisfy the statistical properties given by--
$\langle V_i \rangle_{dis} = 0$, $\langle V_i^2 \rangle_{dis} = V_0^2 \neq 0$, $\langle V_i^3 \rangle_{dis} = V_1^3 \neq 0$, where $\langle \cdots \rangle_{dis} $ denotes the disorder average\cite{sinitsyn2007semiclassical,sinitsyn2007anomalous}.

In the eigenstate representation (Eq.\eqref{eq:basis}) of 2D Dirac Hamiltonian the disorder has the following matrix form,

\begin{eqnarray}\label{eq:disorder_matrix}
    V_{\mathbf{kk'}} &=& \begin{pmatrix}
        V_{\mathbf{kk'}}^{++} & V_{\mathbf{kk'}}^{+-} \\
        V_{\mathbf{kk'}}^{-+} & V_{\mathbf{kk'}}^{--}
    \end{pmatrix} = V_{\mathbf{kk'}}^{0} \begin{pmatrix}
        \langle u_{\mathbf{k}}^{+}|u_{\mathbf{k'}}^{+}\rangle & \langle u_{\mathbf{k}}^{+}|u_{\mathbf{k'}}^{-}\rangle \\
        \langle u_{\mathbf{k}}^{-}|u_{\mathbf{k'}}^{+}\rangle & \langle u_{\mathbf{k}}^{-}|u_{\mathbf{k'}}^{-}\rangle
    \end{pmatrix},
\end{eqnarray}

where the superscript $\pm$ denotes the conduction (\(+\)) and valence (\(-\)) bands. The components of the matrix are given as follows:
\begin{eqnarray*}
    \langle u_{\mathbf{k}}^{+}|u_{\mathbf{k'}}^{+}\rangle &=& \cos\frac{\beta}{2}\cos\frac{\beta'}{2} + \sin\frac{\beta}{2} \sin\frac{\beta'}{2} e^{i(\alpha' - \alpha)} \\
    \langle u_{\mathbf{k}}^{+}|u_{\mathbf{k'}}^{-}\rangle &=& \cos\frac{\beta}{2} \sin\frac{\beta'}{2} - \sin\frac{\beta}{2} \cos\frac{\beta'}{2} e^{i(\alpha' - \alpha)} \\
    \langle u_{\mathbf{k}}^{-}|u_{\mathbf{k'}}^{+}\rangle &=& \sin\frac{\beta}{2}\cos\frac{\beta'}{2} - \cos\frac{\beta}{2}\sin\frac{\beta'}{2}e^{i(\alpha' - \alpha)} \\
    \langle u_{\mathbf{k}}^{-}|u_{\mathbf{k'}}^{-}\rangle &=& \sin\frac{\beta}{2}\sin\frac{\beta'}{2} + \cos\frac{\beta}{2}\cos\frac{\beta'}{2} e^{i(\alpha' - \alpha)} \\
    V_{\mathbf{kk'}}^{0} &=& \sum_{i} V_{i} e^{i\mathbf{(k'-k)\cdot R_i}}
\end{eqnarray*}

Here, \((\alpha, \beta)\) and \((\alpha', \beta')\) are parameters that depend on the wavevectors $\mathbf{k}$ and $\mathbf{k'}$ respectively.

\vspace{0.2cm}
{\par} Considering the Fermi level lies in the conduction band, the second-order scattering rate is given by:

\begin{eqnarray}
    \varpi_{\mathbf{kk'}}^{(2)} &=& 2\pi \langle V_{\mathbf{kk'}}^{++} (V_{\mathbf{kk'}}^{++})^\ast \rangle_{dis}\,\delta\left(\ve_{\mathbf{k}}^{+} - \ve_{\mathbf{k'}}^{+}\right) \notag \\
    &=& \frac{\pi}{2} \langle V_{\mathbf{kk'}}^{0} (V_{\mathbf{kk'}}^{0})^\ast \rangle_{dis} \left[1 + \cos\beta\cos\beta' + \sin\beta\sin\beta'\cos(\alpha'-\alpha)\right] \,\delta\left(\ve_{\mathbf{k}}^{+} - \ve_{\mathbf{k'}}^{+}\right) \notag \\
    &=& \frac{\pi N_i V_0^2}{2} \left[1 + \cos\beta\cos\beta' + \sin\beta\sin\beta'\cos(\alpha'-\alpha)\right] \,\delta\left(\ve_{\mathbf{k}}^{+} - \ve_{\mathbf{k'}}^{+}\right),
\end{eqnarray}

where $\langle V_{\mathbf{kk'}}^{0} (V_{\mathbf{kk'}}^{0})^\ast \rangle_{dis} \simeq N_iV_0^2$ with $N_i$ is the impurity concentration \cite{sinitsyn2007semiclassical,sinitsyn2007anomalous}.

The skew scattering contribution to the Hall effect arises from the antisymmetric part of the scattering rate. The third-order scattering rate, $\varpi_{ll'}^{(3)}$ comprises both symmetric and antisymmetric parts. However, the symmetric part is not crucial, as it only serves to renormalize the second-order scattering rate, $\varpi_{\mathbf{kk'}}^{(2)} $ \cite{sinitsyn2007anomalous}. The antisymmetric part ( $\varpi_{ll'}^{(3as)}$) of third-order scattering rate can be derived as,

\begin{eqnarray} \label{eq:omega3a-1}
     \varpi_{ll'}^{(3as)} &=& \frac{1}{2}\left(\varpi_{ll'}^{(3)} - \varpi_{l'l}^{(3)}\right) \notag \\
     &=& \frac{2\pi}{2}\sum_{l''} 2i\zeta\,\left(\frac{\langle V_{ll'} V_{l'l''} V_{l''l} \rangle_{dis}}{\left(\ve_{l'}-\ve_{l''}\right)^2 + \zeta^2} - \frac{\langle V_{ll'}^\ast V_{l'l''}^\ast V_{l''l}^\ast \rangle_{dis}}{\left(\ve_{l'}-\ve_{l''}\right)^2 + \zeta^2}  \right) \,\delta\left(\ve_l - \ve_{l'}\right) \notag \\
     &=&  2i\pi^2 \,\sum_{l''} \left( \langle V_{ll'} V_{l'l''} V_{l''l} \rangle_{dis} - \langle V_{ll'}^\ast V_{l'l''}^\ast V_{l''l}^\ast \rangle_{dis} \right) \,\delta\left(\ve_{l'} - \ve_{l''}\right) \,\delta\left(\ve_l - \ve_{l'}\right) \notag \\
     &=&  -(2\pi)^2 \,\sum_{l''} \text{Im}\left[ \langle V_{ll'} V_{l'l''} V_{l''l} \rangle_{dis} \right] \,\delta\left(\ve_{l'} - \ve_{l''}\right) \,\delta\left(\ve_l - \ve_{l'}\right) \\
     &=& -(2\pi)^2 \,\sum_{n'',\mathbf{k''}} \text{Im}\left[ \langle V_{\mathbf{kk'}}^{nn'} V_{\mathbf{k'k''}}^{n'n''} V_{\mathbf{k''k}}^{n''n} \rangle_{dis} \right] \,\delta\left(\ve_{\mathbf{k'}}^{n'} - \ve_{\mathbf{k''}}^{n''}\right) \,\delta\left(\ve_{\mathbf{k}}^{n} - \ve_{\mathbf{k'}}^{n'}\right)
\end{eqnarray}

Here $n, n', n''$ serve as indices representing both the conduction and valence band. Assuming the Fermi level lies in the conduction band, the third-order antisymmetric scattering rate between momenta $\mathbf{k}$ to $\mathbf{k''}$ yields:

\begin{eqnarray}\label{eq:omega3a}
    \varpi_{\mathbf{kk'}}^{(3as)} &=& -(2\pi)^2 \,\sum_{n,\mathbf{k''}} \text{Im}\left[ \langle V_{\mathbf{kk'}}^{++} V_{\mathbf{k'k''}}^{+n} V_{\mathbf{k''k}}^{n+} \rangle_{dis} \right] \,\delta\left(\ve_{\mathbf{k'}}^{+} - \ve_{\mathbf{k''}}^{n}\right) \,\delta\left(\ve_{\mathbf{k}}^{+} - \ve_{\mathbf{k'}}^{+}\right),
\end{eqnarray}

where $n\in \lbrace\pm\rbrace $ corresponds to the conduction and valence bands. Assuming the Fermi level lies in the conduction band and that intraband contributions prevail while interband contributions are negligible, substituting Eq.\eqref{eq:disorder_matrix} into Eq.\eqref{eq:omega3a} yields the simplified form of the third-order antisymmetric scattering rate as:

\begin{eqnarray} \label{eq:omega3a_final}
    \varpi_{\mathbf{kk'}}^{(3as)} &=& \frac{N_i V_1^3}{4}\int[d\mathbf{k''}] \left[\sin\beta\sin\beta'\cos\beta''\sin(\alpha - \alpha') + \cos\beta\sin\beta'\sin\beta''\sin(\alpha'-\alpha'') \right. \notag \\
    && \left. + \sin\beta\cos\beta'\sin\beta''\sin(\alpha''-\alpha) \right]  \,\delta\left(\ve_{\mathbf{k'}}^{+} - \ve_{\mathbf{k''}}^{+}\right) \,\delta\left(\ve_{\mathbf{k}}^{+} - \ve_{\mathbf{k'}}^{+}\right),
\end{eqnarray}

where $\langle V_{\mathbf{kk'}}^{0} V_{\mathbf{k'k''}}^{0} V_{\mathbf{k''k}}^{0}\rangle_{dis} \simeq N_iV_1^3$ with $N_i$ is the impurity concentration concentration and $\int [d\mathbf{k''}] = \int_0^\infty k''\,dk'' \int_0^{2\pi}d\alpha''$.

 {\par}
Following a similar rigorous derivation as also outlined in Ref. \cite{xiao2017role,Du_2019}, the antisymmetric component of the fourth-order scattering rate can be expressed as: 

\begin{eqnarray}
    \varpi_{\mathbf{kk'}}^{(4as)} &=& \frac{N_i^2 V_0^4}{8} \int[d\mathbf{k''}] \left( \cos\beta + \cos\beta' + \cos\beta'' \right) \left[ \sin\beta \sin\beta' \cos\beta'' \sin(\alpha - \alpha') \right. \notag \\
    && \left. + \sin\beta' \sin\beta'' \cos\beta \sin(\alpha' - \alpha'') +\sin\beta\sin\beta'' \cos\beta' \sin(\alpha'' - \alpha) \right]\,\delta\left(\ve_{\mathbf{k'}}^{+} - \ve_{\mathbf{k''}}^{+}\right) \,\delta\left(\ve_{\mathbf{k}}^{+} - \ve_{\mathbf{k'}}^{+}\right)
\end{eqnarray}

Here, $\langle V_{\mathbf{kk'}}^{0} V_{\mathbf{k'k''}}^{0} V_{\mathbf{k''k'''}}^{0} V_{\mathbf{k'''k}}^{0} \rangle \simeq N_i V_0^4$. This expression for $\varpi_{\mathbf{kk'}}^{(4as)}$. represents the antisymmetric part of the fourth-order scattering rate, which is involved in higher-order contributions to the Hall effect due to skew scattering.

\subsection{Supplementary Note 4: Coordinate shift due to side-jump scattering}

Wave-packet undergoing scattering due to impurities or defects in a material undergoes a lateral displacement (side jump) that is perpendicular to its velocity and the force acting on it. This displacement represents the coordinate shift\,\cite{Sinitsyn2006coordinate,Berger_prb70} associated with the scattering process that transitions the Wave-packet from one state to another. For spin-independent scalar disorder under the lowest Born approximation, the coordinate shift in the scattering process that transitions an electron from state \( l \) to state \( l' \), is given by:

\[
\delta \mathbf{r}_{l, l'} =  \langle u_{l} | i \nabla_{\mathbf{k}} | u_{l} \rangle - \langle u_{l'} | i \nabla_{\mathbf{k}} | u_{l'} \rangle  - 
 \left(\nabla_{\mathbf{k}} + \nabla_{\mathbf{k'}}\right) \, \arg \langle u_{l} | u_{l'} \rangle,
\]

For the 2D Dirac gapped model described in Eq.~\eqref{diracHam}, considering $\delta-$correlated scalar impurities and assuming the Fermi level lies in the conduction band, the coordinate shift from momenta \(k\) to \(k'\) is given by:

\begin{eqnarray}\label{eq:side-jump}
\delta \mathbf{r}_{k, k'} &=&  \langle u_{k}^{+} | i \nabla_{\mathbf{k}} | u_{k}^{+} \rangle - \langle u_{k'}^{+} | i \nabla_{\mathbf{k}} | u_{k'}^{+} \rangle  - 
\left(\nabla_{\mathbf{k}} + \nabla_{\mathbf{k'}}\right) \, \arg \langle u_{k}^{+} | u_{k'}^{+} \rangle
\end{eqnarray}

Considering $k_x = k \cos \phi$, $k_y = k \sin \phi$ and hence, $k=\sqrt{k_x^2 + k_y^2}$, we evaluate the individual terms in Eq.~\eqref{eq:side-jump}.

\begin{eqnarray*}
    \frac{\partial}{\partial k_x } \equiv \cos \phi \frac{\partial}{\partial k} - \frac{\sin\phi}{k} \frac{v_y \cos^2\alpha}{v_x \cos^2\phi} \frac{\partial}{\partial \alpha} \\
    \frac{\partial}{\partial k_y } \equiv \sin \phi \frac{\partial}{\partial k} + \frac{\cos\phi}{k} \frac{v_y \cos^2\alpha}{v_x \cos^2\phi} \frac{\partial}{\partial \alpha}
\end{eqnarray*}

For $\mathbf{k}\longrightarrow\mathbf{k'}$, $\left\lbrace\alpha, \,\beta, \,\phi \right\rbrace \longrightarrow \left\lbrace\alpha', \,\beta', \,\phi' \right\rbrace$. Using Eq.\,\eqref{eq:basis} for $|u_{k}^{+}\rangle$ we now evaluate,

\begin{eqnarray*}
    \langle u_{k}^{+} | i \frac{\partial}{\partial k_x} | u_{k}^{+} \rangle &=&  \frac{\sin\phi}{k} \frac{v_y \cos^2\alpha}{v_x \cos^2\phi} \frac{\left(1 - \cos\beta\right)}{2},\quad \langle u_{k}^{+} | i \frac{\partial}{\partial k_y} | u_{k}^{+} \rangle = - \frac{\cos\phi}{k} \frac{v_y \cos^2\alpha}{v_x \cos^2\phi} \frac{\left(1 - \cos\beta\right)}{2}
\end{eqnarray*}

\begin{eqnarray*}
    \left\langle u_{k}^{+}|u_{k'}^{+}\right\rangle &=& \cos\frac{\beta}{2}\cos\frac{\beta'}{2} + \sin\frac{\beta}{2}\sin\frac{\beta'}{2}e^{i\left(\alpha' - \alpha\right)} \\ 
    \left\vert\left\langle u_{k}^{+}|u_{k'}^{+}\right\rangle \right\vert^2 &=& \frac{1}{2}\left[1 + \cos\beta \cos\beta' + \sin\beta \sin\beta' \cos\left(\alpha' - \alpha\right)\right]
\end{eqnarray*}

Thus, for this complex number $\mathcal{Z}=\left\langle u_{k}^{+}|u_{k'}^{+}\right\rangle$, the argument of $\mathcal{Z}$ is given by \[arg(\mathcal{Z}) = -i \ln{\frac{\mathcal{Z}}{|\mathcal{Z}|}},\]
and it's derivative with respect to a parameter 
$a$ is:

\[\frac{\partial}{\partial a}arg(\mathcal{Z}) = \frac{-i}{|\mathcal{Z}|^2}\left(\mathcal{Z}^\ast \frac{\partial \mathcal{Z}}{\partial a} - \frac{1}{2}\frac{\partial |\mathcal{Z}|^2}{\partial a}\right)\]

Therefore,
\begin{eqnarray*}
    \frac{\partial}{\partial k}arg(\mathcal{Z}) &=& \frac{1}{4|\mathcal{Z}|^2}\sin\beta'\sin\left(\alpha' - \alpha\right) \frac{d\beta}{dk} \\
    \frac{\partial}{\partial k'}arg(\mathcal{Z}) &=& \frac{1}{4|\mathcal{Z}|^2}\sin\beta\sin\left(\alpha' - \alpha\right) \frac{d\beta'}{dk'} 
\end{eqnarray*}

\begin{eqnarray*}
    \frac{\partial}{\partial \alpha }arg(\mathcal{Z}) &=& \frac{-1}{4|\mathcal{Z}|^2} \left[ \left(1-\cos\beta\right)\left(1-\cos\beta'\right) + \sin\beta\sin\beta'\cos\left(\alpha' - \alpha\right)\right] \\
    \frac{\partial}{\partial \alpha' }arg(\mathcal{Z}) &=& \frac{1}{4|\mathcal{Z}|^2} \left[ \left(1-\cos\beta\right)\left(1-\cos\beta'\right) + \sin\beta\sin\beta'\cos\left(\alpha' - \alpha\right)\right]
\end{eqnarray*}

\begin{eqnarray*}
    \frac{d\beta}{dk} &=& \frac{\sin\beta\cos\beta}{k}, \quad \,\frac{d\beta'}{dk'} = \frac{\sin\beta'\cos\beta'}{k'}
\end{eqnarray*}

Therefore, the coordinate shift for a conduction band electron in the 2D Dirac model, as described by Eq.~\eqref{diracHam}, under a $\delta-$correlated weak impurity potential can be expressed in terms of its $x-$ and $y-$components as follows:

\begin{eqnarray} \label{eq:side-jump-xkkp}
    \delta\,r^x_{\mathbf{kk'}} &=& \langle u_{k}^{+} | i \frac{\partial}{\partial k_x} | u_{k}^{+} \rangle     - \langle u_{k'}^{+} | i \frac{\partial}{\partial k'_x} | u_{k'}^{+} \rangle  -\left(\frac{\partial}{\partial k_x} + \frac{\partial}{\partial k'_x}\right)arg(\mathcal{Z}) =  \frac{1}{4 k k' |\mathcal{Z}|^2} \left[T_{1x} + T_{2x} + T_{3x}\right]
\end{eqnarray}

\begin{eqnarray}\label{eq:side-jump-ykkp}
    \delta\,r^y_{\mathbf{kk'}} &=& \langle u_{k}^{+} | i \frac{\partial}{\partial k_y} | u_{k}^{+} \rangle     - \langle u_{k'}^{+} | i \frac{\partial}{\partial k'_y} | u_{k'}^{+} \rangle  -\left(\frac{\partial}{\partial k_y} + \frac{\partial}{\partial k'_y}\right)arg(\mathcal{Z}) =  \frac{1}{4 k k' |\mathcal{Z}|^2} \left[T_{1y} + T_{2y} + T_{3y}\right]
\end{eqnarray}

\begin{eqnarray}
    T_{1x} &=& \sin (\beta ) \sin (\text{$\beta'$}) \sin (\alpha -\text{$\alpha'$}) \left[k \cos (\text{$\beta' $}) \cos (\text{$\phi' $})+ k' \cos (\beta ) \cos (\phi )\right] \\
    T_{2x} &=& \frac{v_y}{v_x} \sin (\beta ) \sin (\text{$\beta' $}) \cos (\alpha -\text{$\alpha' $}) \left[k \cos ^2(\text{$\alpha' $}) \cos (\text{$\beta' $}) \tan (\text{$\phi' $}) \sec (\text{$\phi' $})-k' \cos ^2(\alpha ) \cos (\beta ) \tan (\phi ) \sec (\phi )\right] \\
    T_{3x} &=& \frac{v_y}{v_x} \left[k' \cos ^2(\alpha ) \sin ^2(\beta ) \cos (\text{$\beta' $}) \tan (\phi ) \sec (\phi )-k \cos ^2(\text{$\alpha' $}) \cos (\beta ) \sin ^2(\text{$\beta' $}) \tan (\text{$\phi' $}) \sec (\text{$\phi' $})\right] \\
%
%
%
    T_{1y} &=& \sin (\beta ) \sin (\text{$\beta' $}) \sin (\alpha -\text{$\alpha' $}) \left[k \cos (\text{$\beta' $}) \sin (\text{$\phi' $})+k' \cos (\beta ) \sin (\phi )\right] \\
    T_{2y} &=&  \frac{v_y}{v_x} \sin (\beta ) \sin (\text{$\beta' $'}) \cos (\alpha -\text{$\alpha' $}) \left[k' \cos ^2(\alpha ) \cos (\beta ) \sec (\phi )-k \cos ^2(\text{$\alpha' $}) \cos (\text{$\beta' $}) \sec (\text{$\phi' $})\right] \\
    T_{3y} &=& \frac{v_y}{v_x} \left[k \cos ^2(\text{$\alpha' $}) \cos (\beta ) \sin ^2(\text{$\beta' $}) \sec (\text{$\phi' $})-k' \cos ^2(\alpha ) \sin ^2(\beta ) \cos (\text{$\beta' $}) \sec (\phi )\right]
\end{eqnarray}

In the case of isotropic velocity, where $v_y = v_x$, and the angular parameter $\alpha (\alpha') $ is equal to $ \phi (\phi') $, the equations for the coordinate shift in the $x$ and $y-$directions, given by Eq.~\eqref{eq:side-jump-xkkp} and \eqref{eq:side-jump-ykkp}, simplifies to --

\begin{eqnarray} \label{eq:side-jump-xkkp-2}
    \delta\,r^x_{\mathbf{kk'}} &=& \frac{1}{4|\mathcal{Z}|^2}\left(\frac{\sin\beta}{k} + \frac{\sin\beta'}{k'} \right) \left( \sin\beta\cos\beta'\sin\phi - \cos\beta\sin\beta'\sin\phi' \right)
\end{eqnarray}

\begin{eqnarray} \label{eq:side-jump-ykkp-2}
    \delta\,r^y_{\mathbf{kk'}} &=& \frac{1}{4|\mathcal{Z}|^2}\left(\frac{\sin\beta}{k} + \frac{\sin\beta'}{k'} \right) \left(\cos\beta\sin\beta'\cos\phi' - \sin\beta\cos\beta'\cos\phi\right)
\end{eqnarray}

\subsection{Supplementary Note 5: General expression of distribution function for the intrinsic contribution}

The semiclassical equations of electron motion in the absence of an external magnetic field can be written in the following form~\cite{Niu_1999}:
\begin{eqnarray}
\dot {\mathbf r_l} &=& \frac{1}{\hbar}\nabla_{\bm k}\tilde{\ve}_{l}
-\dot{\mathbf{k}} \times {\bm \tilde{\Omega}_{l}}+\mathbf{v}_l^{sj},\quad \hbar \dot {\mathbf k}= e {\mathbf E}~,
\label{eq_rdot}
\end{eqnarray} 
where $e<0$, $\tilde{\ve}_l = \ve_l + \varepsilon^{(2)}_l$, $\varepsilon^{(2)}_l = -\frac{1}{2}E_bG_{bc}E_c$ and ${\bm \tilde{\Omega}_{l}}=\bm{\nabla_{\mathbf{k}}} \times \sum_{i=0}^{1}\bm{A}_{l}^{(i)}$
with ${\ve}_{l}$ and $\bm{A}_{\alpha \delta}^{(0)}(\mathbf{k})$ are the unperturbed band energy and intraband Berry connection, respectively. $G_{bc}$ is the Berry connection polarizabiliy.

The intrinsic distribution function $f_l^{in}$ is then analyzed in the context of the Boltzmann equation, specified in Eq. \eqref{BEqintrinsic}. The Eq.\,\eqref{BEqintrinsic} contains the intrinsic contribution arises from the symmetric scattering part. To solve this equation, the relaxation time approximation is employed, resulting in a collision integral written as $\mathcal{I}_{coll}^{in}\left\lbrace f_l\right\rbrace = (f_l^0 - f_l^{in})/\tau_l$, where $f_l^0$ is the Fermi-Dirac distribution function. 

The subsequent expression for $f_l^{in}$ is derived as an infinite series as follows,
\begin{eqnarray}
&&\frac{\partial f_l^{in}}{\partial t} - \mathbf{\dot{k}}\cdot \nabla_{\mathbf{k}} f_l^{in} = - \frac{f_l^{in}-f_l^{0}(\tilde{\varepsilon})}{\tau} \notag \\
&&f_l^{in} = \sum_{\alpha=0}^{\infty} \left[ -\tau_l \partial_t + \frac{e\tau_l}{2\hbar}\left(\mathcal{\xi}_a e^{i\omega t} + \mathcal{\xi}_a^{\ast} e^{-i\omega t}\right)\partial^{\mathbf{k}}_a  \right]^\alpha f_l^{0}(\tilde{\varepsilon}) \notag\\
&&f_l^{in} = \sum_{\alpha=0}^{\infty} \left[ -\tau_l \partial_t + \frac{e\tau_l}{2\hbar}\left(\mathcal{\xi}_a e^{i\omega t} + \mathcal{\xi}_a^{\ast} e^{-i\omega t}\right)\partial^{\mathbf{k}}_a  \right]^\alpha\left(f_l^{0} (\varepsilon) + \varepsilon^{(2)} {f_l^{0}}^{\prime}(\varepsilon) \right)\notag \\ 
&&f_l^{in} = \sum_{\alpha=0}^{\infty} \left[ -\tau_l \partial_t + \frac{e\tau_l}{2\hbar}\left(\mathcal{\xi}_a e^{i\omega t} + \mathcal{\xi}_a^{\ast} e^{-i\omega t}\right)\partial^{\mathbf{k}}_a  \right]^\alpha\left(f_l^{0} (\varepsilon) + \varepsilon^{(2)} {f_l^{0}}^{\prime}(\varepsilon) \right)\notag \\ 
&&f_l^{in} = \sum_{\alpha=0}^{\infty} \left[ -\tau_l \partial_t + \frac{e\tau_l}{2\hbar}\left(\mathcal{\xi}_a e^{i\omega t} + \mathcal{\xi}_a^{\ast} e^{-i\omega t}\right)\partial^{\mathbf{k}}_a  \right]^\alpha\left(  \varepsilon^{(2)} {f_l^{0}}^{\prime}(\varepsilon) \right) +  \sum_{\alpha=0}^{\infty} \left[ -\tau_l \partial_t + \frac{e\tau_l}{2\hbar}\left(\mathcal{\xi}_a e^{i\omega t} + \mathcal{\xi}_a^{\ast} e^{-i\omega t}\right)\partial^{\mathbf{k}}_a  \right]^\alpha f_l^{0} (\varepsilon)   \label{BCPndf1} \\
&&f_l^{in} \equiv f_l^{0} + \delta^{1}f_l^{in} +  \delta^{2}f_l^{in} +  \delta^{3}f_l^{in} + \cdots \\\notag
\end{eqnarray}

Our primary objective is to analyze the system's response up to the third-order concerning the electric field. Consequently, our expansion strategy focuses exclusively on unfolding the distribution function up to the third-order. 

Here, $f_l^{0}(\varepsilon)  = f_l^{0}(\ve_l)$. For simplicity, we maintain $\frac{\partial}{\partial k_a}\equiv\partial^{k}_a = \partial_a$ unless there is partial derivatives with respect to prime coordinates (e.g., $\partial^{k'}_a$). Whenever, we meet  ${f_l^{0}}^{\prime}(\varepsilon) $ or ${f_l^{0}}^{''}(\varepsilon) $ it means ${f_l^{0}}^{\prime}(\varepsilon)  = \frac{\partial f^{0}_{l}(\tilde{\varepsilon})}{\partial \tilde{\varepsilon}}|_{\tilde{\varepsilon} =\varepsilon}$ or ${f_l^{0}}^{''}(\varepsilon)  = \frac{\partial^2 f^{0}_{l}(\tilde{\varepsilon})}{\partial \tilde{\varepsilon}^2}|_{\tilde{\varepsilon} =\varepsilon}$, unless otherwise it is explicitly mentioned.

\begin{eqnarray}
f_l^{in} &=& f_l^0(\varepsilon) + \left(\frac{e\tau_l}{2\hbar}\right) \Big[ \frac{\xi_b e^{i\omega t}}{\left(1+i\omega\tau_l\right)} + c.c. \Big] \partial_b f_l^0(\varepsilon) + \left(\frac{e\tau_l}{2\hbar}\right)^2  \Big[ \frac{\xi_b\xi_c e^{2i\omega t}}{\left(1+i\omega\tau_l\right)\left(1+2i\omega\tau_l\right)} + \frac{\xi_b\xi^{\ast}_c }{\left(1-i\omega\tau_l\right)} + c.c. \Big] \partial_b \partial_c f_l^0(\varepsilon) \notag \\
&& + \left(\frac{e\tau_l}{2\hbar}\right)^3 \Big[ \frac{\xi_b \xi_c\xi_d e^{3i\omega t}}{\left(1+i\omega\tau_l\right)\left(1+2i\omega\tau_l\right)\left(1+3i\omega\tau_l\right)} + \frac{\xi_b \xi_c\xi^{\ast}_d e^{i\omega t}}{\left(1-i\omega\tau_l\right)\left(1+i\omega\tau_l\right)} + \frac{\xi^{\ast}_b \xi_c\xi_d e^{i\omega t}}{\left(1+i\omega\tau_l\right)^2\left(1+2i\omega\tau_l\right)} \notag \\
&& + \frac{\xi^{\ast}_b \xi_c\xi^{\ast}_d e^{-i\omega t}}{\left(1-i\omega\tau_l\right)^2} + c.c. \Big] \partial_b \partial_c \partial_d f_l^0(\varepsilon) -\frac{1}{8} \left[ \frac{\xi_b\xi_c e^{2i\omega t}}{1+2i\omega \tau_l} + \xi_b\xi^{\ast}_c + c.c. \right] G_{bc} {f_l^{0}}^{\prime}(\varepsilon)  -\frac{1}{8} \left(\frac{e\tau_l}{2\hbar}\right)  \Big[ \frac{\xi_b\xi_c\xi_d e^{3i\omega t}}{\left(1+2i\omega \tau_l\right)\left(1+3i\omega \tau_l\right)} \notag \\
&& + \frac{\xi^{\ast}_b\xi_c\xi_d e^{i\omega t}}{\left(1+i\omega \tau_l\right)\left(1+2i\omega \tau_l\right)} + \frac{\xi_b\xi_c\xi^{\ast}_d e^{i\omega t}}{\left(1+i\omega \tau_l\right)}  + \frac{\xi^{\ast}_b\xi_c\xi^{\ast}_d e^{-i\omega t}}{\left(1-i\omega \tau_l\right)} + c.c.\Big] ~\partial_b \big\lbrace G_{cd} {f_l^{0}}^{\prime}(\varepsilon)\big\rbrace \label{intdist}
\end{eqnarray}

\subsection{Supplementary Note 6: Total intrinsic third order current}

Total intrinsic current due to the intrinsic part of the distribution function $f_l^{in}$ --
\begin{eqnarray}
\mathbf{J}^{in} &=& -e\sum_{l} \left[  \frac{1}{\hbar} \nabla_{\mathbf{k}} \ve_l +  \frac{1}{\hbar} \nabla_{\mathbf{k}} \ve_l^{(2)} +  \frac{e}{2\hbar} \left(\xi e^{i\omega t} + \xi^{\ast} e^{-i\omega t} \right) \times \left( \mathbf{\Omega}_l + \mathbf{\Omega}_l^{(1)}  \right) \right]  f_l^{in} \notag 
\end{eqnarray}

Here, $\sum_{l} \longrightarrow \sum_{n}\int [d\mathbf{k}]$, where $[d\mathbf{k}]$ denotes $d^d\bm \mathbf{k}/(2\pi)^{d}$, with $d$ representing the dimensionality of the system. The $a^{th}$ component of the total intrinsic current is expressed as:
\begin{eqnarray}
J^{in}_a &=& -e\sum_{l} \left[  v_l^a + \partial_a \ve_l^{(2)} +  \frac{e}{2\hbar} \epsilon^{aef} \left(\xi_e e^{i\omega t} + \xi^{\ast}_e e^{-i\omega t} \right) \left( \Omega_l^f + \Omega_l^{(1)f}  \right) \right]  f_l^{in} \notag 
\end{eqnarray}

Now, substituting the expression for $f_l^{in}$ and $\ve_l^{(2)}$ in the expression of $J^{in}_a$, we get the third order component of intrinsic current as,

\begin{eqnarray}
J^{in}_{3,a} &=& -e\sum_{l}\left(\frac{e\tau_l}{2\hbar}\right)^3  v_l^a \Big[ \frac{\xi_b \xi_c\xi_d e^{3i\omega t}}{\left(1+i\omega\tau_l\right)\left(1+2i\omega\tau_l\right)\left(1+3i\omega\tau_l\right)} + \frac{\xi_b \xi_c\xi^{\ast}_d e^{i\omega t}}{\left(1-i\omega\tau_l\right)\left(1+i\omega\tau_l\right)} + \frac{\xi^{\ast}_b \xi_c\xi_d e^{i\omega t}}{\left(1+i\omega\tau_l\right)^2\left(1+2i\omega\tau_l\right)} \notag \\
&& +  \frac{\xi_b \xi^{\ast}_c\xi_d e^{i\omega t}}{\left(1+i\omega\tau_l\right)^2} + c.c. \Big] \partial_b \partial_c \partial_d f_l^0(\varepsilon) \notag \\
&& -e\sum_{l}\left(-\frac{e\tau_l}{16\hbar}\right)     \Big[ \frac{\xi_b\xi_c\xi_d e^{3i\omega t}}{\left(1+2i\omega \tau_l\right)\left(1+3i\omega \tau_l\right)}  + \frac{\xi^{\ast}_b\xi_c\xi_d e^{i\omega t}}{\left(1+i\omega \tau_l\right)\left(1+2i\omega \tau_l\right)} + \frac{\xi_b\xi_c\xi^{\ast}_d e^{i\omega t}}{\left(1+i\omega \tau_l\right)} \notag \\
&& + \frac{\xi_b\xi^{\ast}_c\xi_d e^{i\omega t}}{\left(1+i\omega \tau_l\right)} + c.c.\Big] \Big[ v_l^a v_l^b G_{cd}  {f_l^0}^{\prime\prime} -(\partial_a \partial_b G_{cd}) f_l^0 \Big] \notag \\
&& -e\sum_{l}\left(\frac{e\tau_l}{16\hbar}\right)   \Big[\frac{\xi_b\xi_c\xi_d e^{3i\omega t}}{\left(1+i\omega\tau_l\right)} + \frac{\xi_b \xi_c\xi^{\ast}_d e^{i\omega t}}{\left(1+i\omega\tau_l\right)} +  \frac{\xi^{\ast}_b \xi_c\xi_d e^{i\omega t}}{\left(1-i\omega\tau_l\right)} +  \frac{\xi_b\xi^{\ast}_c\xi_d e^{i\omega t}}{\left(1+i\omega\tau_l\right)} +c.c. \Big] \left(\partial_b \partial_a G_{cd} \right) f_l^0(\varepsilon)  \notag \\
&& -e\sum_{l} \left(\frac{e}{2\hbar}\right)^2 \frac{\tau_l}{2}  \Big[\frac{\xi_b\xi_c\xi_d e^{3i\omega t}}{\left(1+i\omega\tau_l\right)} + \frac{\xi_b\xi_c\xi^{\ast}_d e^{i\omega t}}{\left(1+i\omega\tau_l\right)} + \frac{\xi^{\ast}_b\xi_c\xi_d e^{i\omega t}}{\left(1-i\omega\tau_l\right)} + \frac{\xi_b \xi^{\ast}_c\xi_d e^{i\omega t}}{\left(1+i\omega\tau_l\right)} + c.c. \Big] \times \notag \\ 
&& \Big[(\partial_b \partial_d G_{ac}) f_l^0(\varepsilon) -(\partial_b\partial_a G_{cd}) f_l^0(\varepsilon)   \Big] \notag \\
&& -e\sum_{l}\left(\frac{e}{2\hbar}\right)^3 \tau_l^2  \epsilon^{ade} \Omega_l^e \left(\partial_b \partial_c f_l^0(\varepsilon) \right) \Big[ \frac{\xi_b\xi_c \xi_d e^{3i\omega t}}{\left(1+i\omega\tau_l\right)\left(1+2i\omega\tau_l\right)} + \frac{\xi_b\xi^{\ast}_c \xi_d e^{i\omega t}}{\left(1-i\omega\tau_l\right)} + \frac{\xi_b\xi_c \xi^{\ast}_d e^{i\omega t}}{\left(1+i\omega\tau_l\right)\left(1+2i\omega\tau_l\right)} \notag \\
&& + \frac{\xi^{\ast}_b\xi_c \xi_d e^{i\omega t}}{\left(1+i\omega\tau_l\right)}  + c.c. \Big] \notag \\
&& -e\sum_{l}\left(-\frac{e}{16\hbar}\right)   \epsilon^{ade}  \Omega_l^e G_{bc} {f_l^{0}}^{\prime}(\varepsilon)  \left[ \frac{\xi_b\xi_c\xi_d e^{3i\omega t}}{1+2i\omega \tau_l} + \frac{\xi_b\xi_c\xi^{\ast}_d e^{i\omega t}}{1+2i\omega \tau_l} + \xi_b\xi^{\ast}_c \xi_d e^{i\omega t}  + \xi^{\ast}_b\xi_c \xi_d e^{i\omega t}  + c.c. \right] \notag\\
&\equiv & \text{Re}\left[\mathcal{J}_{3,a}^{in}\left(\omega\right) e^{i\omega t} + \mathcal{J}_{3,a}^{in}\left(3\omega\right) e^{3i\omega t}\right]
\end{eqnarray}

\subsection{Supplementary Note 7: Frequency-dependent decomposition of $\chi_{abcd}$ for third-order current}

 The total third-order current can be grouped into two parts based on frequency dependence, specifically at the fundamental frequency $\omega$ and its third harmonic $3\omega$. This can be expressed as:
 \begin{eqnarray*}
 	\mathcal{J}_{3,a} &=& \mathcal{J}_{3,a}\left(\omega\right) + \mathcal{J}_{3,a}\left(3\omega\right) \\
 	&=& \chi_{abcd}\left(\omega\right)  ~\xi_b \xi_c\xi_d  \cos\left(\omega t\right) + \chi_{abcd}\left(3\omega\right)  ~\xi_b \xi_c\xi_d  \cos\left(3\omega t\right)
 \end{eqnarray*}
 
 In this context, $\chi_{abcd}\left(\omega\right)$ and $\chi_{abcd}\left(3\omega\right)$ represent the third-order conductivity with dependencies on $\omega$ and $3\omega$ respectively. This formulation provides a comprehensive expression for the third-order current, demonstrating its decomposition into distinct frequency components.  The variables $\xi_b$, $\xi_c$, $\xi_d$ are the components of AC electric fields. The terms 
 $\cos\left(\omega t\right)$ and 
$\cos\left(3\omega t\right)$ account for the time-dependent variations at the fundamental and third harmonic frequencies, respectively.

\subsection{Supplementary Note 8: Third order current and $\chi_{abcd}$ for the intrinsic contribution under Time-Reversal Symmetry ($\mathcal{T}$)}

\begin{eqnarray}
\mathcal{J}_{3,a}^{in}\left(\omega\right)  &=& \left(\frac{e^2\tau_l}{8\hbar}\right)\sum_{l}   \Big[ v_l^a v_l^b G_{cd}  {f_l^0}^{\prime\prime} -(\partial_a \partial_b G_{cd}) f_l^0 \Big]    \Big[ 2 + \frac{1}{\left(1+2i\omega \tau_l\right)}\Big]\frac{\xi_b \xi_c\xi_d}{\left(1+i\omega \tau_l\right)}  \notag \\
&& -\left(\frac{e^2\tau_l}{8\hbar}\right)\sum_{l}   \left(\partial_b \partial_a G_{cd} \right) f_l^0(\varepsilon) \Big[ \frac{2}{\left(1+i\omega\tau_l\right)} +  \frac{1}{\left(1-i\omega\tau_l\right)}  \Big] \xi_b \xi_c\xi_d \notag \\
&& -\left(\frac{e^3\tau_l}{4\hbar^2}\right)  \sum_{l}   \Big[(\partial_b \partial_d G_{ac}) f_l^0(\varepsilon) -(\partial_b\partial_a G_{cd}) f_l^0(\varepsilon)   \Big] \Big[ \frac{2}{\left(1+i\omega\tau_l\right)} + \frac{1}{\left(1-i\omega\tau_l\right)}  \Big] \xi_b\xi_c\xi_d \notag \\
&& -\left(\frac{e^4\tau_l^3}{4\hbar^3}\right)\sum_{l}  v_l^a \big\lbrace\partial_b \partial_c \partial_d f_l^0(\varepsilon)\big\rbrace \Big[\frac{1}{\left(1-i\omega\tau_l\right)} + \frac{1}{\left(1+i\omega\tau_l\right)\left(1+2i\omega\tau_l\right)}  +  \frac{1}{\left(1+i\omega\tau_l\right)}    \Big] \frac{\xi_b\xi_c\xi_d}{\left(1+i\omega\tau_l\right)} \notag \\
\end{eqnarray}

\begin{eqnarray}
\mathcal{J}_{3,a}^{in}\left(3\omega\right)   &=&  \left(\frac{e^2\tau_l}{8\hbar}\right)\sum_{l}   \Big[ v_l^a v_l^b G_{cd}  {f_l^0}^{\prime\prime} -(\partial_a \partial_b G_{cd}) f_l^0 \Big]   \frac{\xi_b\xi_c\xi_d }{\left(1+2i\omega \tau_l\right)\left(1+3i\omega \tau_l\right)}   \notag \\
&& -\left(\frac{e^2\tau_l}{8\hbar}\right)\sum_{l}   \left(\partial_b \partial_a G_{cd} \right) f_l^0(\varepsilon) \frac{\xi_b\xi_c\xi_d }{\left(1+i\omega\tau_l\right)}   \notag \\
&& -\left(\frac{e^3\tau_l}{4\hbar^2}\right)  \sum_{l}   \Big[(\partial_b \partial_d G_{ac}) f_l^0(\varepsilon) -(\partial_b\partial_a G_{cd}) f_l^0(\varepsilon)   \Big] \frac{\xi_b\xi_c\xi_d }{\left(1+i\omega\tau_l\right)}  \notag \\
&& -\left(\frac{e^4\tau_l^3}{4\hbar^3}\right)\sum_{l}  v_l^a \big\lbrace\partial_b \partial_c \partial_d f_l^0(\varepsilon)\big\rbrace \frac{\xi_b \xi_c\xi_d }{\left(1+i\omega\tau_l\right)\left(1+2i\omega\tau_l\right)\left(1+3i\omega\tau_l\right)} 
\end{eqnarray}

Further considering $e=1$ and $\hbar=1$ we have-

\begin{eqnarray}
\chi_{abcd}^{in}\left(\omega\right) &=& \sum_{l}  ~\frac{\tau_l}{8}  \Big[ \frac{2}{\left(1+i\omega \tau_l\right)} + \frac{1}{\left(1+i\omega \tau_l\right)\left(1+2i\omega \tau_l\right)}\Big] \left\lbrace v_l^a v_l^b G_{cd}  {f_l^0}^{\prime\prime} -(\partial_a \partial_b G_{cd}) f_l^0 \right\rbrace  \notag \\
&& -\sum_{l}   ~\frac{\tau_l}{8} \Big[ \frac{2}{\left(1+i\omega\tau_l\right)} +  \frac{1}{\left(1-i\omega\tau_l\right)}  \Big] \left(\partial_b \partial_a G_{cd} \right) f_l^0(\varepsilon)  \notag \\
&& -\sum_{l}   ~\frac{\tau_l}{4} \Big[ \frac{2}{\left(1+i\omega\tau_l\right)} + \frac{1}{\left(1-i\omega\tau_l\right)}  \Big] \left\lbrace(\partial_b \partial_d G_{ac}) f_l^0(\varepsilon) -(\partial_b\partial_a G_{cd}) f_l^0(\varepsilon)   \right\rbrace 
\end{eqnarray}

\begin{eqnarray}
\chi_{abcd}^{in}\left(3\omega\right) &=& \sum_{l}  \Bigg[ \frac{\tau_l}{8} \frac{\left\lbrace v_l^a v_l^b G_{cd}  {f_l^0}^{\prime\prime} -(\partial_a \partial_b G_{cd}) f_l^0 \right\rbrace }{\left(1+2i\omega \tau_l\right)\left(1+3i\omega \tau_l\right)} -\frac{\tau_l}{8} \frac{\left(\partial_b \partial_a G_{cd} \right) f_l^0(\varepsilon) }{\left(1+i\omega\tau_l\right)} -\frac{\tau_l}{4}  \frac{(\partial_b \partial_d G_{ac}) f_l^0(\varepsilon) -(\partial_b\partial_a G_{cd}) f_l^0(\varepsilon)   }{\left(1+i\omega\tau_l\right)} \Bigg] \notag\\
\end{eqnarray}

\begin{eqnarray}
\tilde{\chi}_{abcd}^{in}\left(\omega\right) &=& -\sum_{l}  ~\frac{\tau_l^3}{4}  \Big[\frac{1}{\left(1-i\omega\tau_l\right)} + \frac{1}{\left(1+i\omega\tau_l\right)\left(1+2i\omega\tau_l\right)}  +  \frac{1}{\left(1+i\omega\tau_l\right)}    \Big] \frac{v_l^a \big\lbrace\partial_b \partial_c \partial_d f_l^0(\varepsilon)\big\rbrace}{\left(1+i\omega\tau_l\right)} \notag \\
\end{eqnarray}

\begin{eqnarray*}
\tilde{\chi}_{abcd}^{in}\left(3\omega\right) &=&  -\sum_{l}   \frac{\tau_l^3}{4} \frac{v_l^a \big\lbrace\partial_b \partial_c \partial_d f_l^0(\varepsilon)\big\rbrace}{\left(1+i\omega\tau_l\right)\left(1+2i\omega\tau_l\right)\left(1+3i\omega\tau_l\right)} 
\end{eqnarray*}

 Here, $\chi_{abcd}^{in}$ and $\tilde{\chi}_{abcd}^{in}$ represent the linear and cubic dependencies in $\tau_l$ respectively.
 
\subsubsection{Isotropic and low frequency limit}

In the isotropic constant relaxation approximation ( $\tau_l = \tau $) where we neglect the angular dependency in relaxation time and in the low frequency limit $(\omega \tau \ll 1)$ we have total intrinsic Hall conductivity is given by --

\begin{eqnarray}
\chi_{abcd}^{in} &&= \lim_{\omega\tau\rightarrow 0} \left(\chi_{abcd}^{in}\left(3\omega\right)+\chi_{abcd}^{in}\left(\omega\right)\right) \notag\\
&&= \tau\sum_{l}  \Bigg[ \frac{1}{2} \left\lbrace v_l^a v_l^b G_{cd}  {f_l^0}^{\prime\prime} -(\partial_a \partial_b G_{cd}) f_l^0 \right\rbrace -\frac{1}{2}\left(\partial_b \partial_a G_{cd} \right) f_l^0- \left\lbrace \partial_b \partial_d G_{ac} -\partial_b\partial_a G_{cd}  \right\rbrace f_l^0 \Bigg]
\end{eqnarray}

\begin{eqnarray}
\tilde{\chi}_{abcd}^{in} &=& \lim_{\omega\tau\rightarrow 0} \left(\tilde{\chi}_{abcd}^{in}\left(3\omega\right)+\tilde{\chi}_{abcd}^{in}\left(\omega\right)\right) = -\tau^3 \sum_{l} ~v_l^a \big\lbrace\partial_b \partial_c \partial_d f_l^0\big\rbrace
\end{eqnarray}

 Here, $f^{0\prime \prime}_{l}=\frac{\partial^{2} f_{l}^{0} }{\partial^2 \ve_{\bm{l}}}$ with $f_{l}^{0}$  be the equilibrium Fermi distribution function and $v_{l}^{a}=\frac{\partial \ve_l }{\partial {\bm{k_a}}}$.  

\begin{center}
    \underline{Calculation of $\chi_{abcd}^{in}$ for the intrinsic part}
\end{center}

Considering $\lambda = \frac{v_y}{v_x}$, we calculate various components of $\chi_{abcd}^{in}$ and $\tilde{\chi}_{abcd}^{in}$.

\begin{eqnarray}
\chi_{xxxx}^{in} &=& \frac{\tau~v_x^2 \left(2 \mu^4 - 3\Delta^4 + 3\mu^2 \Delta^2 \right)}{ 32\pi \lambda  \mu^7}  + \frac{\tau~t^2 \left(4\mu^6 + 140\Delta^6 - 165\mu^2 \Delta^4 + 20\mu^4 \Delta^2\right)}{64\pi \lambda  \mu ^9} \notag
\end{eqnarray}

\begin{eqnarray}
\chi_{yyyy}^{in} &=& \frac{\tau~\lambda^3 v_x^2 \left(2 \mu^4 - 3\Delta^4 + 3\mu^2 \Delta^2 \right)}{32 \pi  \mu ^7 }  + \frac{\tau~\lambda ^3 t^2 \left(2\mu^6 +28\Delta^6 -15\mu^2 \Delta^4 -12\mu^4 \Delta^2\right)}{64 \pi  \mu ^9} \notag
\end{eqnarray}

\begin{eqnarray*}
\chi_{xyxy}^{in} &=& \frac{\tau~\lambda v_x^2 \left(4 \mu^4 - \Delta^4 - 3\mu^2 \Delta^2 \right)}{32 \pi  \mu ^7 }  + \frac{\tau~\lambda t^2 \left(3\mu^6 +28\Delta^6 -20\mu^4 \Delta^2\right)}{64 \pi  \mu ^9} 
\end{eqnarray*}

\begin{eqnarray*}
\chi_{xyyx}^{in} &=& -\frac{\tau~\lambda v_x^2 \left(4 \mu^4 + \Delta^4 - 5\mu^2 \Delta^2 \right)}{32 \pi  \mu ^7 }  + \frac{\tau~\lambda t^2 \left(28\Delta^6 - 21\mu^6  - 120\mu^2\Delta^4 + 108 \mu^4 \Delta^2\right)}{64 \pi  \mu ^9} 
\end{eqnarray*}

\begin{eqnarray*}
\chi_{xxyy}^{in} &=& -\frac{\tau~\lambda v_x^2 \left(\Delta^4 - \mu^2 \Delta^2 - 2\mu^4\right)}{32 \pi  \mu ^7 }  + \frac{\tau~\lambda t^2 \left(28\Delta^6 + 2\mu^6  - 15\mu^2\Delta^4 - 20 \mu^4 \Delta^2\right)}{64 \pi  \mu ^9} 
\end{eqnarray*}

\begin{eqnarray*}
\chi_{yxxy}^{in} &=& -\frac{\tau~\lambda v_x^2 \left(4 \mu^4 + \Delta^4 - 5\mu^2 \Delta^2 \right)}{32 \pi  \mu ^7 }  + \frac{\tau~\lambda t^2 \left(\mu^6 -28\Delta^6 -28\mu^4 \Delta^2 + 60 \mu^2\Delta^4\right)}{64 \pi  \mu ^9} 
\end{eqnarray*}

\begin{eqnarray*}
\chi_{yxyx}^{in} &=& -\frac{\tau~\lambda v_x^2 \left(\Delta^4 -4\mu^4 + 3\mu^2 \Delta^2 \right)}{32 \pi  \mu ^7 }  + \frac{\tau~\lambda t^2 \left(23\mu^6 +28\Delta^6 -100\mu^4 \Delta^2 + 60 \mu^2\Delta^4\right)}{64 \pi  \mu ^9} 
\end{eqnarray*}

\begin{eqnarray*}
\chi_{yyxx}^{in} &=& -\frac{\tau~\lambda v_x^2 \left(\Delta^4 -2\mu^4 - \mu^2 \Delta^2 \right)}{32 \pi  \mu ^7 }  + \frac{\tau~\lambda t^2 \left(28\Delta^6 +12\mu^4 \Delta^2 - 45 \mu^2\Delta^4\right)}{64 \pi  \mu ^9} 
\end{eqnarray*}

\begin{eqnarray}
\chi_{12}^{in} &=& \frac{1}{3}\left(\chi_{xxyy}^{in} + \chi_{xyxy}^{in} + \chi_{xyyx}^{in}\right) \notag \\
&=& \frac{\tau~\lambda v_x^2 \left(2 \mu^4 - 3\Delta^4 + 3\mu^2 \Delta^2 \right)}{ 96\pi \mu^7} + \frac{\tau~\lambda  t^2 \left(-16\mu^6 + 84 \Delta^6 - 135\mu^2 \Delta^4 + 68\mu^4 \Delta^2\right)}{192 \pi  \mu^9} \notag
\end{eqnarray}

\begin{eqnarray}
\chi_{21}^{in} &=& \frac{1}{3}\left(\chi_{yyxx}^{in} + \chi_{yxyx}^{in} + \chi_{yxxy}^{in}\right) \notag \\
&=& \frac{\tau~\lambda v_x^2 \left(2 \mu^4 - 3\Delta^4 + 3\mu^2 \Delta^2 \right)}{ 96\pi \mu^7}  + \frac{\tau~\lambda t^2 \left(22\mu^6 + 84 \Delta^6 - 45\mu^2 \Delta^4 - 60\mu^4 \Delta^2\right)}{192 \pi  \mu^9} \notag
\end{eqnarray}

   \begin{eqnarray*}
       \left(-\chi_{xxxx}^{in} + 3\chi_{21}^{in}\right) &=& \frac{\tau~\left(\lambda^2-1\right) v_x^2 \left(2\mu^4 - 3 \Delta^4 + 3\mu^2 \Delta^2\right)}{32 \pi  \lambda  \mu^7} \\
       && + \frac{\tau~t^2 \left(2 \left(11 \lambda^2-2\right) \mu^6 + 28\left(3\lambda^2-5\right) \Delta^6 + 15\left(11-3\lambda^2\right) \mu^2 \Delta^4 - 20\left(3\lambda^2 + 1\right) \mu^4 \Delta^2\right)}{64 \pi  \lambda  \mu^9}
   \end{eqnarray*}

   \begin{eqnarray*}
       \left(\chi_{yyyy}^{in} - 3\chi_{12}^{in} \right) &=& \frac{\tau~\lambda  \left(\lambda^2-1\right) v_x^2 \left(2\mu^4 - 3 \Delta^4 + 3\mu ^2 \Delta^2\right)}{32 \pi \mu ^7}  \\
       && + \frac{\tau~\lambda  t^2 \left(2\left(\lambda^2 + 8\right) \mu^6 + 28\left(\lambda^2 - 3\right) \Delta^6 - 15\left(\lambda^2 - 9\right) \mu^2 \Delta^4 - 4\left(3\lambda^2 +17\right) \mu^4 \Delta^2\right)}{64 \pi  \mu^9}
   \end{eqnarray*}

Expressing $\tilde{v}_x = v_x/\mu$, $\Delta = \tilde{\Delta}/\mu$, and $\tilde{t}=t/\mu$,  we have --

\begin{eqnarray*}
	\left(-\chi_{xxxx}^{in} + 3\chi_{21}^{in}\right) &=& \frac{2\tau~\left(\lambda^2-1\right) \tilde{v}_x^2 \left(2 + 3\tilde{\Delta}^2 - 3 \tilde{\Delta}^4 \right)}{64 \pi  \lambda  \mu} \\
	&& + \frac{\tau~\tilde{t}^2 \left\lbrace-4 - 5\tilde{\Delta}^2\left(4-33\tilde{\Delta}^2 + 28 \tilde{\Delta}^4 \right) + \lambda^2 \left(22 - 60 \tilde{\Delta}^2 -45\tilde{\Delta}^4 + 84 \tilde{\Delta}^6 \right) \right\rbrace }{64 \pi  \lambda  \mu}
\end{eqnarray*}

\begin{eqnarray*}
	\left(\chi_{yyyy}^{in} - 3\chi_{12}^{in} \right) &=& \frac{2\tau ~\lambda  \left(\lambda^2-1\right) \tilde{v}_x^2 \left(2 + 3 \tilde{\Delta}^2 - 3 \tilde{\Delta}^4 \right)}{64 \pi \mu}  \\
	&& + \frac{\tau~\lambda  \tilde{t}^2 \left\lbrace 2\left(\lambda^2 + 8\right) + 28\left(\lambda^2 - 3\right) \tilde{\Delta}^6 - 15\left(\lambda^2 - 9\right) \tilde{\Delta}^4 - 4\left(3\lambda^2 +17\right) \tilde{\Delta}^2\right\rbrace }{64 \pi  \mu}
\end{eqnarray*}

\begin{eqnarray*}   \chi_{\perp}^{in}\left(\theta\right)  &=& \left(-\chi_{xxxx}^{in} + 3\chi_{21}^{in}\right)\cos^3(\theta)\sin(\theta)  + \left(\chi_{yyyy}^{in} - 3\chi_{12}^{in} \right)\cos(\theta)\sin^3(\theta) \\\\
&=& \frac{\tau}{\lambda  \mu}\left[\left(\lambda^2-1\right) \tilde{v}_x^2 \mathcal{F}_1(\tilde{\Delta}) + \tilde{t}^2 \mathcal{G}_1(\lambda,\tilde{\Delta})\right] \cos^3(\theta)\sin(\theta)  + \frac{\tau}{  \mu}\left[\lambda\left(\lambda^2-1\right) \tilde{v}_x^2 \mathcal{F}_2(\tilde{\Delta}) +  \lambda~\tilde{t}^2 \mathcal{G}_2(\lambda,\tilde{\Delta}) \right]
\end{eqnarray*}

Here, $\mathcal{F}_1(\tilde{\Delta})$ is the polynomial of $\tilde{\Delta}$. $\mathcal{G}_1(\lambda,\tilde{\Delta})$ and $\mathcal{G}_2(\lambda,\tilde{\Delta})$ both are polynomials of $\lambda$ and $\tilde{\Delta}$. The expression for these polynomials are given below,

\begin{eqnarray*}
\mathcal{F}_1(\tilde{\Delta}) &=& \mathcal{F}_2(\tilde{\Delta}) = \frac{\left(4 + 6 \tilde{\Delta}^2 - 6 \tilde{\Delta}^4 \right)}{64 \pi } \\
\mathcal{G}_1(\lambda,\tilde{\Delta}) &=& \frac{ \left( -4 - 20\tilde{\Delta}^2 +165 \tilde{\Delta}^4 -140 \tilde{\Delta}^6 \right)  + \lambda^2 \left(22 - 60 \tilde{\Delta}^2 -45\tilde{\Delta}^4 + 84 \tilde{\Delta}^6 \right) }{64 \pi } \\
\mathcal{G}_2(\lambda,\tilde{\Delta})
&=& \frac{ \left(16 + 68 \tilde{\Delta}^2 + 135 \tilde{\Delta}^4 -84 \tilde{\Delta}^6\right) + \lambda^2 \left(2-12\tilde{\Delta}^2 -15 \tilde{\Delta}^4 +28\tilde{\Delta}^6 \right) }{64 \pi  }
\end{eqnarray*}


\subsection{Supplementary Note 9: Third order current and $\chi_{abcd}$ associated with side-jump velocity}

The total current attributed to the side-jump velocity is expressed as

\begin{eqnarray}
J^{sj,1}_a &=& -e\sum_{l} ~  v_{sj}^a \delta f_l^{in}
\end{eqnarray}

Here, $\mathbf{v}_l^{sj}$ denotes the side-jump velocity, which describes the transverse velocity resulting from the transverse coordinate shift of the wave packet in response to scattering by an impurity potential. The side-jump velocity is mathematically defined as, $\mathbf{v}_l^{sj} = \sum_{l'}\varpi_{ll'}^{sym} \delta \mathbf{r}_{l'l}$ \cite{sinitsyn2007anomalous}.  

By substituting $\delta f_l^{in}$ from Eq. \eqref{intdist} and consolidating the third-order component of the current, we obtain the following expression.



\begin{eqnarray}\label{THOE_sidejump_velocity}
J^{sj,1}_{3,a} &=& -e\sum_{l} ~  v_{sj}^a \delta f_l^{in} \notag \\
&=& -\sum_{l}  \Bigg[\left(\frac{e^4\tau_l^3}{8\hbar^3}\right) ~v_{sj}^a  \Big\lbrace \frac{\xi_b \xi_c\xi_d e^{3i\omega t}}{\left(1+i\omega\tau_l\right)\left(1+2i\omega\tau_l\right)\left(1+3i\omega\tau_l\right)} + \frac{\xi_b \xi_c\xi^{\ast}_d e^{i\omega t}}{\left(1-i\omega\tau_l\right)\left(1+i\omega\tau_l\right)} + \frac{\xi^{\ast}_b \xi_c\xi_d e^{i\omega t}}{\left(1+i\omega\tau_l\right)^2\left(1+2i\omega\tau_l\right)} \notag \\
&& + \frac{\xi^{\ast}_b \xi_c\xi^{\ast}_d e^{-i\omega t}}{\left(1-i\omega\tau_l\right)^2} + c.c. \Big\rbrace \partial_b \partial_c \partial_d f_l^0(\varepsilon) \notag \\
&-& \left(\frac{e^2\tau_l}{16\hbar}\right) ~v_{sj}^a    \Big\lbrace \frac{\xi_b\xi_c\xi_d e^{3i\omega t}}{\left(1+2i\omega \tau_l\right)\left(1+3i\omega \tau_l\right)} + \frac{\xi^{\ast}_b\xi_c\xi_d e^{i\omega t}}{\left(1+i\omega \tau_l\right)\left(1+2i\omega \tau_l\right)} + \frac{\xi_b\xi_c\xi^{\ast}_d e^{i\omega t}}{\left(1+i\omega \tau_l\right)}  + \frac{\xi^{\ast}_b\xi_c\xi^{\ast}_d e^{-i\omega t}}{\left(1-i\omega \tau_l\right)} + c.c.\Big\rbrace ~\partial_b \big\lbrace G_{cd} {f_l^{0}}^{\prime}(\varepsilon)\big\rbrace \Bigg] \notag \\
\end{eqnarray}


Examining the expression for $J^{sj,1}_{3,a}$ in Eq. \eqref{THOE_sidejump_velocity}, it becomes apparent that the third-order current under time reversal symmetry \emph{vanishes}.

\subsection{Supplementary Note 10: Modification of the distribution function due to side-jump scattering}

The modification to the distribution function induced by the side-jump phenomenon can be derived by substituting Eq.~\eqref{intdist} into Eq.~\eqref{BEqSJ}, we get the expression for $\delta f^{sj}_l$--

\begin{eqnarray*}
\left( \partial_t + \dot{\mathbf{k}}\cdot \partial_{\mathbf{k}} \right) \delta f_l^{sj} &=& \mathcal{I}^{in}_{coll} \left\lbrace \delta f_l^{sj}\right\rbrace + \mathcal{I}^{sj}_{coll} \left\lbrace f_l^{in}\right\rbrace \\
&=& -\frac{\delta f_l^{sj}}{\tau_l}  - e \mathbf{E}\cdot\sum_{l'} \mathbf{O}_{ll'} \left(f_l^{in} - f_{l'}^{in}\right) \notag \\
\Rightarrow \left[ \partial_t -\frac{e}{2\hbar} \left(\xi_a e^{i\omega t} + \xi^{\ast}_a e^{-i\omega t}\right) \partial_{\mathbf{k}}^a \right] \delta f_l^{sj} &=& -\frac{\delta f_l^{sj}}{\tau_l}  - \frac{e}{2} \left(\xi_b e^{i\omega t} + \xi^{\ast}_b e^{-i\omega t}\right) \sum_{l'} \mathbf{O}_{ll'}^{b} \left(f_l^{in} - f_{l'}^{in}\right) \notag \\
\Rightarrow  \left[ 1 + \tau_l\partial_t -\frac{e\tau_l}{2\hbar} \left(\xi_a e^{i\omega t} + \xi^{\ast}_a e^{-i\omega t}\right) \partial_{\mathbf{k}}^a \right] \delta f_l^{sj} &=& - \frac{e\tau_l}{2} \left(\xi_b e^{i\omega t} + \xi^{\ast}_b e^{-i\omega t}\right) \sum_{l'} \mathbf{O}_{ll'}^{b} \left(f_l^{in} - f_{l'}^{in}\right) \notag 
\end{eqnarray*}

\begin{eqnarray}
\delta f_l^{sj} &=& - \frac{\left(e\tau_l/2\right)\left(\xi_b e^{i\omega t} + \xi^{\ast}_b e^{-i\omega t}\right) \sum_{l'} \mathbf{O}_{ll'}^{b} \left(f_l^{in} - f_{l'}^{in}\right)}{\left[ 1 + \tau_l\partial_t -\frac{e\tau_l}{2\hbar} \left(\xi_a e^{i\omega t} + \xi^{\ast}_a e^{-i\omega t}\right) \partial_{\mathbf{k}}^a \right]} \notag \\
\delta f_l^{sj} &=& \sum_{\alpha = 0}^{\infty} \left[ - \tau_l\partial_t +\frac{e\tau_l}{2\hbar} \left(\xi_a e^{i\omega t} + \xi^{\ast}_a e^{-i\omega t}\right) \partial_{\mathbf{k}}^a \right]^{\alpha} \left(-\frac{e\tau_l}{2}\right)\left(\xi_b e^{i\omega t} + \xi^{\ast}_b e^{-i\omega t}\right) \sum_{l'} \mathbf{O}_{ll'}^{b} \left(f_l^{in} - f_{l'}^{in}\right) \notag \\
&\equiv& \delta^{1} f_l^{sj} + \delta^{2} f_l^{sj} + \delta^{3} f_l^{sj} \label{deltaSJ1}
\end{eqnarray}

After tedious algebra we have --

\begin{eqnarray*}
	\delta^1 f_l^{sj} &=& \left(-\frac{e\tau_l}{2}\right) \Big[\frac{\xi_b e^{i\omega t}}{1+i\omega \tau_l} + c.c. \Big]  P_{b,l}
\end{eqnarray*}

\begin{eqnarray*}
	\delta^2 f_l^{sj} &=& -\frac{e^2\tau_l}{4\hbar} \left[ \Big\lbrace \frac{\xi_b\xi_c e^{2i\omega t}}{\left(1+2i\omega \tau_l\right)} + \xi^{\ast}_b\xi_c \Big\rbrace \left(Q_{bc,l} - \overline{Q}_{bc,l}\right) + c.c. \right] -\frac{e^2\tau_l^2}{4\hbar}  \Big[\frac{\xi_b\xi_c e^{2i\omega t}}{\left(1+i\omega \tau_l\right)\left(1+2i\omega \tau_l\right)} + \frac{\xi_b\xi^{\ast}_c }{\left(1-i\omega \tau_l\right)}  +c.c. \Big] \partial_b P_{c,l}
\end{eqnarray*}

\begin{eqnarray*}
	\delta^3 f_l^{sj} &=& -\frac{e^3\tau_l^3}{8\hbar^2}  \Big[\frac{\xi_b\xi_c\xi_d e^{3i\omega t}}{\left(1+i\omega \tau_l\right)\left(1+2i\omega \tau_l\right)\left(1+3i\omega \tau_l\right)}  + \frac{\xi_b\xi_c\xi^{\ast}_d e^{i\omega t}}{\left(1-i\omega \tau_l\right)\left(1+i\omega \tau_l\right)}   + \frac{\xi_b\xi^{\ast}_c\xi_d e^{i\omega t}}{\left(1+i\omega \tau_l\right)^2}  +  \frac{\xi^{\ast}_b \xi_c\xi_d e^{i\omega t}}{\left(1+i\omega \tau_l\right)^2\left(1+2i\omega \tau_l\right)} \\
    && + c.c. \Big] \partial_b \partial_c P_{d,l}   \\
	&& -\frac{e^3\tau_l^2}{8\hbar^2} \Big[\Big\lbrace \frac{\xi_b\xi_c\xi_d e^{3i\omega t}}{\left(1+2i\omega \tau_l\right)\left(1+3i\omega \tau_l\right)} + \frac{\xi_b \xi^{\ast}_c\xi_d e^{i\omega t}}{\left(1+i\omega \tau_l\right)} +  \frac{\xi^{\ast}_b\xi_c\xi_d e^{i\omega t}}{\left(1+i\omega \tau_l\right)\left(1+2i\omega \tau_l\right)} + \frac{\xi^{\ast}_b \xi^{\ast}_c\xi_d e^{-i\omega t}}{\left(1-i\omega \tau_l\right)} \Big\rbrace \partial_b \left(Q_{cd,l} - \overline{Q}_{cd,l}\right) + c.c. \Big] \notag \\
	&&+ \frac{e\tau_l}{2} \Big[ \Big\lbrace \frac{\xi_b \xi_c\xi^{\ast}_d e^{i\omega t}}{\left(1+i\omega \tau_l\right)} + \frac{\xi^{\ast}_b \xi_c\xi^{\ast}_d e^{-i\omega t}}{\left(1-i\omega \tau_l\right)}  \Big\rbrace \left(R_{bcd,l} - \overline{R}_{bcd,l}\right) + c.c. \Big] +  \frac{e\tau_l}{2}  \Big[\left\lbrace \frac{\xi_b\xi_c\xi_d e^{3i\omega t}}{\left(1+3i\omega \tau_l\right)} + \frac{\xi^{\ast}_b \xi_c\xi_d e^{i\omega t}}{\left(1+i\omega \tau_l\right)} \right\rbrace  \left(S_{bcd,l} - \overline{S}_{bcd,l}\right) \notag \\
	&& + c.c. \Big]  -\frac{e^3\tau_l}{8\hbar^2} \Big[ \Big\lbrace \frac{\xi_b \xi_c\xi^{\ast}_d e^{i\omega t}}{\left(1+i\omega \tau_l\right)} + \frac{\xi^{\ast}_b \xi_c\xi^{\ast}_d e^{-i\omega t}}{\left(1-i\omega \tau_l\right)}  \Big\rbrace \left(Q_{bcd,l} - \overline{Q}_{bcd,l}\right) + c.c. \Big] -\frac{e^3\tau_l}{8\hbar^2} \Big[ \left\lbrace\frac{\xi_b\xi_c\xi_d e^{3i\omega t}}{\left(1+3i\omega \tau_l\right)} + \frac{\xi^{\ast}_b \xi_c\xi_d e^{i\omega t}}{\left(1+i\omega \tau_l\right)} \right\rbrace \times \\ \notag && \left(W_{bcd,l} - \overline{W}_{bcd,l}\right) + c.c. \Big]
\end{eqnarray*}

Here, $P_{b,l} = \sum_{l'} \mathbf{O}_{ll'}^{b}\left\lbrace f_l^0(\varepsilon)-f_{l'}^0(\varepsilon)\right\rbrace$, $Q_{bc,l} = \sum_{l'} \mathbf{O}_{ll'}^{b}\frac{\tau_l \partial_{c}^{k} f_{l}^0(\varepsilon)}{\left(1+i\omega\tau_{l}\right)}$, $\overline{Q}_{bc,l} =\sum_{l'} \mathbf{O}_{ll'}^{b} \frac{\tau_{l'}\partial_{c}^{k'} f_{l'}^0(\varepsilon)}{\left(1+i\omega\tau_{l'}\right)}$, $R_{bcd,l} = \sum_{l'} \mathbf{O}_{ll'}^{b}\frac{G_{cd,l} {f_{l}^{0}}^{\prime}(\varepsilon)}{8}$, $\overline{R}_{bcd,l} =\sum_{l'} \mathbf{O}_{ll'}^{b} \frac{G_{cd,{l'}} {f_{l'}^{0}}^{\prime}(\varepsilon)}{8}$, $ S_{bcd,l} = \frac{1}{8}\sum_{l'} \mathbf{O}_{ll'}^{b} \frac{G_{cd,l} {f_{l}^{0}}^{\prime}(\varepsilon)}{1+2i\omega \tau_l} $, $ \overline{S}_{bcd,l} = \frac{1}{8} \sum_{l'} \mathbf{O}_{ll'}^{b}\frac{G_{cd,{l'}} {f_{l'}^{0}}^{\prime}(\varepsilon)}{1+2i\omega \tau_{l'}} $, $ Q_{bcd,l} = \sum_{l'} \mathbf{O}_{ll'}^{b}\frac{\tau_{l}^2 \partial_{c}^{k} \partial_{d}^{k}f_{l}^0(\varepsilon)}{\left(1-i\omega\tau_l\right)} $, $ \overline{Q}_{bcd,l} = \sum_{l'} \mathbf{O}_{ll'}^{b}\frac{\tau_{l'}^2 \partial_{c}^{k'} \partial_{d}^{k'}f_{l'}^0(\varepsilon)}{\left(1-i\omega\tau_{l'}\right)}$, $W_{bcd,l} = \sum_{l'} \mathbf{O}_{ll'}^{b} \frac{\tau_{l}^2 \partial_{c}^{k} \partial_{d}^{k}f_{l}^0(\varepsilon)}{\left(1+i\omega\tau_l\right)\left(1+2i\omega\tau_l\right)}$, $\overline{W}_{bcd,l} = \sum_{l'} \mathbf{O}_{ll'}^{b} \frac{\tau_{l'}^2 \partial_{c}^{k'} \partial_{d}^{k'}f_{l'}^0(\varepsilon)}{\left(1+i\omega\tau_{l'}\right)\left(1+2i\omega\tau_{l'}\right)} $


\begin{table}[h!]
\begin{ruledtabular}
\caption {Terms which are even or odd under time reversal symmetry (TRS). }
\label{tab:table-name}
\begin{tabular}{c c }
\text{Even terms} & $\mathbf{O}_{ll'}$, $P_{b,l}$, $R_{bcd,l}$, $\overline{R}_{bcd,l}$, $S_{bcd,l}$, $\overline{S}_{bcd,l}$, $Q_{bcd,l}$, $\overline{Q}_{bcd,l}$, $W_{bcd,l}$, $\overline{W}_{bcd,l}$ \\
\text{Odd terms} &  $Q_{bc,l}$, $\overline{Q}_{bc,l}$
\end{tabular}
\end{ruledtabular}
\end{table}

Summing up $\delta^1 f_l^{sj}$, $\delta^2 f_l^{sj}$, and $\delta^3 f_l^{sj}$, we can find $\delta f_l^{sj}$--
\begin{eqnarray*}
\delta f_l^{sj} &\equiv & \underbrace{\delta^1 f_l^{sj}}_{\text{TRS even}} +  \underbrace{\delta^2 f_l^{sj}}_{\text{TRS odd}} + \underbrace{\delta^3 f_l^{sj}}_{\text{TRS even}}
\end{eqnarray*}

\subsection{Supplementary Note 11: Third order current due to side-jump induced modifications of the distribution function}

Using the expression for $\delta f_l^{sj}$, we find the expression for current as --

\begin{eqnarray}
J^{sj,2}_a &=& -e\sum_{l} \left[  v_l^a + \partial_a \ve_l^{(2)} +  \frac{e}{2\hbar} \epsilon^{aef} \left(\xi_e e^{i\omega t} + \xi^{\ast}_e e^{-i\omega t} \right) \left( \Omega_l^f + \Omega_l^{(1)f}  \right) \right]  \delta f_l^{sj} \notag \\
&\equiv& -e\sum_{l} \left[  v_l^a + \partial_a \ve_l^{(2)} +  \frac{e}{2\hbar} \epsilon^{aef} \left(\xi_e e^{i\omega t} + \xi^{\ast}_e e^{-i\omega t} \right) \left( \Omega_l^f + \Omega_l^{(1)f}  \right) \right]  \left(\delta^1 f_l^{sj} +  \delta^2 f_l^{sj} + \delta^3 f_l^{sj}\right) \notag 
\end{eqnarray}

From the above expression, we now collect the non-vanishing component of third order current under time reversal symmetry, which is given as --
\begin{eqnarray}
J^{sj,2}_{3,a} &=& -e\sum_{l} \left[  \frac{e}{2\hbar} \epsilon^{aef} \left(\xi_e e^{i\omega t} + \xi^{\ast}_e e^{-i\omega t} \right)  \Omega_l^f \right]  \delta^2 f_l^{sj} \notag \\
 &=& -e\sum_{l} \left[  \frac{e}{2\hbar} \epsilon^{ade} \left(\xi_d e^{i\omega t} + \xi^{\ast}_d e^{-i\omega t} \right)  \Omega_l^e \right] \left(-\frac{e^2\tau_l}{4\hbar}\right) \left[ \Big\lbrace \frac{\xi_b\xi_c e^{2i\omega t}}{\left(1+2i\omega \tau_l\right)} + \xi^{\ast}_b\xi_c \Big\rbrace \left(Q_{bc,l} - \overline{Q}_{bc,l}\right) + c.c. \right] \notag \\
 && -e\sum_{l} \left[  \frac{e}{2\hbar} \epsilon^{ade} \left(\xi_d e^{i\omega t} + \xi^{\ast}_d e^{-i\omega t} \right)  \Omega_l^e \right] \left( -\frac{e^2\tau_l^2}{4\hbar} \right) \Big[\frac{\xi_b\xi_c e^{2i\omega t}}{\left(1+i\omega \tau_l\right)\left(1+2i\omega \tau_l\right)} + \frac{\xi_b\xi^{\ast}_c }{\left(1-i\omega \tau_l\right)}  +c.c. \Big] \partial_b P_{c,l} \notag \\
 &=& \frac{e^4\tau_l}{8\hbar^2} \sum_{l} \epsilon^{ade}  \Omega_l^e \left[ \Big\lbrace \frac{\xi_b\xi_c\xi_d e^{3i\omega t}}{1+2i\omega \tau_l} + \xi^{\ast}_b\xi_c \xi_d e^{i\omega t} + \frac{\xi_b\xi_c\xi^{\ast}_d e^{i\omega t}}{1+2i\omega \tau_l} + \xi^{\ast}_b\xi_c \xi^{\ast}_d e^{-i\omega t}\Big\rbrace \left(Q_{bc,l} - \overline{Q}_{bc,l}\right) + c.c. \right] + \notag \\
 &&  \left(\frac{e^4\tau_l^2}{8\hbar^2} \right) \sum_{l} \epsilon^{ade} \Omega_l^e \Big[\frac{\xi_b\xi_c \xi_d e^{3i\omega t}}{\left(1+i\omega \tau_l\right)\left(1+2i\omega \tau_l\right)} + \frac{\xi_b\xi^{\ast}_c \xi_d e^{i\omega t} }{\left(1-i\omega \tau_l\right)} + \frac{\xi_b\xi_c \xi_d^{\ast} e^{i\omega t}}{\left(1+i\omega \tau_l\right)\left(1+2i\omega \tau_l\right)} + \frac{\xi_b^{\ast}\xi_c \xi_d e^{i\omega t} }{\left(1+i\omega \tau_l\right)} + c.c. \Big] \partial_b P_{c,l} \notag\\
 &=&  J^{sj,2}_{3,a}\left(\omega\right) + J^{sj,2}_{3,a}\left(3\omega\right) \notag \\
 &\equiv & \text{Re}\left[\mathcal{J}_{3,a}^{sj,2}\left(\omega\right) e^{i\omega t} + \mathcal{J}_{3,a}^{sj,2}\left(3\omega\right) e^{3i\omega t}\right]
\end{eqnarray}

\begin{eqnarray}
J^{sj,2}_{3,a}\left(\omega\right) &=& \frac{e^4}{8\hbar^2} \sum_{l} \epsilon^{ade}  \Omega_l^e \Bigg[\tau_l\left\lbrace \Big(  \xi^{\ast}_b\xi_c \xi_d e^{i\omega t} + \frac{\xi_b\xi_c\xi^{\ast}_d e^{i\omega t}}{1+2i\omega \tau_l} + \xi^{\ast}_b\xi_c \xi^{\ast}_d e^{-i\omega t}\Big) \left(Q_{bc,l} - \overline{Q}_{bc,l}\right) + c.c. \right\rbrace  \notag \\
&& + \tau_l^2\left\lbrace \frac{\xi_b\xi^{\ast}_c \xi_d e^{i\omega t} }{\left(1-i\omega \tau_l\right)} + \frac{\xi_b\xi_c \xi_d^{\ast} e^{i\omega t}}{\left(1+i\omega \tau_l\right)\left(1+2i\omega \tau_l\right)} + \frac{\xi_b^{\ast}\xi_c \xi_d e^{i\omega t} }{\left(1+i\omega \tau_l\right)} + c.c.\right\rbrace \Bigg] \partial_b P_{c,l} \notag\\
&\equiv & \text{Re}\left[\mathcal{J}_{3,a}^{sj,2}\left(\omega\right) e^{i\omega t} \right]
\end{eqnarray}

\begin{eqnarray}
\mathcal{J}_{3,a}^{sj,2}\left(\omega\right) &=& \frac{e^4}{4\hbar^2} \sum_{l} \epsilon^{ade}  \Omega_l^e \left[ \frac{\tau_l \left(Q_{bc,l} - \overline{Q}_{bc,l}\right)}{\left(1+2i\omega \tau_l\right)} + \tau_l \left(Q_{bc,l} - \overline{Q}_{bc,l}\right) + \tau_l \left(Q_{bc,l} - \overline{Q}_{bc,l}\right)^\ast\right]\xi_b\xi_c\xi_d \notag \\
&& + \frac{e^4}{4\hbar^2}\sum_{l} \epsilon^{ade}  \Omega_l^e \left[ \frac{\tau_l^2}{\left(1+i\omega \tau_l\right)\left(1+2i\omega \tau_l\right)} + \frac{\tau_l^2}{\left(1-i\omega \tau_l\right)} + \frac{\tau_l^2}{\left(1+i\omega \tau_l\right)} \right] \left(\partial_b P_{c,l} \right) ~\xi_b\xi_c\xi_d
\end{eqnarray}

\begin{eqnarray}
J^{sj,2}_{3,a}\left(3\omega\right) &=& \frac{e^4}{8\hbar^2} \sum_{l} \epsilon^{ade}  \Omega_l^e \Bigg[ \tau_l\left\lbrace\frac{\xi_b\xi_c\xi_d e^{3i\omega t}}{1+2i\omega \tau_l} \left(Q_{bc,l} - \overline{Q}_{bc,l}\right) + c.c. \right\rbrace + \tau_l^2\left\lbrace \frac{\xi_b\xi_c \xi_d e^{3i\omega t}}{\left(1+i\omega \tau_l\right)\left(1+2i\omega \tau_l\right)} + c.c. \right\rbrace \partial_b P_{c,l} \Bigg] \notag\\
&\equiv & \text{Re}\left[ \mathcal{J}_{3,a}^{sj,2}\left(3\omega\right) e^{3i\omega t}\right]
\end{eqnarray}

\begin{eqnarray}
\mathcal{J}_{3,a}^{sj,2}\left(3\omega\right) &=& \frac{e^4}{4\hbar^2}\sum_{l} \epsilon^{ade}  \Omega_l^e \left[\tau_l  \left(Q_{bc,l} - \overline{Q}_{bc,l}\right) + \frac{\tau_l^2  \partial_b P_{c,l} }{\left(1+i\omega \tau_l\right)} \right] \frac{\xi_b\xi_c\xi_d}{\left(1+2i\omega \tau_l\right)}
\end{eqnarray}


Therefore, conductivities are--

\begin{eqnarray*}
\chi^{sj,2}_{abcd}\left(\omega\right) &=& \frac{e^4}{4\hbar^2}\sum_{l} \epsilon^{ade}  \Omega_l^e \left[ \frac{\tau_l \left(Q_{bc,l} - \overline{Q}_{bc,l}\right)}{\left(1+2i\omega \tau_l\right)} + \tau_l \left(Q_{bc,l} - \overline{Q}_{bc,l}\right) + \tau_l \left(Q_{bc,l} - \overline{Q}_{bc,l}\right)^\ast\right] \notag \\
&& + \frac{e^4}{4\hbar^2}\sum_{l} \epsilon^{ade}  \Omega_l^e \left[ \frac{\tau_l^2}{\left(1+i\omega \tau_l\right)\left(1+2i\omega \tau_l\right)} + \frac{\tau_l^2}{\left(1-i\omega \tau_l\right)} + \frac{\tau_l^2}{\left(1+i\omega \tau_l\right)} \right] \partial_b P_{c,l} 
\end{eqnarray*}

\begin{eqnarray}
\chi^{sj,2}_{abcd}\left(3\omega\right) &=& \frac{e^4}{4\hbar^2}\sum_{l} \epsilon^{ade}  \frac{\Omega_l^e}{\left(1+2i\omega \tau_l\right)} \left[\tau_l  \left(Q_{bc,l} - \overline{Q}_{bc,l}\right) + \frac{\tau_l^2  \partial_b P_{c,l} }{\left(1+i\omega \tau_l\right)} \right]
\end{eqnarray}

\subsubsection{Isotropic limit}
In the isotropic limit, the angular dependence in $\tau_l$ can be neglected. Therefore, considering $\tau_l = \tau_{l'} = \tau$, we can reformulate the expression for $\chi_{abc}$.

\begin{eqnarray}
\chi^{sj,2}_{abcd}\left(3\omega\right) &=& \frac{e^4}{4\hbar^2}\sum_{l} \epsilon^{ade}  \frac{\Omega_l^e}{\left(1+2i\omega \tau\right)} \left[\tau  \left(Q_{bc,l} - \overline{Q}_{bc,l}\right) + \frac{\tau^2  \partial_b P_{c,l} }{\left(1+i\omega \tau\right)} \right] \notag \\
&=& \frac{e^4}{4\hbar^2}\sum_{l,l'}  \epsilon^{ade}  \frac{\Omega_l^e}{\left(1+2i\omega \tau\right)} \frac{\tau^2}{\left(1+i\omega\tau\right)} \left[\mathcal{O}_{ll'}^b \left\lbrace\partial_{c}^{k} f_{l}^0(\varepsilon) - \partial_{c}^{k'} f_{l'}^0(\varepsilon)\right\rbrace + \partial_b^k \left\lbrace\mathcal{O}_{ll'}^{c} \left( f_l^0-f_{l'}^0\right)\right\rbrace \right] \notag \\
\end{eqnarray}

\begin{eqnarray}
\chi^{sj,2}_{abcd}\left(\omega\right) &=& \frac{e^4}{4\hbar^2}\sum_{l} \epsilon^{ade}  \Omega_l^e \sum_{l'} \left[ \frac{\tau^2}{\left(1+i\omega\tau\right)\left(1+2i\omega\tau\right)} + \frac{\tau^2}{\left(1+i\omega\tau\right)}  + \frac{\tau^2}{\left(1-i\omega\tau\right)}\right] \mathcal{O}_{ll'}^b \left\lbrace\partial_{c}^{k} f_{l}^0 - \partial_{c}^{k'} f_{l'}^0\right\rbrace \notag \\
&& + \frac{e^4}{4\hbar^2}\sum_{l} \epsilon^{ade}  \Omega_l^e \left[ \frac{\tau^2}{\left(1+i\omega\tau\right)\left(1+2i\omega\tau\right)} + \frac{\tau^2}{\left(1+i\omega\tau\right)}  + \frac{\tau^2}{\left(1-i\omega\tau\right)}\right] \sum_{l'}  \partial_b^k \left\lbrace\mathcal{O}_{ll'}^{c} \left( f_l^0-f_{l'}^0\right)\right\rbrace \notag \\
&=& \frac{e^4}{4\hbar^2} \sum_{l,l'}\, \epsilon^{ade}  \Omega_l^e  \left[ \frac{\tau^2}{\left(1+i\omega\tau\right)\left(1+2i\omega\tau\right)} + \frac{\tau^2}{\left(1+i\omega\tau\right)}  + \frac{\tau^2}{\left(1-i\omega\tau\right)}\right] \times\notag \\
&& \hspace{6cm} \left[ \mathcal{O}_{ll'}^b \left\lbrace\partial_{c}^{k} f_{l}^0 - \partial_{c}^{k'} f_{l'}^0\right\rbrace  + \partial_b^k \left\lbrace\mathcal{O}_{ll'}^{c} \left( f_l^0-f_{l'}^0\right)\right\rbrace\right]
\end{eqnarray}





Here,
\begin{eqnarray*}
\mathcal{O}_{ll'}^b = \frac{2\pi}{\hbar} \vert T_{ll'}\vert^2 \delta r_{ll'}^b \left\lbrace\frac{\partial}{\partial \ve_l} \delta \left(\ve_l - \ve_{l'} \right) \right\rbrace
\end{eqnarray*}

\begin{eqnarray*}
  && \sum_{ll'}  \mathbf{O}_{ll'}^{b} \left[\tau_l \partial_{c}^{k} f_{l}^0(\varepsilon) - \tau_{l'}\partial_{c}^{k'} f_{l'}^0(\varepsilon) \right] \\
  &=& \frac{2\pi}{\hbar} \sum_{ll'} \tau_l \Omega_l^e \vert T_{ll'}\vert^2 \delta r_{ll'}^b \left\lbrace\frac{\partial}{\partial \ve_l} \delta \left(\ve_l - \ve_{l'} \right)\right\rbrace \left[\tau_l \partial_{c}^{k} f_{l}^0(\varepsilon) - \tau_{l'}\partial_{c}^{k'} f_{l'}^0(\varepsilon) \right] \\
\end{eqnarray*}

\subsubsection{Low frequency limit}

Further, considering $e = 1$ and $\hbar = 1$ and in the low-frequency limit, $(\omega \tau \ll 1)$ we have total side jump contribution in the Hall conductivity ($\chi^{sj,2}$) due
to side jump scattering is --


\begin{eqnarray}
\chi^{sj,2}_{abcd} &=& \lim_{\omega\tau\rightarrow 0} \left(\chi^{sj,2}_{abcd}\left(3\omega\right)+\chi^{sj,2}_{abcd}\left(\omega\right)\right) \notag\\
&=&\tau^2~\sum_{l,l'}\, \epsilon^{ade}  \Omega_l^e   \left[ \mathcal{O}_{ll'}^b \left\lbrace\partial_{c}^{k} f_{l}^0 - \partial_{c}^{k'} f_{l'}^0\right\rbrace  + \partial_b^k \left\lbrace\mathcal{O}_{ll'}^{c} \left( f_l^0-f_{l'}^0\right)\right\rbrace\right] \label{chisj2}
\end{eqnarray}

\subsection{Supplementary Note 12: Modification of the distribution function due to skew-scattering contribution}

Substituting the Eq.\,\eqref{intdist} into Eq.\,\eqref{BEqSK}, we find the expression for $\delta f_l^{sk}$ as follows,

\begin{eqnarray}
\left( \partial_t + \dot{\mathbf{k}}\cdot \partial_{\mathbf{k}} \right) \delta f_l^{sk} &=& \mathcal{I}^{in}_{coll} \left\lbrace \delta f_l^{sk}\right\rbrace + \mathcal{I}^{sk}_{coll} \left\lbrace f_l^{in}\right\rbrace  \notag\\
\Rightarrow \left[ \partial_t -\frac{e}{2\hbar} \left(\xi_a e^{i\omega t} + \xi^{\ast}_a e^{-i\omega t}\right) \partial_a \right] \delta f_l^{sk} &=& -\frac{\delta f_l^{sk}}{\tau_l}  - \sum_{l'} \varpi_{l'l}^{asym} \left(f_l^{in} + f_{l'}^{in}\right) \notag \\
\end{eqnarray}

\begin{eqnarray*}
\delta f_l^{sk} &=& \frac{- \tau_l\sum_{l'} \varpi_{l'l}^{asym} \left(f_l^{in} + f_{l'}^{in}\right)}{1+ \tau_l\partial_t - \left(\frac{e\tau_l}{2\hbar}\right)\left(\xi_a e^{i\omega t} + \xi^{\ast}_a e^{-i\omega t}\right) \partial_a }   \notag \\
&=& -\sum_{n=0}^{\infty} \left[-\tau_l\partial_t + \left(\frac{e\tau_l}{2\hbar}\right)\left(\xi_d e^{i\omega t} + \xi^{\ast}_d e^{-i\omega t}\right) \partial_d \right]^n \tau_l\sum_{l'} \varpi_{l'l}^{asym} \left(f_l^{in} + f_{l'}^{in}\right) \notag \\
&=& -\tau_l\sum_{n=0}^{\infty} \left[-\tau_l\partial_t + \left(\frac{e\tau_l}{2\hbar}\right)\left(\xi_d e^{i\omega t} + \xi^{\ast}_d e^{-i\omega t}\right) \partial_d \right]^n \sum_{l'} \varpi_{l'l}^{asym} \left(f_l^{in} + f_{l'}^{in}\right) \notag \\
&\equiv& \delta^1 f_l^{sk} + \delta^2 f_l^{sk}  + \delta^3 f_l^{sk} + \cdots
\end{eqnarray*}

Note that the equilibrium distribution function does not contribute to the scattering,
$ \sum_{l'} \varpi_{l'l}^{asym}\Big\lbrace f_l^0(\varepsilon)+f_{l'}^0(\varepsilon) \Big\rbrace = 0 $

 After performing the tedious algebra we have,

\begin{eqnarray}
\delta^1_{1st} f_l^{sk} &=& -\tau_l \sum_{n=0}^{\infty} \left(-\tau_l \partial_t \right)^n  \sum_{l'} \varpi_{l'l}^{asym}\Big\lbrace  \frac{e}{2\hbar}\Big[\frac{\tau_l \partial_{b}^{k} f_{l}^0(\varepsilon)}{\left(1+i\omega\tau_{l}\right)}+\frac{\tau_{l'}\partial_{b}^{k'} f_{l'}^0(\varepsilon)}{\left(1+i\omega\tau_{l'}\right)}\Big] \xi_b e^{i\omega t} + c.c \Big \rbrace \notag \\
& =& -\tau_l \sum_{l'} \varpi_{l'l}^{asym}\Big\lbrace  \frac{e}{2\hbar}\Big[\frac{\tau_l \partial_{b}^{k} f_{l}^0(\varepsilon)}{\left(1+i\omega\tau_{l}\right)}+\frac{\tau_{l'}\partial_{b}^{k'} f_{l'}^0(\varepsilon)}{\left(1+i\omega\tau_{l'}\right)}\Big] \frac{\xi_b e^{i\omega t}}{\left(1+i\omega \tau_l \right)} + c.c \Big \rbrace  \notag \\
& =& -\tau_l \sum_{l'} \varpi_{l'l}^{asym}\Big\lbrace  \frac{e}{2\hbar}\Big[\frac{\tau_l \partial_{b}^{k} f_{l}^0(\varepsilon)}{\left(1+i\omega\tau_{l}\right)^2}+\frac{\tau_{l'}\partial_{b}^{k'} f_{l'}^0(\varepsilon)}{\left(1+i\omega \tau_l \right) \left(1+i\omega\tau_{l'}\right)}\Big] \xi_b e^{i\omega t} + c.c \Big \rbrace \notag
\end{eqnarray}

\begin{eqnarray}
\delta^2 f_l^{sk} &=& -\tau_l \sum_{l'} \varpi_{l'l}^{asym}\Big\lbrace - \frac{1}{8}\Big[ G_{bc,l} {f_{l}^{0}}^{\prime}(\varepsilon) + G_{bc,{l'}} {f_{l'}^{0}}^{\prime}(\varepsilon)\Big]\xi_b\xi^{\ast}_c   + c.c. \Big\rbrace \notag \\
&& -\tau_l \sum_{l'} \varpi_{l'l}^{asym}\Big\lbrace -\frac{1}{8}\Big[ \frac{G_{bc,l} {f_{l}^{0}}^{\prime}(\varepsilon)}{\left(1+2i\omega \tau_l\right)^2} + \frac{G_{bc,{l'}} {f_{l'}^{0}}^{\prime}(\varepsilon)}{\left(1+2i\omega \tau_{l'}\right)\left(1+2i\omega \tau_l\right)} \Big] \xi_b\xi_c e^{2i\omega t}  + c.c. \Big\rbrace \notag \\
&& -\tau_l \sum_{l'} \varpi_{l'l}^{asym}\Big\lbrace \left(\frac{e^2\tau_l}{4\hbar^2}\right) \Big[\frac{\tau_l \partial_b^{k} \partial_{c}^{k} f_{l}^0(\varepsilon)}{\left(1+i\omega\tau_{l}\right)^2}+\frac{\tau_{l'} \partial_b^{k} \partial_{c}^{k'} f_{l'}^0(\varepsilon)}{\left(1+i\omega \tau_l \right) \left(1+i\omega\tau_{l'}\right)}\Big]  \Big[\frac{\xi_b\xi_c e^{2i\omega t}}{\left(1+2i\omega\tau_l\right)} + \xi^{\ast}_b \xi_c\Big] + c.c \Big \rbrace \notag \\
&&  -\tau_l   \sum_{l'} \varpi_{l'l}^{asym}\Big\lbrace \left(\frac{e}{2\hbar}\right)^2 \Big[ \frac{\tau_{l}^2 \partial_{b}^{k} \partial_{c}^{k}f_{l}^0(\varepsilon)}{\left(1+i\omega\tau_l\right)\left(1+2i\omega\tau_l\right)^2} + \frac{\tau_{l'}^2 \partial_{b}^{k'} \partial_{c}^{k'}f_{l'}^0(\varepsilon)}{\left(1+i\omega\tau_{l'}\right)\left(1+2i\omega\tau_{l'}\right)\left(1+2i\omega\tau_l\right)} \Big] \xi_b\xi_c e^{2i\omega t} + c.c. \Big\rbrace \notag \\
&& -\tau_l \sum_{l'} \varpi_{l'l}^{asym}\Big\lbrace \left(\frac{e}{2\hbar}\right)^2  \Big[ \frac{\tau_{l}^2 \partial_{b}^{k} \partial_{c}^{k}f_{l}^0(\varepsilon)}{\left(1-i\omega\tau_l\right)} + \frac{\tau_{l'}^2 \partial_{b}^{k'} \partial_{c}^{k'}f_{l'}^0(\varepsilon)}{\left(1-i\omega\tau_{l'}\right)} \Big] \xi_b\xi^{\ast}_c   + c.c. \Big\rbrace 
\end{eqnarray}

\begin{eqnarray}
\delta^3 f_l^{sk} &=& -\tau_l  \sum_{l'} \varpi_{l'l}^{asym}\Big\lbrace \left(\frac{e^3\tau_l^2}{8\hbar^3}\right) \Big[\frac{\tau_l \partial_b^{k} \partial_c^{k} \partial_{d}^{k} f_{l}^0(\varepsilon)}{\left(1+i\omega\tau_{l}\right)^2}+\frac{\tau_{l'} \partial_b^{k} \partial_c^{k} \partial_{d}^{k'} f_{l'}^0(\varepsilon)}{\left(1+i\omega \tau_l \right) \left(1+i\omega\tau_{l'}\right)}\Big] \notag  \\ \notag 
&& \times \Big[\frac{\xi_b\xi_c\xi_d e^{3i\omega t}}{\left(1+2i\omega\tau_l\right)\left(1+3i\omega\tau_l\right)} + \frac{\xi^{\ast}_b\xi_c\xi_d e^{i\omega t}}{\left(1+i\omega\tau_l\right)\left(1+2i\omega\tau_l\right)} + \frac{\xi_b \xi^{\ast}_c \xi_d e^{i\omega t}}{\left(1+i\omega\tau_l\right)} + \frac{\xi^{\ast}_b \xi^{\ast}_c \xi_d e^{-i\omega t}}{\left(1-i\omega\tau_l\right)}\Big] + c.c \Big \rbrace \notag \\
&&  -\tau_l \sum_{l'} \varpi_{l'l}^{asym}\Big\lbrace \left(-\frac{e\tau_l}{16\hbar}\right) \Big[ \frac{\partial_b^{k} \big\lbrace G_{cd,l} {f_{l}^{0}}^{\prime}(\varepsilon)\big\rbrace }{\left(1+2i\omega \tau_l\right)^2} + \frac{\partial_b^{k} \big\lbrace G_{cd,{l'}} {f_{l'}^{0}}^{\prime}(\varepsilon)\big\rbrace}{\left(1+2i\omega \tau_{l'}\right)\left(1+2i\omega \tau_l\right)} \Big]  \Big[\frac{\xi_b \xi_c\xi_d e^{3i\omega t}}{\left(1+3i\omega \tau_l\right)} + \frac{\xi^{\ast}_b \xi_c\xi_d e^{i\omega t}}{\left(1+i\omega \tau_l\right)} \Big]  + c.c. \Big\rbrace \notag \\
&&  -\tau_l  \sum_{l'} \varpi_{l'l}^{asym}\Big\lbrace \left(-\frac{e\tau_l}{16\hbar}\right) \Big[ \partial_b^{k}\big\lbrace G_{cd,l} {f_{l}^{0}}^{\prime}(\varepsilon) \big\rbrace +  \partial_b^{k} \big\lbrace G_{cd,{l'}} {f_{l'}^{0}}^{\prime}(\varepsilon)\big\rbrace \Big]  \Big[\frac{\xi_b \xi_c\xi^{\ast}_d e^{i\omega t}}{\left(1+i\omega \tau_l\right)} + \frac{\xi^{\ast}_b \xi_c\xi^{\ast}_d  e^{-i\omega t}}{\left(1-i\omega \tau_l\right)}\Big] + c.c. \Big\rbrace \notag \\
&& -\tau_l \sum_{l'} \varpi_{l'l}^{asym}\Big\lbrace \left(\frac{e^3\tau_l}{8\hbar^3}\right) \Big[ \frac{\tau_{l}^2 \partial_b^{k} \partial_{c}^{k} \partial_{d}^{k}f_{l}^0(\varepsilon)}{\left(1+i\omega\tau_l\right)\left(1+2i\omega\tau_l\right)^2} + \frac{\tau_{l'}^2 \partial_b^{k} \partial_{c}^{k'} \partial_{d}^{k'}f_{l'}^0(\varepsilon)}{\left(1+i\omega\tau_{l'}\right)\left(1+2i\omega\tau_{l'}\right)\left(1+2i\omega\tau_l\right)} \Big] \notag \\ 
&& \times \Big[\frac{\xi_b \xi_c\xi_d e^{3i\omega t}}{\left(1+3i\omega \tau_l\right)}  + \frac{\xi^{\ast}_b \xi_c\xi_d e^{i\omega t}}{\left(1+i\omega \tau_l\right)} \Big]   + c.c. \Big\rbrace \notag \\
&&  -\tau_l \sum_{l'} \varpi_{l'l}^{asym}\Big\lbrace \left(\frac{e^3\tau_l}{8\hbar^3}\right)  \Big[ \frac{\tau_{l}^2 \partial_b^{k} \partial_{c}^{k} \partial_{d}^{k}f_{l}^0(\varepsilon)}{\left(1-i\omega\tau_l\right)} + \frac{\tau_{l'}^2 \partial_b^{k} \partial_{c}^{k'} \partial_{d}^{k'}f_{l'}^0(\varepsilon)}{\left(1-i\omega\tau_{l'}\right)} \Big] \Big[\frac{\xi_b\xi_c\xi^{\ast}_d e^{i\omega t}}{\left(1+i\omega\tau_l\right)} + \frac{\xi^{\ast}_b\xi_c\xi^{\ast}_d e^{-i\omega t}}{\left(1-i\omega\tau_l\right)}\Big]     + c.c. \Big\rbrace \notag \\
&& -\tau_l\sum_{l'} \varpi_{l'l}^{asym}\Big\lbrace  \left(-\frac{e}{16\hbar}\right) \Big[ \frac{\tau_l \partial^{k}_b \big\lbrace G_{cd,l} {f_l^{0}}^{\prime}(\varepsilon)\big\rbrace }{\left(1+2i\omega \tau_l\right)\left(1+3i\omega \tau_l\right)} + \frac{\tau_{l'} \partial^{k'}_b \big\lbrace G_{cd,l'} {f_{l'}^{0}}^{\prime}(\varepsilon)\big\rbrace}{\left(1+2i\omega \tau_{l'}\right)\left(1+3i\omega \tau_{l'}\right)}  \Big] \frac{\xi_b\xi_c\xi_d e^{3i\omega t}}{\left(1+3i\omega \tau_l\right)} \notag  \\
 && + \left(-\frac{e}{16\hbar}\right) \Big[ \frac{\tau_l \partial^{k}_b \big\lbrace G_{cd,l} {f_l^{0}}^{\prime}(\varepsilon)\big\rbrace }{\left(1+i\omega \tau_l\right)\left(1+2i\omega \tau_l\right)} + \frac{\tau_{l'} \partial^{k'}_b \big\lbrace G_{cd,l'} {f_{l'}^{0}}^{\prime}(\varepsilon)\big\rbrace}{\left(1+i\omega \tau_{l'}\right)\left(1+2i\omega \tau_{l'}\right)}  \Big]  \frac{\xi^{\ast}_b\xi_c\xi_d e^{i\omega t}}{\left(1+i\omega \tau_l\right)} \notag \\
 && + \left(-\frac{e}{16\hbar}\right) \Big[ \frac{\tau_l \partial^{k}_b \big\lbrace G_{cd,l} {f_l^{0}}^{\prime}(\varepsilon)\big\rbrace }{\left(1+i\omega \tau_l\right)} + \frac{\tau_{l'} \partial^{k'}_b \big\lbrace G_{cd,l'} {f_{l'}^{0}}^{\prime}(\varepsilon)\big\rbrace}{\left(1+i\omega \tau_{l'}\right)}  \Big]\frac{ \xi_b\xi_c\xi^{\ast}_d e^{i\omega t}}{\left(1+i\omega \tau_l\right)} \notag \\
 && + \left(-\frac{e}{16\hbar}\right) \Big[ \frac{\tau_l \partial^{k}_b \big\lbrace G_{cd,l} {f_l^{0}}^{\prime}(\varepsilon)\big\rbrace }{\left(1+i\omega \tau_l\right)} + \frac{\tau_{l'} \partial^{k'}_b \big\lbrace G_{cd,l'} {f_{l'}^{0}}^{\prime}(\varepsilon)\big\rbrace}{\left(1+i\omega \tau_{l'}\right)}  \Big] \frac{\xi_b\xi^{\ast}_c\xi_d e^{i\omega t}}{\left(1+i\omega \tau_l\right)} \notag \\
 && + \left(\frac{e}{2\hbar}\right)^3 \Big[ \frac{\tau_l^2 \partial^k_b \partial^k_c \partial^k_d f_l^0(\varepsilon)}{\left(1+i\omega\tau_l\right)\left(1+2i\omega\tau_l\right)\left(1+3i\omega\tau_l\right)} + \frac{\tau_{l'}^2 \partial^{k'}_b \partial^{k'}_c \partial^{k'}_d f_{l'}^0(\varepsilon)}{\left(1+i\omega\tau_{l'}\right)\left(1+2i\omega\tau_{l'}\right)\left(1+3i\omega\tau_{l'}\right)} \Big] \frac{\xi_b \xi_c\xi_d e^{3i\omega t}}{\left(1+3i\omega \tau_l\right)} \notag\\
&& + \left(\frac{e}{2\hbar}\right)^3 \Big[ \frac{\tau_l^2 \partial^k_b \partial^k_c \partial^k_d f_l^0(\varepsilon)}{\left(1-i\omega\tau_l\right)\left(1+i\omega\tau_l\right)} + \frac{\tau_{l'}^2 \partial^{k'}_b \partial^{k'}_c \partial^{k'}_d f_{l'}^0(\varepsilon)}{\left(1-i\omega\tau_{l'}\right)\left(1+i\omega\tau_{l'}\right)} \Big] \frac{\xi_b \xi_c\xi^{\ast}_d e^{i\omega t}}{\left(1+i\omega \tau_l\right)}  \notag\\
&& + \left(\frac{e}{2\hbar}\right)^3 \Big[ \frac{\tau_l^2 \partial^k_b \partial^k_c \partial^k_d f_l^0(\varepsilon)}{\left(1+i\omega\tau_l\right)^2\left(1+2i\omega\tau_l\right)} + \frac{\tau_{l'}^2 \partial^{k'}_b \partial^{k'}_c \partial^{k'}_d f_{l'}^0(\varepsilon)}{\left(1+i\omega\tau_{l'}\right)^2\left(1+2i\omega\tau_{l'}\right)} \Big] \frac{\xi^{\ast}_b \xi_c\xi_d e^{i\omega t}}{\left(1+i\omega \tau_l\right)} \notag\\
&& + \left(\frac{e}{2\hbar}\right)^3 \Big[ \frac{\tau_l^2 \partial^k_b \partial^k_c \partial^k_d f_l^0(\varepsilon)}{\left(1+i\omega\tau_l\right)^2} + \frac{\tau_{l'}^2 \partial^{k'}_b \partial^{k'}_c \partial^{k'}_d f_{l'}^0(\varepsilon)}{\left(1+i\omega\tau_l\right)^2} \Big] \frac{\xi_b \xi^{\ast}_c\xi_d e^{i\omega t}}{\left(1+i\omega \tau_l\right)}  + c.c. \Big\rbrace
\end{eqnarray}

\subsection{Supplementary Note 13: Third order current and $\chi_{abcd}$ due to skew-scattering contribution }

The expression for current due to $\delta f_l^{sk}$ is given by,

\begin{eqnarray}
J^{sk}_a &=& -e\sum_{l}\, \left[  v_l^a -\frac{1}{8} \left[ \xi_c\xi_d e^{2i\omega t} + \xi_c\xi^{\ast}_d + c.c. \right] \partial_a G_{cd} +  \frac{e}{2\hbar} \epsilon^{aef} \left(\xi_e e^{i\omega t} + \xi^{\ast}_e e^{-i\omega t} \right) \left( \Omega_l^f + \Omega_l^{(1)f}  \right) \right]  \delta f_l^{sk} \notag 
\end{eqnarray}

Under time reversal symmetry ($\mathcal{T}$), both $\delta^1 f_l^{sk}$ and $\delta^3 f_l^{sk}$ are even function of momenta while $\delta^2 f_l^{sk}$ is odd function of momenta. Thus, the non-vanishing third order current under $\mathcal{T}$ symmetry due to $\delta f_l^{sk}$ is now given by,

\begin{eqnarray*}
J^{sk}_{3,a} &=& J^{sk}_{3,a}(\omega) + J^{sk}_{3,a}(3\omega),
\end{eqnarray*}
where,
\begin{eqnarray*}
J^{sk}_{3,a}(\omega) &=& -\frac{e^2}{2\hbar} \sum_{l}  \, \epsilon^{ade} ~\Omega_l^e \sum_{l'} \varpi_{l'l}^{asym} \Bigg[  \frac{\tau_l}{8} \left\lbrace G_{bc,l} {f_{l}^{0}}^{\prime}(\varepsilon) + G_{bc,{l'}} {f_{l'}^{0}}^{\prime}(\varepsilon)\right\rbrace \left\lbrace \xi_b\xi^{\ast}_c \xi_d e^{i\omega t} + \xi^{\ast}_b\xi_c \xi_d e^{i\omega t} + c.c. \right\rbrace \notag \\
&&  + \frac{\tau_l}{8}  \left\lbrace \Big[ \frac{G_{bc,l} {f_{l}^{0}}^{\prime}(\varepsilon)}{\left(1+2i\omega \tau_l\right)^2} + \frac{G_{bc,{l'}} {f_{l'}^{0}}^{\prime}(\varepsilon)}{\left(1+2i\omega \tau_{l'}\right)\left(1+2i\omega \tau_l\right)} \Big]\xi_b\xi_c\xi^{\ast}_d e^{i\omega t} + c.c. \right\rbrace \notag \\
&& -\tau_l \left(\frac{e^2\tau_l}{4\hbar^2}\right) \Bigg\lbrace \Big[\frac{\tau_l \partial_b^{k} \partial_{c}^{k} f_{l}^0(\varepsilon)}{\left(1+i\omega\tau_{l}\right)^2}+\frac{\tau_{l'} \partial_b^{k'} \partial_{c}^{k'} f_{l'}^0(\varepsilon)}{\left(1+i\omega \tau_l \right) \left(1+i\omega\tau_{l'}\right)}\Big]\frac{\xi_b\xi_c \xi^{\ast}_d e^{i\omega t}}{\left(1+2i\omega\tau_l\right)} + c.c. \Bigg\rbrace \notag \\ 
&& -\tau_l \left(\frac{e^2\tau_l}{4\hbar^2}\right) \Bigg\lbrace \Big[\frac{\tau_l \partial_b^{k} \partial_{c}^{k} f_{l}^0(\varepsilon)}{\left(1+i\omega\tau_{l}\right)^2}+\frac{\tau_{l'} \partial_b^{k'} \partial_{c}^{k'} f_{l'}^0(\varepsilon)}{\left(1+i\omega \tau_l \right) \left(1+i\omega\tau_{l'}\right)}\Big]\xi^{\ast}_b \xi_c \xi_d e^{i\omega t} + c.c. \Bigg\rbrace \notag \\ 
&& -\tau_l \left(\frac{e^2\tau_l}{4\hbar^2}\right) \Bigg\lbrace \Big[\frac{\tau_l \partial_b^{k} \partial_{c}^{k} f_{l}^0(\varepsilon)}{\left(1+i\omega\tau_{l}\right)^2}+\frac{\tau_{l'} \partial_b^{k'} \partial_{c}^{k'} f_{l'}^0(\varepsilon)}{\left(1+i\omega \tau_l \right) \left(1+i\omega\tau_{l'}\right)}\Big]\xi^{\ast}_b \xi_c \xi^{\ast}_d e^{-i\omega t}  + c.c. \Bigg\rbrace \notag \\ 
&& -\tau_l \left(\frac{e}{2\hbar}\right)^2   \left\lbrace\Big[ \frac{\tau_{l}^2 \partial_{b}^{k} \partial_{c}^{k}f_{l}^0(\varepsilon)}{\left(1+i\omega\tau_l\right)\left(1+2i\omega\tau_l\right)^2} + \frac{\tau_{l'}^2 \partial_{b}^{k'} \partial_{c}^{k'}f_{l'}^0(\varepsilon)}{\left(1+i\omega\tau_{l'}\right)\left(1+2i\omega\tau_{l'}\right)\left(1+2i\omega\tau_l\right)} \Big]  \xi_b\xi_c\xi^{\ast}_d e^{i\omega t} + c.c. \right\rbrace \notag \\
&& -\tau_l \left(\frac{e}{2\hbar}\right)^2 \left\lbrace  \Big[ \frac{\tau_{l}^2 \partial_{b}^{k} \partial_{c}^{k}f_{l}^0(\varepsilon)}{\left(1-i\omega\tau_l\right)} + \frac{\tau_{l'}^2 \partial_{b}^{k'} \partial_{c}^{k'}f_{l'}^0(\varepsilon)}{\left(1-i\omega\tau_{l'}\right)} \Big] \xi_b\xi^{\ast}_c \xi_d e^{i\omega t}   + c.c. \right\rbrace  \notag \\
&& -\tau_l \left(\frac{e}{2\hbar}\right)^2 \left\lbrace  \Big[ \frac{\tau_{l}^2 \partial_{b}^{k} \partial_{c}^{k}f_{l}^0(\varepsilon)}{\left(1-i\omega\tau_l\right)} + \frac{\tau_{l'}^2 \partial_{b}^{k'} \partial_{c}^{k'}f_{l'}^0(\varepsilon)}{\left(1-i\omega\tau_{l'}\right)} \Big] \xi_b\xi^{\ast}_c \xi^{\ast}_d e^{-i\omega t}  + c.c. \right\rbrace  \Bigg] \notag\\
&\equiv & \text{Re}\left[\mathcal{J}_{3,a}^{sk}\left(\omega\right) e^{i\omega t} \right]
\end{eqnarray*}

\begin{eqnarray*}
\mathcal{J}_{3,a}^{sk}\left(\omega\right) &=& -\frac{e^2}{4\hbar} \sum_{l,l'}  \, \epsilon^{ade} ~\tau_l\Omega_l^e \varpi_{l'l}^{asym} \left[ G_{bc,l} {f_{l}^{0}}^{\prime}(\varepsilon) + G_{bc,{l'}} {f_{l'}^{0}}^{\prime}(\varepsilon)\right] \xi_b\xi_c \xi_d \notag \\
&& -\frac{e^2}{8\hbar} \sum_{l,l'}  \, \epsilon^{ade} ~\frac{\tau_l\Omega_l^e \varpi_{l'l}^{asym}}{\left(1+2i\omega \tau_l\right)} \Bigg[ \frac{G_{bc,l} {f_{l}^{0}}^{\prime}(\varepsilon)}{\left(1+2i\omega \tau_l\right)} + \frac{G_{bc,{l'}} {f_{l'}^{0}}^{\prime}(\varepsilon)}{\left(1+2i\omega \tau_{l'}\right)} \Bigg]\xi_b\xi_c\xi_d  \notag \\
&& + \left(\frac{e^4}{4\hbar^3}\right) \sum_{l,l'}  \, \epsilon^{ade} ~\frac{\tau_l^2 \Omega_l^e \varpi_{l'l}^{asym}}{\left(1+2i\omega\tau_l\right)} \Bigg[\frac{\tau_l \partial_b^{k} \partial_{c}^{k} f_{l}^0(\varepsilon)}{\left(1+i\omega\tau_{l}\right)^2}+\frac{\tau_{l'} \partial_b^{k'} \partial_{c}^{k'} f_{l'}^0(\varepsilon)}{\left(1+i\omega \tau_l \right) \left(1+i\omega\tau_{l'}\right)}\Bigg]\xi_b\xi_c \xi_d  \notag \\ 
&& + \left(\frac{e^4}{4\hbar^3}\right) \sum_{l,l'}  \, \epsilon^{ade} ~\frac{\tau_l^2 \Omega_l^e \varpi_{l'l}^{asym}}{\left(1+i\omega \tau_l \right)} \Bigg[\frac{\tau_l \partial_b^{k} \partial_{c}^{k} f_{l}^0(\varepsilon)}{\left(1+i\omega\tau_{l}\right)}+\frac{\tau_{l'} \partial_b^{k'} \partial_{c}^{k'} f_{l'}^0(\varepsilon)}{ \left(1+i\omega\tau_{l'}\right)}\Bigg]\xi_b \xi_c \xi_d   \notag \\ 
&& + \left(\frac{e^4}{4\hbar^3}\right) \sum_{l,l'}  \, \epsilon^{ade} ~\frac{\tau_l^2 \Omega_l^e \varpi_{l'l}^{asym}}{\left(1-i\omega \tau_l \right)} \Bigg[\frac{\tau_l \partial_b^{k} \partial_{c}^{k} f_{l}^0(\varepsilon)}{\left(1-i\omega\tau_{l}\right)} + \frac{\tau_{l'} \partial_b^{k'} \partial_{c}^{k'} f_{l'}^0(\varepsilon)}{ \left(1-i\omega\tau_{l'}\right)}\Bigg]\xi_b \xi_c \xi_d  \notag \\ 
&& + \left(\frac{e^4}{4\hbar^3}\right)  \sum_{l,l'}  \, \epsilon^{ade} ~\frac{\tau_l \Omega_l^e \varpi_{l'l}^{asym}}{\left(1+2i\omega\tau_l\right)} \Bigg[ \frac{\tau_{l}^2 \partial_{b}^{k} \partial_{c}^{k}f_{l}^0(\varepsilon)}{\left(1+i\omega\tau_l\right)\left(1+2i\omega\tau_l\right)} + \frac{\tau_{l'}^2 \partial_{b}^{k'} \partial_{c}^{k'}f_{l'}^0(\varepsilon)}{\left(1+i\omega\tau_{l'}\right)\left(1+2i\omega\tau_{l'}\right)} \Bigg]  \xi_b\xi_c\xi_d \notag \\
&& + \left(\frac{e^4}{4\hbar^3}\right) \sum_{l,l'}  \, \epsilon^{ade} ~\tau_l \Omega_l^e \varpi_{l'l}^{asym} \Bigg[ \frac{\tau_{l}^2 \partial_{b}^{k} \partial_{c}^{k}f_{l}^0(\varepsilon)}{\left(1-i\omega\tau_l\right)} + \frac{\tau_{l'}^2 \partial_{b}^{k'} \partial_{c}^{k'}f_{l'}^0(\varepsilon)}{\left(1-i\omega\tau_{l'}\right)} \Bigg] \xi_b\xi_c \xi_d  \notag \\
&& + \left(\frac{e^4}{4\hbar^3}\right) \sum_{l,l'}  \, \epsilon^{ade} ~\tau_l\Omega_l^e \varpi_{l'l}^{asym} \Bigg[ \frac{\tau_{l}^2 \partial_{b}^{k} \partial_{c}^{k}f_{l}^0(\varepsilon)}{\left(1+i\omega\tau_l\right)} + \frac{\tau_{l'}^2 \partial_{b}^{k'} \partial_{c}^{k'}f_{l'}^0(\varepsilon)}{\left(1+i\omega\tau_{l'}\right)} \Bigg] \xi_b\xi_c \xi_d  
\end{eqnarray*}

\begin{eqnarray*}
J^{sk}_{3,a}(3\omega) &=& -\frac{e^2}{2\hbar} \sum_{l}  \, \epsilon^{ade} ~\Omega_l^e \sum_{l'} \varpi_{l'l}^{asym} \Bigg[ \frac{\tau_l}{8}\left\lbrace \Big[ \frac{G_{bc,l} {f_{l}^{0}}^{\prime}(\varepsilon)}{\left(1+2i\omega \tau_l\right)^2} + \frac{G_{bc,{l'}} {f_{l'}^{0}}^{\prime}(\varepsilon)}{\left(1+2i\omega \tau_{l'}\right)\left(1+2i\omega \tau_l\right)} \Big]\xi_b\xi_c\xi_d e^{3i\omega t} + c.c. \right\rbrace \notag \\
&& -\tau_l \left(\frac{e^2\tau_l}{4\hbar^2}\right) \Bigg\lbrace \Big[\frac{\tau_l \partial_b^{k} \partial_{c}^{k} f_{l}^0(\varepsilon)}{\left(1+i\omega\tau_{l}\right)^2}+\frac{\tau_{l'} \partial_b^{k'} \partial_{c}^{k'} f_{l'}^0(\varepsilon)}{\left(1+i\omega \tau_l \right) \left(1+i\omega\tau_{l'}\right)}\Big] \frac{\xi_b\xi_c \xi_d e^{3i\omega t}}{\left(1+2i\omega\tau_l\right)} + c.c. \Bigg\rbrace \notag \\ 
&& -\tau_l \left(\frac{e}{2\hbar}\right)^2 \Bigg\lbrace \Big[ \frac{\tau_{l}^2 \partial_{b}^{k} \partial_{c}^{k}f_{l}^0(\varepsilon)}{\left(1+i\omega\tau_l\right)\left(1+2i\omega\tau_l\right)^2} + \frac{\tau_{l'}^2 \partial_{b}^{k'} \partial_{c}^{k'}f_{l'}^0(\varepsilon)}{\left(1+i\omega\tau_{l'}\right)\left(1+2i\omega\tau_{l'}\right)\left(1+2i\omega\tau_l\right)} \Big]  \xi_b\xi_c\xi_d e^{3i\omega t} + c.c. \Bigg\rbrace \Bigg]\notag\\
&\equiv & \text{Re}\left[\mathcal{J}_{3,a}^{sk}\left(3\omega\right) e^{3i\omega t} \right]
\end{eqnarray*}

\begin{eqnarray*}
\mathcal{J}_{3,a}^{sk}\left(3\omega\right) &=& -\frac{e^2 }{8\hbar} \sum_{l,l'}  \, \epsilon^{ade} ~\tau_l~\Omega_l^e \varpi_{l'l}^{asym}  \Bigg[ \frac{G_{bc,l} {f_{l}^{0}}^{\prime}(\varepsilon)}{\left(1+2i\omega \tau_l\right)^2} + \frac{G_{bc,{l'}} {f_{l'}^{0}}^{\prime}(\varepsilon)}{\left(1+2i\omega \tau_{l'}\right)\left(1+2i\omega \tau_l\right)} \Bigg]\xi_b\xi_c\xi_d \notag \\
&& + \left(\frac{e^4}{4\hbar^3}\right)\sum_{l,l'}  \, \epsilon^{ade} ~\tau_l^2~\Omega_l^e \varpi_{l'l}^{asym} \Bigg[\frac{\tau_l \partial_b^{k} \partial_{c}^{k} f_{l}^0(\varepsilon)}{\left(1+i\omega\tau_{l}\right)^2}+\frac{\tau_{l'} \partial_b^{k'} \partial_{c}^{k'} f_{l'}^0(\varepsilon)}{\left(1+i\omega \tau_l \right) \left(1+i\omega\tau_{l'}\right)}\Bigg] \frac{\xi_b\xi_c \xi_d }{\left(1+2i\omega\tau_l\right)} \notag \\ 
&&  + \left(\frac{e^4}{4\hbar^3}\right)\sum_{l,l'}  \, \epsilon^{ade} ~\tau_l~\Omega_l^e \varpi_{l'l}^{asym} \Bigg[ \frac{\tau_{l}^2 \partial_{b}^{k} \partial_{c}^{k}f_{l}^0(\varepsilon)}{\left(1+i\omega\tau_l\right)\left(1+2i\omega\tau_l\right)^2} + \frac{\tau_{l'}^2 \partial_{b}^{k'} \partial_{c}^{k'}f_{l'}^0(\varepsilon)}{\left(1+i\omega\tau_{l'}\right)\left(1+2i\omega\tau_{l'}\right)\left(1+2i\omega\tau_l\right)} \Bigg]  \xi_b\xi_c\xi_d \notag
\end{eqnarray*}

\begin{eqnarray*}
\chi^{sk}_{abcd}\left(\omega\right) &=& -\frac{e^2}{4\hbar} \sum_{l,l'}  \, \epsilon^{ade} ~\tau_l\Omega_l^e \varpi_{l'l}^{asym} \left[ G_{bc,l} {f_{l}^{0}}^{\prime}(\varepsilon) + G_{bc,{l'}} {f_{l'}^{0}}^{\prime}(\varepsilon)\right]  \notag \\
&& -\frac{e^2}{8\hbar} \sum_{l,l'}  \, \epsilon^{ade} ~\frac{\tau_l\Omega_l^e \varpi_{l'l}^{asym}}{\left(1+2i\omega \tau_l\right)} \Bigg[ \frac{G_{bc,l} {f_{l}^{0}}^{\prime}(\varepsilon)}{\left(1+2i\omega \tau_l\right)} + \frac{G_{bc,{l'}} {f_{l'}^{0}}^{\prime}(\varepsilon)}{\left(1+2i\omega \tau_{l'}\right)} \Bigg]  \notag \\
&& + \left(\frac{e^4}{4\hbar^3}\right) \sum_{l,l'}  \, \epsilon^{ade} ~\frac{\tau_l^2 \Omega_l^e \varpi_{l'l}^{asym}}{\left(1+2i\omega\tau_l\right)} \Bigg[\frac{\tau_l \partial_b^{k} \partial_{c}^{k} f_{l}^0(\varepsilon)}{\left(1+i\omega\tau_{l}\right)^2}+\frac{\tau_{l'} \partial_b^{k'} \partial_{c}^{k'} f_{l'}^0(\varepsilon)}{\left(1+i\omega \tau_l \right) \left(1+i\omega\tau_{l'}\right)}\Bigg]  \notag \\ 
&& + \left(\frac{e^4}{4\hbar^3}\right) \sum_{l,l'}  \, \epsilon^{ade} ~\frac{\tau_l^2 \Omega_l^e \varpi_{l'l}^{asym}}{\left(1+i\omega \tau_l \right)} \Bigg[\frac{\tau_l \partial_b^{k} \partial_{c}^{k} f_{l}^0(\varepsilon)}{\left(1+i\omega\tau_{l}\right)}+\frac{\tau_{l'} \partial_b^{k'} \partial_{c}^{k'} f_{l'}^0(\varepsilon)}{ \left(1+i\omega\tau_{l'}\right)}\Bigg]   \notag \\ 
&& + \left(\frac{e^4}{4\hbar^3}\right) \sum_{l,l'}  \, \epsilon^{ade} ~\frac{\tau_l^2 \Omega_l^e \varpi_{l'l}^{asym}}{\left(1-i\omega \tau_l \right)} \Bigg[\frac{\tau_l \partial_b^{k} \partial_{c}^{k} f_{l}^0(\varepsilon)}{\left(1-i\omega\tau_{l}\right)} + \frac{\tau_{l'} \partial_b^{k'} \partial_{c}^{k'} f_{l'}^0(\varepsilon)}{ \left(1-i\omega\tau_{l'}\right)}\Bigg] \notag \\ 
&& + \left(\frac{e^4}{4\hbar^3}\right)  \sum_{l,l'}  \, \epsilon^{ade} ~\frac{\tau_l \Omega_l^e \varpi_{l'l}^{asym}}{\left(1+2i\omega\tau_l\right)} \Bigg[ \frac{\tau_{l}^2 \partial_{b}^{k} \partial_{c}^{k}f_{l}^0(\varepsilon)}{\left(1+i\omega\tau_l\right)\left(1+2i\omega\tau_l\right)} + \frac{\tau_{l'}^2 \partial_{b}^{k'} \partial_{c}^{k'}f_{l'}^0(\varepsilon)}{\left(1+i\omega\tau_{l'}\right)\left(1+2i\omega\tau_{l'}\right)} \Bigg]   \notag \\
&& + \left(\frac{e^4}{4\hbar^3}\right) \sum_{l,l'}  \, \epsilon^{ade} ~\tau_l \Omega_l^e \varpi_{l'l}^{asym} \Bigg[ \frac{\tau_{l}^2 \partial_{b}^{k} \partial_{c}^{k}f_{l}^0(\varepsilon)}{\left(1-i\omega\tau_l\right)} + \frac{\tau_{l'}^2 \partial_{b}^{k'} \partial_{c}^{k'}f_{l'}^0(\varepsilon)}{\left(1-i\omega\tau_{l'}\right)} \Bigg]   \notag \\
&& + \left(\frac{e^4}{4\hbar^3}\right) \sum_{l,l'}  \, \epsilon^{ade} ~\tau_l\Omega_l^e \varpi_{l'l}^{asym} \Bigg[ \frac{\tau_{l}^2 \partial_{b}^{k} \partial_{c}^{k}f_{l}^0(\varepsilon)}{\left(1+i\omega\tau_l\right)} + \frac{\tau_{l'}^2 \partial_{b}^{k'} \partial_{c}^{k'}f_{l'}^0(\varepsilon)}{\left(1+i\omega\tau_{l'}\right)} \Bigg]  
\end{eqnarray*}

\begin{eqnarray*}
\chi^{sk}_{abcd}\left(3\omega\right) &=& -\frac{e^2 }{8\hbar} \sum_{l,l'}  \, \epsilon^{ade} ~\tau_l~\Omega_l^e \varpi_{l'l}^{asym}  \Bigg[ \frac{G_{bc,l} {f_{l}^{0}}^{\prime}(\varepsilon)}{\left(1+2i\omega \tau_l\right)^2} + \frac{G_{bc,{l'}} {f_{l'}^{0}}^{\prime}(\varepsilon)}{\left(1+2i\omega \tau_{l'}\right)\left(1+2i\omega \tau_l\right)} \Bigg] \notag \\
&& + \left(\frac{e^4}{4\hbar^3}\right)\sum_{l,l'}  \, \epsilon^{ade}  \frac{\tau_l^2~\Omega_l^e \varpi_{l'l}^{asym}}{\left(1+2i\omega\tau_l\right)} \Bigg[\frac{\tau_l \partial_b^{k} \partial_{c}^{k} f_{l}^0(\varepsilon)}{\left(1+i\omega\tau_{l}\right)^2}+\frac{\tau_{l'} \partial_b^{k'} \partial_{c}^{k'} f_{l'}^0(\varepsilon)}{\left(1+i\omega \tau_l \right) \left(1+i\omega\tau_{l'}\right)}\Bigg] \notag \\ 
&&  + \left(\frac{e^4}{4\hbar^3}\right)\sum_{l,l'}  \, \epsilon^{ade} \frac{\tau_l~\Omega_l^e \varpi_{l'l}^{asym}}{\left(1+2i\omega\tau_l\right)} \Bigg[ \frac{\tau_{l}^2 \partial_{b}^{k} \partial_{c}^{k}f_{l}^0(\varepsilon)}{\left(1+i\omega\tau_l\right)\left(1+2i\omega\tau_l\right)} + \frac{\tau_{l'}^2 \partial_{b}^{k'} \partial_{c}^{k'}f_{l'}^0(\varepsilon)}{\left(1+i\omega\tau_{l'}\right)\left(1+2i\omega\tau_{l'}\right)} \Bigg]  
\end{eqnarray*}

\subsubsection{Isotropic limit}
In the isotropic limit, the angular dependence in $\tau_l$ can be neglected. Therefore, considering $\tau_l = \tau_{l'} = \tau$, we can reformulate the expression for $\chi_{abc}$.

\begin{eqnarray}
\chi^{sk}_{abcd}\left(\omega\right) &=& -\frac{e^2}{4\hbar} \sum_{l,l'}  \, \epsilon^{ade} ~\Omega_l^e \varpi_{l'l}^{asym} \left[ G_{bc,l} {f_{l}^{0}}^{\prime}(\varepsilon) + G_{bc,{l'}} {f_{l'}^{0}}^{\prime}(\varepsilon)\right]~\tau  \notag \\
&& -\frac{e^2}{8\hbar} \sum_{l,l'}  \, \epsilon^{ade} ~\Omega_l^e \varpi_{l'l}^{asym} \Big[ G_{bc,l} {f_{l}^{0}}^{\prime}(\varepsilon) + G_{bc,{l'}} {f_{l'}^{0}}^{\prime}(\varepsilon) \Big] \frac{\tau}{\left(1+2i\omega \tau\right)^2} \notag \\
&& + \left(\frac{e^4}{4\hbar^3}\right) \sum_{l,l'}  \, \epsilon^{ade} ~\Omega_l^e \varpi_{l'l}^{asym} \Big[\partial_b^{k} \partial_{c}^{k} f_{l}^0(\varepsilon) + \partial_b^{k'} \partial_{c}^{k'} f_{l'}^0(\varepsilon)\Big]  \frac{\tau^3}{\left(1+i\omega \tau\right)^2\left(1+2i\omega \tau\right)}\notag \\ 
&& + \left(\frac{e^4}{4\hbar^3}\right) \sum_{l,l'}  \, \epsilon^{ade} ~\Omega_l^e \varpi_{l'l}^{asym} \Big[\partial_b^{k} \partial_{c}^{k} f_{l}^0(\varepsilon) + \partial_b^{k'} \partial_{c}^{k'} f_{l'}^0(\varepsilon)\Big]  \frac{\tau^3}{\left(1+i\omega \tau\right)^2}  \notag \\ 
&& + \left(\frac{e^4}{4\hbar^3}\right) \sum_{l,l'}  \, \epsilon^{ade}  ~\Omega_l^e \varpi_{l'l}^{asym} \Big[\partial_b^{k} \partial_{c}^{k} f_{l}^0(\varepsilon) + \partial_b^{k'} \partial_{c}^{k'} f_{l'}^0(\varepsilon)\Big]  \frac{\tau^3}{\left(1-i\omega \tau\right)^2}  \notag \\ 
&& + \left(\frac{e^4}{4\hbar^3}\right)  \sum_{l,l'}  \, \epsilon^{ade} ~\Omega_l^e \varpi_{l'l}^{asym} \Big[\partial_b^{k} \partial_{c}^{k} f_{l}^0(\varepsilon) + \partial_b^{k'} \partial_{c}^{k'} f_{l'}^0(\varepsilon)\Big]  \frac{\tau^3}{\left(1+i\omega \tau\right)\left(1+2i\omega \tau\right)^2}\notag \\ 
&& + \left(\frac{e^4}{4\hbar^3}\right) \sum_{l,l'}  \, \epsilon^{ade} ~\Omega_l^e \varpi_{l'l}^{asym} \Big[\partial_b^{k} \partial_{c}^{k} f_{l}^0(\varepsilon) + \partial_b^{k'} \partial_{c}^{k'} f_{l'}^0(\varepsilon)\Big]  \frac{\tau^3}{\left(1-i\omega \tau\right)}  \notag \\ 
&& + \left(\frac{e^4}{4\hbar^3}\right) \sum_{l,l'}  \, \epsilon^{ade} ~\Omega_l^e \varpi_{l'l}^{asym} \Big[\partial_b^{k} \partial_{c}^{k} f_{l}^0(\varepsilon) + \partial_b^{k'} \partial_{c}^{k'} f_{l'}^0(\varepsilon)\Big]  \frac{\tau^3}{\left(1+i\omega \tau\right)^2}  \notag 
\end{eqnarray}

\begin{eqnarray*}
\chi^{sk}_{abcd}\left(\omega\right) &=& -\frac{e^2}{4\hbar} \sum_{l,l'}  \, \epsilon^{ade} ~\tau_l\Omega_l^e \varpi_{l'l}^{asym} \left[ G_{bc,l} {f_{l}^{0}}^{\prime}(\varepsilon) + G_{bc,{l'}} {f_{l'}^{0}}^{\prime}(\varepsilon)\right]  \notag \\
&& -\frac{e^2}{8\hbar} \sum_{l,l'}  \, \epsilon^{ade} ~\frac{\tau_l\Omega_l^e \varpi_{l'l}^{asym}}{\left(1+2i\omega \tau_l\right)} \Bigg[ \frac{G_{bc,l} {f_{l}^{0}}^{\prime}(\varepsilon)}{\left(1+2i\omega \tau_l\right)} + \frac{G_{bc,{l'}} {f_{l'}^{0}}^{\prime}(\varepsilon)}{\left(1+2i\omega \tau_{l'}\right)} \Bigg]  \notag \\
&& + \left(\frac{e^4}{4\hbar^3}\right) \sum_{l,l'}  \, \epsilon^{ade} ~\frac{\tau_l^2 \Omega_l^e \varpi_{l'l}^{asym}}{\left(1+2i\omega\tau_l\right)} \Bigg[\frac{\tau_l \partial_b^{k} \partial_{c}^{k} f_{l}^0(\varepsilon)}{\left(1+i\omega\tau_{l}\right)^2}+\frac{\tau_{l'} \partial_b^{k'} \partial_{c}^{k'} f_{l'}^0(\varepsilon)}{\left(1+i\omega \tau_l \right) \left(1+i\omega\tau_{l'}\right)}\Bigg]  \notag \\ 
&& + \left(\frac{e^4}{4\hbar^3}\right) \sum_{l,l'}  \, \epsilon^{ade} ~\frac{\tau_l^2 \Omega_l^e \varpi_{l'l}^{asym}}{\left(1+i\omega \tau_l \right)} \Bigg[\frac{\tau_l \partial_b^{k} \partial_{c}^{k} f_{l}^0(\varepsilon)}{\left(1+i\omega\tau_{l}\right)}+\frac{\tau_{l'} \partial_b^{k'} \partial_{c}^{k'} f_{l'}^0(\varepsilon)}{ \left(1+i\omega\tau_{l'}\right)}\Bigg]   \notag \\ 
&& + \left(\frac{e^4}{4\hbar^3}\right) \sum_{l,l'}  \, \epsilon^{ade} ~\frac{\tau_l^2 \Omega_l^e \varpi_{l'l}^{asym}}{\left(1-i\omega \tau_l \right)} \Bigg[\frac{\tau_l \partial_b^{k} \partial_{c}^{k} f_{l}^0(\varepsilon)}{\left(1-i\omega\tau_{l}\right)} + \frac{\tau_{l'} \partial_b^{k'} \partial_{c}^{k'} f_{l'}^0(\varepsilon)}{ \left(1-i\omega\tau_{l'}\right)}\Bigg] \notag \\ 
&& + \left(\frac{e^4}{4\hbar^3}\right)  \sum_{l,l'}  \, \epsilon^{ade} ~\frac{\tau_l \Omega_l^e \varpi_{l'l}^{asym}}{\left(1+2i\omega\tau_l\right)} \Bigg[ \frac{\tau_{l}^2 \partial_{b}^{k} \partial_{c}^{k}f_{l}^0(\varepsilon)}{\left(1+i\omega\tau_l\right)\left(1+2i\omega\tau_l\right)} + \frac{\tau_{l'}^2 \partial_{b}^{k'} \partial_{c}^{k'}f_{l'}^0(\varepsilon)}{\left(1+i\omega\tau_{l'}\right)\left(1+2i\omega\tau_{l'}\right)} \Bigg]   \notag \\
&& + \left(\frac{e^4}{4\hbar^3}\right) \sum_{l,l'}  \, \epsilon^{ade} ~\tau_l \Omega_l^e \varpi_{l'l}^{asym} \Bigg[ \frac{\tau_{l}^2 \partial_{b}^{k} \partial_{c}^{k}f_{l}^0(\varepsilon)}{\left(1-i\omega\tau_l\right)} + \frac{\tau_{l'}^2 \partial_{b}^{k'} \partial_{c}^{k'}f_{l'}^0(\varepsilon)}{\left(1-i\omega\tau_{l'}\right)} \Bigg]   \notag \\
&& + \left(\frac{e^4}{4\hbar^3}\right) \sum_{l,l'}  \, \epsilon^{ade} ~\tau_l\Omega_l^e \varpi_{l'l}^{asym} \Bigg[ \frac{\tau_{l}^2 \partial_{b}^{k} \partial_{c}^{k}f_{l}^0(\varepsilon)}{\left(1+i\omega\tau_l\right)} + \frac{\tau_{l'}^2 \partial_{b}^{k'} \partial_{c}^{k'}f_{l'}^0(\varepsilon)}{\left(1+i\omega\tau_{l'}\right)} \Bigg]  
\end{eqnarray*}

\begin{eqnarray*}
\chi^{sk}_{abcd}\left(3\omega\right) &=& -\frac{e^2 }{8\hbar} \sum_{l,l'}  \, \epsilon^{ade} ~\Omega_l^e \varpi_{l'l}^{asym}  \Big[ G_{bc,l} {f_{l}^{0}}^{\prime}(\varepsilon) + G_{bc,{l'}} {f_{l'}^{0}}^{\prime}(\varepsilon) \Big] \frac{\tau}{\left(1+2i\omega \tau\right)^2} \notag \\
&& + \left(\frac{e^4}{4\hbar^3}\right)\sum_{l,l'}  \, \epsilon^{ade}  ~\Omega_l^e \varpi_{l'l}^{asym} \Big[\partial_b^{k} \partial_{c}^{k} f_{l}^0(\varepsilon) + \partial_b^{k'} \partial_{c}^{k'} f_{l'}^0(\varepsilon) \Big] \frac{\tau^3}{\left(1+i\omega \tau\right)^2 \left(1+2i\omega \tau\right)}  \notag \\ 
&&  + \left(\frac{e^4}{4\hbar^3}\right)\sum_{l,l'}  \, \epsilon^{ade}  ~\Omega_l^e \varpi_{l'l}^{asym} \Big[\partial_b^{k} \partial_{c}^{k} f_{l}^0(\varepsilon) + \partial_b^{k'} \partial_{c}^{k'} f_{l'}^0(\varepsilon) \Big] \frac{\tau^3}{\left(1+i\omega \tau\right) \left(1+2i\omega \tau\right)^2}   
\end{eqnarray*}

\subsubsection{Low frequency limit}

In the the low frequency limit $(\omega \tau \ll 1)$ we have total 
skew scattering dependent 3$^{rd}$ Hall conductivity ($\chi^{sk}$) is given by--

\begin{eqnarray}
\chi^{sk}_{abcd} &=& \lim_{\omega\tau\rightarrow 0} \left(\chi^{sk}_{abcd}\left(3\omega\right)+\chi^{sk}_{abcd}\left(\omega\right)\right) \notag\\
&=& -\frac{\tau e^2 }{2\hbar} \sum_{l,l'}  \, \epsilon^{ade} ~\Omega_l^e \varpi_{l'l}^{asym}  \Big[ G_{bc,l} {f_{l}^{0}}^{\prime}(\varepsilon) + G_{bc,{l'}} {f_{l'}^{0}}^{\prime}(\varepsilon) \Big]  \notag \\
&&  + \left(\frac{8\tau^3 e^4}{4\hbar^3}\right)\sum_{l,l'}  \, \epsilon^{ade}  ~\Omega_l^e \varpi_{l'l}^{asym} \Big[\partial_b^{k} \partial_{c}^{k} f_{l}^0(\varepsilon) + \partial_b^{k'} \partial_{c}^{k'} f_{l'}^0(\varepsilon) \Big]    
\end{eqnarray}

Further simplification of $\chi^{sk}_{abcd}$ gives,-
\begin{eqnarray}
\chi^{sk}_{abcd} &=&  \frac{\tau e^2 }{2\hbar} \sum_{l,l'}\, \epsilon^{ade}~\varpi_{ll'}^{asym}\left\lbrace \Omega_l^e     - \Omega_{l'}^e  \right\rbrace G_{bc,{l}} ~\frac{\partial f_{l}^0(\ve_l)}{\partial \ve_l}   \notag \\ 
&&  + \left(\frac{2\tau^3 e^4}{\hbar^3}\right)\sum_{l,l'}   \, \epsilon^{ade}~ (\partial_b^{k} \left\lbrace \varpi_{ll'}^{asym} \left( \Omega_l^e  - \Omega_{l'}^e  \right)\right\rbrace)  \frac{\partial \ve_l}{\partial k_c} \frac{\partial f_{l}^0(\ve_l)}{\partial \ve_l}    
\end{eqnarray}

\begin{center}
    \underline{Calculation of skew scattering related conductivities $\chi_{abcd}^{sk}$}
\end{center}

The anti-symmetric scattering rate ($\varpi_{ll'}^{asym}$) includes leading-order contributions from both third-order ($\varpi_{ll'}^{3asym}$) and fourth-order ($\varpi_{ll'}^{4asym}$)terms. Assuming $e=\hbar=1$ and considering a simpler case where the velocities satisfy $v_y=v_x$, we compute various components of $\chi^{sk}_{abcd}$. For $t < v_x$,  interband scattering is energetically forbidden in the weak disorder limit of our model, allowing us to omit the band index in the subsequent calculation of $\chi^{sk}_{abcd}$.

\begin{eqnarray}
\chi^{sk,3asym}_{abcd} &=&  \frac{\tau }{2}   \int[d\mathbf{k}]\int[d\mathbf{k'}] \epsilon^{ade}~\varpi_{kk'}^{3asym}\left\lbrace \Omega_k^e     - \Omega_{k'}^e  \right\rbrace G_{bc,{k}} ~\frac{\partial f_{k}^0(\ve_k)}{\partial \ve_k}   \notag \\ 
&&  + 2\tau^3 \int[d\mathbf{k}]\int[d\mathbf{k'}] \epsilon^{ade}~ (\partial_b^{k} \left\lbrace \varpi_{kk'}^{3asym} \left( \Omega_k^e  - \Omega_{k'}^e  \right)\right\rbrace)  \frac{\partial \ve_k}{\partial k_c} \frac{\partial f_{k}^0(\ve_k)}{\partial \ve_k}    
\end{eqnarray}

$$\chi^{sk,3asym}_{xxxx} = \chi^{sk,3asym}_{yyyy} = \chi^{sk,3asym}_{xyyx} = \chi^{sk,3asym}_{yxxy} = 0$$

\begin{eqnarray*}
\chi^{sk,3asym}_{xxyy} &=&  \frac{\tau }{2}  ~\frac{t^2 \pi ^3 \Delta^2 \left(\Delta^2-\mu ^2\right)^2 \left(5 \Delta^2-\mu ^2\right) N_i^2  V_1^3}{(2\pi)^2~ 64 \mu^8 v_x^4 }     - 2\tau^3 ~\frac{t^2 \pi^3 \Delta^4 \left(11 \mu^4+7\Delta^4 -18 \mu ^2 \Delta^2\right) N_i^2 V_1^3 }{(2\pi)^2~16 \mu^6 v_x^4}
\end{eqnarray*}

\begin{eqnarray*}
\chi^{sk,3asym}_{xyxy} &=&  \frac{\tau }{2}  ~\frac{t^2 \pi ^3 \Delta^2 \left(\Delta^2-\mu ^2\right)^2 \left(5 \Delta^2-\mu ^2\right) N_i^2  V_1^3}{(2\pi)^2~64 \mu^8 v_x^4 }    - 2\tau^3 ~\frac{t^2 \pi^3 \Delta^4 \left(11 \mu^4+7\Delta^4 -18 \mu ^2 \Delta^2\right) N_i^2 V_1^3 }{(2\pi)^2~16 \mu^6 v_x^4}
\end{eqnarray*}

\begin{eqnarray*}
\chi^{sk,3asym}_{yyxx} &=&  -\frac{\tau }{2}  ~\frac{t^2 \pi ^3 \Delta^2 \left(\Delta^2-\mu ^2\right)^2 \left(5 \Delta^2-\mu ^2\right) N_i^2  V_1^3}{(2\pi)^2~ 64 \mu^8 v_x^4 }    + 2\tau^3 ~\frac{t^2 \pi^3 \Delta^4 \left(11 \mu^4+7\Delta^4 -18 \mu ^2 \Delta^2\right) N_i^2 V_1^3 }{(2\pi)^2~ 16 \mu^6 v_x^4}
\end{eqnarray*}

\begin{eqnarray*}
\chi^{sk,3asym}_{yxyx} &=&  -\frac{\tau }{2}  ~\frac{t^2 \pi ^3 \Delta^2 \left(\Delta^2-\mu ^2\right)^2 \left(5 \Delta^2-\mu ^2\right) N_i^2  V_1^3}{(2\pi)^2~ 64 \mu^8 v_x^4 }     + 2\tau^3 ~\frac{t^2 \pi^3 \Delta^4 \left(11 \mu^4+7\Delta^4 -18 \mu ^2 \Delta^2\right) N_i^2 V_1^3 }{(2\pi)^2~ 16 \mu^6 v_x^4}
\end{eqnarray*}

\begin{eqnarray*}
\chi^{sk,4asym}_{abcd} &=&  \frac{\tau }{2}   \int[d\mathbf{k}]\int[d\mathbf{k'}] \epsilon^{ade}~\varpi_{kk'}^{4asym}\left\lbrace \Omega_k^e     - \Omega_{k'}^e  \right\rbrace G_{bc,{k}} ~\frac{\partial f_{k}^0(\ve_k)}{\partial \ve_k}  \notag \\ 
&&  + 2\tau^3 \int[d\mathbf{k}]\int[d\mathbf{k'}] \epsilon^{ade}~ (\partial_b^{k} \left\lbrace \varpi_{kk'}^{4asym} \left( \Omega_k^e  - \Omega_{k'}^e  \right)\right\rbrace)  \frac{\partial \ve_k}{\partial k_c} \frac{\partial f_{k}^0(\ve_k)}{\partial \ve_k}    
\end{eqnarray*}

$$\chi^{sk,4asym}_{xxxx} = \chi^{sk,4asym}_{yyyy} = \chi^{sk,4asym}_{xyyx} = \chi^{sk,4asym}_{yxxy} = 0$$

\begin{eqnarray*}
\chi^{sk,4asym}_{xxyy} &=&  \frac{\tau }{2}   ~\frac{ t^2 \pi^3 \Delta^2 (\Delta^2-\mu^2 )^2  \left(4\Delta^2-\mu ^2\right) N_i^2 V_0^4 }{(2\pi)^2~ 32 \mu^9 v_x^4  }    - 2\tau^3  ~\frac{t^2 \pi^3 \Delta^2 \left(16\Delta^6 -43 \Delta^4 \mu^2 + 32 \Delta^2 \mu^4 -5 \mu^6 \right) N_i^2 V_0^4}{(2\pi)^2~ 16 \mu^7 v_x^4}
\end{eqnarray*}

\begin{eqnarray*}
\chi^{sk,4asym}_{xyxy} &=&  \frac{\tau }{2}   ~\frac{ t^2 \pi^3 \Delta^2 (\Delta^2-\mu^2 )^2  \left(4\Delta^2-\mu ^2\right) N_i^2 V_0^4 }{(2\pi)^2~ 32 \mu^9 v_x^4  }    - 2\tau^3  ~\frac{t^2 \pi^3 \Delta^2 \left(16\Delta^6 -43 \Delta^4 \mu^2 + 32 \Delta^2 \mu^4 -5 \mu^6 \right) N_i^2 V_0^4}{(2\pi)^2~ 16 \mu^7 v_x^4}
\end{eqnarray*}

\begin{eqnarray*}
\chi^{sk,4asym}_{yyxx} &=&  -\frac{\tau }{2}   ~\frac{ t^2 \pi^3 \Delta^2 (\Delta^2-\mu^2 )^2  \left(4\Delta^2-\mu ^2\right) N_i^2 V_0^4 }{(2\pi)^2~ 32 \mu^9 v_x^4  }    + 2\tau^3   ~\frac{t^2 \pi^3 \Delta^2 \left(16\Delta^6 -43 \Delta^4 \mu^2 + 32 \Delta^2 \mu^4 -5 \mu^6 \right) N_i^2 V_0^4}{(2\pi)^2~ 16 \mu^7 v_x^4}
\end{eqnarray*}

\begin{eqnarray*}
\chi^{sk,4asym}_{yxyx} &=&  -\frac{\tau }{2}   ~\frac{ t^2 \pi^3 \Delta^2 (\Delta^2-\mu^2 )^2  \left(4\Delta^2-\mu ^2\right) N_i^2 V_0^4 }{(2\pi)^2~ 32 \mu^9 v_x^4  }    + 2\tau^3   ~\frac{t^2 \pi^3 \Delta^2 \left(16\Delta^6 -43 \Delta^4 \mu^2 + 32 \Delta^2 \mu^4 -5 \mu^6 \right) N_i^2 V_0^4}{(2\pi)^2~ 16 \mu^7 v_x^4}
\end{eqnarray*}

\begin{eqnarray*}
    &&\chi_{xxxx}^{sk} = \chi_{xxxx}^{sk,3asym} +  \chi_{xxxx}^{sk,4asym},\, \chi_{yyyy}^{sk} = \chi_{yyyy}^{sk,3asym} +  \chi_{yyyy}^{sk,4asym}, \\\\
    &&\chi_{21}^{sk} = \chi_{21}^{sk,3asym} + \chi_{21}^{sk,4asym},\, \chi_{12}^{sk} = \chi_{12}^{sk,3asym} + \chi_{12}^{sk,4asym}
\end{eqnarray*}

\begin{eqnarray*}
   \chi_{\perp}^{sk}\left(\theta\right)  &=& \left(-\chi_{xxxx}^{sk} + 3\chi_{21}^{sk}\right)\cos^3(\theta)\sin(\theta)  + \left(\chi_{yyyy}^{sk} - 3\chi_{12}^{sk} \right)\cos(\theta)\sin^3(\theta) \notag\\
   &=& \chi_{\perp}^{sk,1}\left(\theta\right) +\chi_{\perp}^{sk,2}\left(\theta\right)  
\end{eqnarray*}

\begin{eqnarray*}
\frac{\chi_{\perp}^{sk,1}\left(\theta\right)}{\tau}  &=& - \frac{t^2 \Delta^2 \left(\Delta^2 - \mu^2\right)^2 \left(5\Delta^2 - \mu^2\right)\sin(2\theta)~N_iV_1^3} {2048\pi ~v_x^4 \mu^8} - \frac{t^2 \Delta^2 \left(\Delta^2 - \mu^2\right)^2 \left(4\Delta^2 - \mu^2\right)\sin(2\theta)~N_i^2V_0^4} {1024\pi ~v_x^4 \mu^9} \\
 &=& - \frac{\tilde{t}^2 \tilde{\Delta}^2 \left(\tilde{\Delta}^2 - 1\right)^2 \sin(2\theta)} {1024\pi ~\tilde{v}_x^4 }\left[ \frac{\left(5\tilde{\Delta}^2 - 1\right)~N_iV_1^3} {2 \mu^2} + \frac{\left(4\tilde{\Delta}^2 - 1\right)~N_i^2V_0^4} {\mu^3} \right]
\end{eqnarray*}


\begin{eqnarray*}
	\frac{\chi_{\perp}^{sk,2}\left(\theta\right)}{\tau^3 }  &=& \frac{\pi t^2 \Delta^4 \left(7\Delta^4 - 18 \Delta^2\mu^2 +11\mu^4\right)\sin(2\theta)~N_iV_1^3} {32 ~v_x^4 \mu^6} +  \frac{\pi t^2 \Delta^2 \left(16\Delta^6 - 43 \Delta^4\mu^2 + 32\Delta^2\mu^4 -5\mu^6 \right)\sin(2\theta)~N_i^2V_0^4} {32 ~v_x^4 \mu^7} \\
 &=& \frac{\pi \tilde{t}^2 \tilde{\Delta}^4 \left(11 - 18 \tilde{\Delta}^2 + 7\tilde{\Delta}^4\right)\sin(2\theta)~N_iV_1^3} {32 ~\tilde{v}_x^4 } +  \frac{\pi \tilde{t}^2 \tilde{\Delta}^2 \left(-5 + 32\tilde{\Delta}^2 - 43 \tilde{\Delta}^4 + 16\tilde{\Delta}^6\right)\sin(2\theta)~N_i^2V_0^4} {32 ~\tilde{v}_x^4 \mu}
\end{eqnarray*}


Here, we have renormalized $t$, $\Delta$, and $v_x$ as $\tilde{t}=t/\mu$, $\tilde{\Delta} = \Delta/\mu$, and $\tilde{v}_x = v_x/\mu$ respectively for simplicity.

\bibliography{TOHE_Disorder}